\input amstexl
% LAMSTEX.TEX   VERSION 2.01
% COPYRIGHT (C) 1989, 1990, 1991 BY THE TEXPLORATORS CORPORATION
%  3701 W. ALABAMA, SUITE 450-273, HOUSTON, TX 77027
% ALL RIGHTS RESERVED

% ABSOLUTELY NO CHANGES SHOULD BE MADE TO THIS FILE;
% CHANGES SHOULD BE MADE ONLY IN STYLE FILES.

\catcode`\@=11
\ifx\amstexloaded@\relax\else
 \errmessage{AmS-TeX must be loaded before LamS-TeX}\fi
\ifx\laxread@\undefined\else\catcode`\@=\active \fi
\def\err@#1{\errmessage{LamS-TeX error: #1}}
\def^^L{\par}
\let\+\tabalign
\def\newcount{\alloc@0\count\countdef\insc@unt}
\def\newdimen{\alloc@1\dimen\dimendef\insc@unt}
\def\newskip{\alloc@2\skip\skipdef\insc@unt}
\def\newmuskip{\alloc@3\muskip\muskipdef\@cclvi}
\def\newbox{\alloc@4\box\chardef\insc@unt}
\let\newtoks\relax
\def\newhelp#1#2{\newtoks#1#1\expandafter{\csname#2\endcsname}}
\def\newtoks{\alloc@5\toks\toksdef\@cclvi}
\def\newread{\alloc@6\read\chardef\sixt@@n}
\def\newwrite{\alloc@7\write\chardef\sixt@@n}
\def\newfam{\alloc@8\fam\chardef\sixt@@n}
\def\newlanguage{\alloc@9\language\chardef\@cclvi}
\def\newinsert#1{\global\advance\insc@unt by\m@ne
  \ch@ck0\insc@unt\count
  \ch@ck1\insc@unt\dimen
  \ch@ck2\insc@unt\skip
  \ch@ck4\insc@unt\box
  \allocationnumber=\insc@unt
  \global\chardef#1=\allocationnumber
  \wlog{\string#1=\string\insert\the\allocationnumber}}
\def\newif#1{\count@\escapechar \escapechar\m@ne
  \expandafter\expandafter\expandafter
   \edef\@if#1{true}{\let\noexpand#1=\noexpand\iftrue}%
  \expandafter\expandafter\expandafter
   \edef\@if#1{false}{\let\noexpand#1=\noexpand\iffalse}%
  \@if#1{false}\escapechar\count@}

\def\Err@#1{\errhelp\defaulthelp@\err@{#1}}
{\catcode`\@=\active
 \edef\next{\gdef\noexpand@{\futurelet\noexpand\next
  \csname at\string@\endcsname}}
 \next
}
\def\at@{\ifcat\noexpand\next a\let\next@\at@@\else
 \ifcat\noexpand\next0\let\next@\at@@\else
 \ifcat\noexpand\next\relax\let\next@\at@@\else
 \let\next@\at@@@\fi\fi\fi\next@}
\def\at@@@{\errhelp\athelp@\err@{Invalid use of @}}
\def\at@@#1{\expandafter
 \ifx\csname\string#1@at\endcsname\relax\let\next@\at@@@\else
 \DN@{\csname\string#1@at\endcsname}\fi\next@}
\def\atdef@#1{\expandafter\def\csname\string#1@at\endcsname}
\newif\iftest@
\def\tagin@#1{\tagin@false
 \DN@##1\tag##2##3\next@{\test@true\ifx\tagin@##2\test@false\fi}%
 \next@#1\tag\tagin@\next@\tagin@false\iftest@\tagin@true\fi}
\let\lkerns@\relax
\def\nolinebreak{\RIfM@\mathmodeerr@\nolinebreak\else
 \ifhmode\saveskip@\lastskip\unskip
 \nobreak\ifdim\saveskip@>\z@\hskip\saveskip@\fi\lkerns@
 \else\vmodeerr@\nolinebreak\fi\fi}
\def\allowlinebreak{\RIfM@\mathmodeerr@\allowlinebreak\else
 \ifhmode\saveskip@\lastskip\unskip
 \allowbreak\ifdim\saveskip@>\z@\hskip\saveskip@\fi\lkerns@
 \else\vmodeerr@\allowlinebreak\fi\fi}
\def\linebreak{\RIfM@\mathmodeerr@\linebreak\else
 \ifhmode\unskip\unkern\break\lkerns@
 \else\vmodeerr@\linebreak\fi\fi}
\let\nkerns@\relax
\def\newline{\RIfM@\mathmodeerr@\newline\else
 \ifhmode\unskip\unkern\null\hfill\break\nkerns@
 \else\vmodeerr@\newline\fi\fi}%
\def\newbox@{\alloc@@4\box\chardef\insc@unt}
\def\newcount@{\alloc@@0\count\countdef\insc@unt}
\def\accentedsymbol#1#2{\expandafter\newbox@\csname\exstring@#1@box\endcsname
 \setbox\csname\exstring@#1@box\endcsname\hbox{$\m@th#2$}%
 \define#1{\copy\csname\exstring@#1@box\endcsname{}}}
\def\rightadd@#1\to#2{\toks@{\\#1}\toks@@\expandafter{#2}\xdef#2{\the\toks@@
 \the\toks@}\toks@{}\toks@@{}}
\def\fontlist@{\\\tenrm\\\sevenrm\\\fiverm\\\teni\\\seveni\\\fivei
 \\\tensy\\\sevensy\\\fivesy\\\tenex\\\tenbf\\\sevenbf\\\fivebf
 \\\tensl\\\tenit}
\def\font@#1=#2 {\rightadd@#1\to\fontlist@\font#1=#2 }
\def\ismember@#1#2{\global\let\Next@ F\let\next@= #2%
 {\def\\##1{\let\nextii@##1\ifx\nextii@\next@\global\let\Next@ T\fi}#1}%
 \test@false\ifx\Next@ T\test@true\fi\let\next@\relax}
\def\FNSS@#1{\let\FNSS@@#1\FN@\FNSS@@@}
\def\FNSS@@@{\ifx\next\space@\def\FNSS@@@@. {\FN@\FNSS@@@}\else
 \def\FNSS@@@@.{\FNSS@@}\fi\FNSS@@@@.}
\atdef@"{\unskip
 \DN@{\ifx\next`\DN@`{\FN@\nextii@}%
  \else\ifx\next\lq\DN@\lq{\FN@\nextii@}%
  \else\DN@####1{\FN@\nextiii@}\fi\fi
  \next@}%
 \DNii@{\ifx\next`\DN@`{\sldl@``}%
  \else\ifx\next\lq\DN@\lq{\sldl@``}%
  \else\DN@{\dlsl@`}\fi\fi\next@}%
 \def\nextiii@{\ifx\next'\DN@'{\srdr@''}%
  \else\ifx\next\rq\DN@\rq{\srdr@''}%
  \else\DN@{\drsr@'}\fi\fi\next@}%
 \FNSS@\next@}
\def\root{%
  \DN@{\ifx\next\uproot\let\next@\nextii@\else
   \ifx\next\leftroot\let\next@\nextiii@\else
   \let\next@\plainroot@\fi\fi\next@}%
  \DNii@\uproot##1{\uproot@##1\relax\FNSS@\nextiv@}%
  \def\nextiv@{\ifx\next\leftroot\let\next@\nextv@\else
   \let\next@\plainroot@\fi\next@}%
  \def\nextv@\leftroot##1{\leftroot@##1\relax\plainroot@}%
  \def\nextiii@\leftroot##1{\leftroot@##1\relax\FNSS@\nextvi@}%
  \def\nextvi@{\ifx\next\uproot\let\next@\nextvii@\else
   \let\next@\plainroot@\fi\next@}%
  \def\nextvii@\uproot##1{\uproot@##1\relax\plainroot@}%
  \bgroup\uproot@\z@\leftroot@\z@
 \FNSS@\next@}
\def\loop#1\repeat{\def\iterate{#1\relax\expandafter\iterate\fi}%
 \iterate\let\iterate\relax}
\def\gloop@#1\repeat{\gdef\iterate@{#1\relax\expandafter\iterate@\fi}%
 \iterate@\global\let\iterate@\relax}
\def\printoptions{\W@{Do you want S(yntax check),
  G(alleys) or P(ages)?^^JType S, G or P, follow by <return>: }\loop
 \read\m@ne to\ans@
 \edef\next@{\def\noexpand\Ans@{\ans@}}\uppercase\expandafter{\next@}%
 \ifx\Ans@\S@\test@true\syntax\else
 \ifx\Ans@\G@\test@true\galleys\else
 \ifx\Ans@\P@\test@true\else
 \test@false\fi\fi\fi
 \iftest@\else\W@{Type S, G or P, follow by <return>: }%
 \repeat}
\expandafter\let\csname A@;\endcsname;
\expandafter\let\csname A@:\endcsname:
\expandafter\let\csname A@?\endcsname?
\expandafter\let\csname A@!\endcsname!
\def\APdef#1{\def\next@{\expandafter\let\csname A@\string#1\endcsname#1}%
 \afterassignment\next@\def#1}
\let\fextra@\,
\def\tdots@{\unskip
 \DN@{$\m@th\mathinner{\ldotp\ldotp\ldotp}\,
   \ifx\next,\,$\else\ifx\next.\,$\else
   \ifx\next;\,$\else
   \expandafter\ifx\csname A@\string;\endcsname\next\fextra@$\else
   \ifx\next:\,$\else
   \expandafter\ifx\csname A@\string:\endcsname\next\fextra@$\else
   \ifx\next?\,$\else
   \expandafter\ifx\csname A@\string?\endcsname\next\fextra@$\else
   \ifx\next!\,$\else
   \expandafter\ifx\csname A@\string!\endcsname\next\fextra@$\else
   $ \fi\fi\fi\fi\fi\fi\fi\fi\fi\fi}%
 \ \FN@\next@}
\def\extrap@#1{%
 \ifx\next,\DN@{#1\,}\else
 \ifx\next;\DN@{#1\,}\else
 \expandafter\ifx\csname A@\string;\endcsname\next\DN@{#1\fextra@}\else
 \ifx\next.\DN@{#1\,}\else\extra@
 \ifextra@\DN@{#1\,}\else
 \let\next@#1\fi\fi\fi\fi\fi\next@}
\def\dotsc{\DN@{\ifx\next;\plainldots@\,\else
 \expandafter\ifx\csname A@\string;\endcsname\next\plainldots@\fextra@\else
 \ifx\next.\plainldots@\,\else\extra@\plainldots@
 \ifextra@\,\fi\fi\fi\fi}%
 \FN@\next@}
\def\keybin@{\keybin@true
 \ifx\next+\else\ifx\next=\else\ifx\next<\else\ifx\next>\else\ifx\next-\else
 \ifx\next*\else\ifx\next:\else
 \expandafter\ifx\csname A@\string;\endcsname\next\else
 \keybin@false\fi\fi\fi\fi\fi\fi\fi\fi}
\def\boldkey#1{\ifcat\noexpand#1A%
  \ifcmmibloaded@{\fam\cmmibfam#1}\else
   \Err@{First bold symbol font not loaded}\fi
 \else
 \let\next=#1%
 \ifx#1!\mathchar"5\bffam@21 \else
 \expandafter\ifx\csname A@\string!\endcsname\next\mathchar"5\bffam@21 \else
 \ifx#1(\mathchar"4\bffam@28 \else\ifx#1)\mathchar"5\bffam@29 \else
 \ifx#1+\mathchar"2\bffam@2B \else\ifx#1:\mathchar"3\bffam@3A \else
 \expandafter\ifx\csname A@\string:\endcsname\next\mathchar"3\bffam@3A \else
 \ifx#1;\mathchar"6\bffam@3B \else
 \expandafter\ifx\csname A@\string;\endcsname\next\mathchar"6\bffam@3B \else
 \ifx#1=\mathchar"3\bffam@3D \else
 \ifx#1?\mathchar"5\bffam@3F \else
 \expandafter\ifx\csname A@\string?\endcsname\next\mathchar"5\bffam@3F \else
 \ifx#1[\mathchar"4\bffam@5B \else
 \ifx#1]\mathchar"5\bffam@5D \else
 \ifx#1,\mathchari@63B \else
 \ifx#1-\mathcharii@200 \else
 \ifx#1.\mathchari@03A \else
 \ifx#1/\mathchari@03D \else
 \ifx#1<\mathchari@33C \else
 \ifx#1>\mathchari@33E \else
 \ifx#1*\mathcharii@203 \else
 \ifx#1|\mathcharii@06A \else
 \ifx#10\bold0\else\ifx#11\bold1\else\ifx#12\bold2\else\ifx#13\bold3\else
 \ifx#14\bold4\else\ifx#15\bold5\else\ifx#16\bold6\else\ifx#17\bold7\else
 \ifx#18\bold8\else\ifx#19\bold9\else
  \Err@{\noexpand\boldkey can't be used with #1}%
 \fi\fi\fi\fi\fi\fi\fi\fi\fi\fi\fi\fi\fi\fi\fi
 \fi\fi\fi\fi\fi\fi\fi\fi\fi\fi\fi\fi\fi\fi\fi\fi\fi\fi}
\def\arabic#1{#1}
\def\alph#1{\count@#1\relax\advance\count@96 \ifnum\count@>122
 \Err@{\noexpand\alph invalid for numbers > 26}\else\char\count@\fi}
\def\Alph#1{\count@#1\relax\advance\count@64 \ifnum\count@>90
 \Err@{\noexpand\Alph invalid for numbers > 26}\else\char\count@\fi}

\def\Roman#1{\uppercase\expandafter{\romannumeral#1}}
\def\fnsymbol#1{\count@#1\relax
 \count@@\count@
 \advance\count@\m@ne\divide\count@7
 \count@@@\count@\advance\count@@@\@ne
 \multiply\count@7 \advance\count@@-\count@
 \count@\count@@@
 {\loop
  \ifcase\count@@\or*\or\dag\or\ddag\or\P\or\S\or\text{$\|$}\or\#\fi
  \advance\count@\m@ne\ifnum\count@>\z@\repeat}}
\def\cardnine@#1{\ifcase#1\or one\or two\or three\or four\or five\or
 six\or seven\or eight\or nine\fi}
\let\alloc@\alloc@@
\newcount\ten@
\ten@10
\def\cardinal#1{\count@#1\relax
 \ifnum\count@>99 \number\count@
 \else
  \ifnum\count@=\z@ zero%
  \else
   \ifnum\count@<\ten@\cardnine@\count@
   \else
    \ifnum\count@<20
     \advance\count@-\ten@
     \ifcase\count@ ten\or eleven\or twelve\or thirteen\or fourteen\or
      fifteen\or sixteen\or seventeen\or eighteen\or nineteen\fi
    \else
     \count@@\count@\count@@@\count@@
     \divide\count@\ten@\multiply\count@\ten@
     \advance\count@@@-\count@\divide\count@\ten@
     \ifcase\count@\or\or twenty\or thirty\or forty\or fifty\or sixty\or
      seventy\or eighty\or ninety\fi
     \ifnum\count@@@=\z@\else-\cardnine@\count@@@\fi
    \fi
   \fi
  \fi
 \fi}
\def\ordnine@#1{\ifcase#1\or first\or second\or third\or fourth\or fifth\or
 sixth\or seventh\or eighth\or ninth\fi}
\newcount\count@@@@
\def\ordsuffix@{\count@@@@\count@
 \divide\count@\ten@
 \count@@@\count@\count@@\count@
 \divide\count@@\ten@\multiply\count@@\ten@
 \advance\count@@@-\count@@
 \ifnum\count@@@=\@ne th%
 \else
  \count@@@\count@@@@
  \count@@\count@@@@
  \divide\count@@\ten@\multiply\count@@\ten@
  \advance\count@@@-\count@@
  \ifcase\count@@@ th\or st\or nd\or rd\else th\fi
 \fi}
\def\nordinal#1{\count@#1\relax\number\count@\ordsuffix@}
\def\spordinal#1{\count@#1\relax\number\count@$^{\text{\ordsuffix@}}$}
\def\ordinal#1{\count@#1\relax
 \ifnum\count@>99 \number\count@\ordsuffix@
 \else
   \ifnum\count@=\z@ zeroth%
  \else
    \ifnum\count@<\ten@\ordnine@\count@
    \else
     \ifnum\count@<20 \advance\count@-\ten@
      \ifcase\count@ tenth\or eleventh\or twelfth\or thirteenth\or
       fourteenth\or fifteenth\or sixteenth\or seventeenth\or eighteenth\or
       nineteenth\fi
     \else
      \count@@\count@
      \divide\count@\ten@\multiply\count@\ten@
      \count@@@\count@@\advance\count@@@-\count@
      \divide\count@\ten@
      \ifcase\count@\or\or twent\or thirt\or fort\or fift\or sixt\or sevent\or
       eight\or ninet\fi
      \ifnum\count@@@=\z@ ieth\else y-\ordnine@\count@@@\fi
     \fi
    \fi
  \fi
 \fi}
\font@\tensmc=cmcsc10
\textonlyfont@\smc\tensmc
\newtoks\noexpandtoks@
\noexpandtoks@{\let\arabic\relax\let\alph\relax\let\Alph\relax
 \let\Roman\relax\let\fnsymbol\relax\let\rm\relax
 \let\it\relax\let\bf\relax\let\sl\relax\let\smc\relax
 \let\/\relax\let\null\relax}
\def\noexpands@{\the\noexpandtoks@}
\def\Nonexpanding#1{\global\noexpandtoks@
 \expandafter{\the\noexpandtoks@\let#1\relax}}
\def\prevanish@{\saveskip@\z@\ifhmode\saveskip@\lastskip\unskip\fi}
\def\postvanish@{\ifdim\saveskip@>\z@\hskip\saveskip@\fi\FN@\postvanish@@}
\def\postvanish@@{\DN@.{}%
 \ifx\next\space@\ifdim\saveskip@>\z@\DN@. {}\fi\fi\next@.}
\def\invisible#1{\prevanish@\ignorespaces#1\unskip\postvanish@}
\def\vanishlist@{\\\invisible}
\let\noindent@\noindent
\def\noindent{\par\noindent@\FN@\pretendspace@}
\def\pretendspace@{\ismember@\vanishlist@\next
 \iftest@\nobreak\hskip-\p@\hskip\p@\fi}

\newtoks\everypartoks@
\def\noindent@@{\par\everypartoks@\expandafter{\the\everypar}\everypar{}%
 \noindent@\everypar\expandafter{\the\everypartoks@}}
\def\page{\Err@{\noexpand\page has no meaning by itself}}
\let\page@C\pageno
\let\page@P\empty
\let\page@Q\empty
\def\page@S#1{#1\/}
\def\page@F{\rm}
\def\page@N{\arabic}   % cannot be \let
\newif\ifindexing@
\def\indexfile{\ifindexing@\else
 \alloc@@7\write\chardef\sixt@@n\ndx@
 \immediate\openout\ndx@=\jobname.ndx
 \global\indexing@true\fi}
\global\advance\insc@unt\m@ne
\ch@ck0\insc@unt\count
\ch@ck1\insc@unt\dimen
\ch@ck2\insc@unt\skip
\ch@ck4\insc@unt\box
\allocationnumber\insc@unt
\global\chardef\margin@\allocationnumber
\dimen\margin@\maxdimen
\count\margin@\z@
\skip\margin@\z@
\newif\ifindexproofing@
\def\indexproofing{\indexproofing@true}
\def\noindexproofing{\indexproofing@false}
\def\unmacro@#1:#2->#3\unmacro@{\def\macpar@{#2}\def\macdef@{#3}}
\def\starparts@#1{\def\stari@{#1}\def\starii@{#1}\let\stariii@\empty
 \test@false
 \DN@##1*##2##3\next@{\ifx\starparts@##2\test@false\else\test@true\fi}%
 \next@#1*\starparts@\next@
 \iftest@\DN@{\starparts@@#1\starparts@@}\else\let\next@\relax\fi\next@}
\def\starparts@@#1*#2\starparts@@{\def\starii@{#1}\def\stariii@{*#2}}
\def\windex@{\ifindexing@
 \expandafter\unmacro@\meaning\stari@\unmacro@
 \edef\macdef@{\string"\macdef@\string"}%
 \edef\next@{\write\ndx@{\macdef@}}\next@
 \write\ndx@{{\number\pageno}{\page@N}{\page@P}{\page@Q}}%
 \fi
 \ifindexproofing@
  \ifx\stariii@\empty\else
   \expandafter\unmacro@\meaning\stariii@\unmacro@\fi
  \insert\margin@{\hbox{\rm\vrule\height9\p@\depth2\p@\width\z@\starii@
  \ifx\stariii@\empty\else\tt\macdef@\fi}}\fi}
\catcode`\"=\active
\def"{\FN@\quote@}
\def\quote@{\ifx\next"\expandafter\quote@@\else\expandafter\quote@@@\fi}
\def\quote@@@#1"{\starparts@{#1}\starii@\windex@}
\def\quote@@"#1"{\prevanish@\starparts@{#1}\windex@\FN@\quote@@@@}
\def\quote@@@@{\ifx\next"\DN@"{\postvanish@}\else
 \let\next@\postvanish@\fi\next@}
\rightadd@"\to\vanishlist@
\def\idefine#1{\DN@{#1}\DNii@{\noexpand#1}%
 \afterassignment\idefine@\def\nextiii@}
\def\idefine@{\ifindexing@
 \expandafter\let\next@\nextiii@
 \expandafter\unmacro@\meaning\nextiii@\unmacro@
 \immediate\write\ndx@{\noexpand\define\nextii@\macpar@{\macdef@}}\fi}
\def\iabbrev*#1#2{\ifindexing@\toks@{#2}%
 \immediate\write\ndx@{\noexpand\abbrev*\noexpand#1{\the\toks@}}\fi}
\newread\laxread@
\newwrite\laxwrite@
\let\fnpages@\empty
\def\Finit@#1#2\Finit@{\let\nextii@#1\def\nextiii@{#2}}
\catcode`\~=11
\def\getparts@ @#1~#2~#3~#4~#5~#6{\def\nextiv@{#1}%
 \def\nextiii@{#2~#3~#4~#5~}\count@#6\relax}
\newif\ifdocument@
\def\document{\ifdocument@\else\global\document@true
 \let\fontlist@\empty
 \immediate\openin\laxread@=\jobname.lax\relax
 {\endlinechar\m@ne\noexpands@\catcode`\@=11 \catcode`\~=11
  \loop\ifeof\laxread@\else
   \read\laxread@ to\next@
   \ifx\next@\empty
   \else
    \expandafter\Finit@\next@\Finit@
    \if\nextii@ F%
     \expandafter\rightadd@\nextiii@\to\fnpages@
    \else
     \expandafter\getparts@\next@
     \edef\next@{\gdef\csname\nextiv@ @L\endcsname{\nextiii@\number\count@}}%
     \next@
    \fi
   \fi
  \repeat}%
 \immediate\closein\laxread@
 \immediate\openout\laxwrite@=\jobname.lax\relax\fi}
\let\thelabel@\relax
\def\thelabels@{\thelabel@ ~\thelabel@@ ~\thelabel@@@ ~\thelabel@@@@ ~}
\def\label#1{\prevanish@
 \ifx\thelabel@\relax
  \Err@{There's nothing here to be labelled}%
 \else
  {\noexpands@
  \expandafter\ifx\csname#1@L\endcsname\relax
   \expandafter\xdef\csname#1@L\endcsname{\thelabels@0}%
   \immediate\write\laxwrite@{@#1~\thelabels@1}%
  \else
   \edef\next@{@~\csname#1@L\endcsname}%
    \expandafter\getparts@\next@
    \ifodd\count@
    \expandafter\xdef\csname#1@L\endcsname{\thelabels@0}%
    \immediate\write\laxwrite@{@#1~\thelabels@1}%
   \else
    \Err@{Label #1 already used}%
   \fi
  \fi
  }%
 \fi
 \postvanish@}
\rightadd@\label\to\vanishlist@
\def\thepages@{\page@N{\number\page@C}~%
 \page@S{\page@P\page@N{\number\page@C}\page@Q}~%
 \number\page@C ~\page@P\page@N{\number\page@C}\page@Q ~}
\def\pagelabel#1{\prevanish@
 \expandafter\ifx\csname#1@L\endcsname\relax
  {\noexpands@
  \expandafter\xdef\csname#1@L\endcsname{\thepages@2}}%
  \write\laxwrite@{@#1~\thepages@3}%
 \else
  {\noexpands@
  \edef\next@{@~\csname#1@L\endcsname}%
  \expandafter\getparts@\next@
  \ifodd\count@
   \ifnum\count@=\@ne
    \expandafter\xdef\csname#1@L\endcsname{\thelabels@2}%
   \fi
   \write\laxwrite@{@#1~\thepages@3}%
  \else
   \Err@{Label #1 already used}%
  \fi
  }%
 \fi
 \postvanish@}
\rightadd@\pagelabel\to\vanishlist@
\newif\ifreferr@
\referr@true
\def\RefErrors{\global\referr@true}
\def\RefWarnings{\global\referr@false}
\setbox\z@\hbox{\global\count@=`^^30}
\ifnum\count@=48 \let\versionthree@\relax\fi
\def\nolabel@#1#2#3{\expandafter\ifx\csname#2@L\endcsname\relax
 \ifreferr@\Err@{No \noexpand\label found for #2}\else
 \W@{Warning: No \noexpand\label found for #2.}%
 \ifx\versionthree@\relax\W@{l.\number\inputlineno\space ... \string#1{#2}}\fi
 \fi#3\else}
\def\csL@#1{{\noexpands@\xdef\Next@{\csname#1@L\endcsname}}}
\def\ref#1{\nolabel@\ref{#1}\relax
 \DNii@##1~##2\nextii@{##1}%
 \csL@{#1}\expandafter\nextii@\Next@\nextii@\fi}
\def\Ref#1{\nolabel@\Ref{#1}\relax
 \DNii@##1~##2~##3\nextii@{##2}%
 \csL@{#1}\expandafter\nextii@\Next@\nextii@\fi}
\def\nref#1{\nolabel@\nref{#1}\relax
 \DNii@##1~##2~##3~##4\nextii@{##3}%
 \csL@{#1}\expandafter\nextii@\Next@\nextii@\fi}
\def\pref#1{\nolabel@\pref{#1}\relax
 \DNii@##1~##2~##3~##4~##5\nextii@{##4}%
 \csL@{#1}\expandafter\nextii@\Next@\nextii@\fi}
\let\pref@\pref
\def\Evaluatenref#1{\nolabel@\Evaluatenref{#1}{\gdef\Nref{-10000 }}%
 \DNii@##1~##2~##3~##4\nextii@{\DNii@{##3}}%
 \csL@{#1}\expandafter\nextii@\Next@\nextii@
 \xdef\Nref{\nextii@}\fi}
\def\Evaluatepref#1{\nolabel@\Evaluatepref{#1}{\global\let\Pref\empty}%
 \DNii@##1~##2~##3~##4~##5\nextii@{\DNii@{##4}}%
 \csL@{#1}\expandafter\nextii@\Next@\nextii@
 \xdef\Pref{\nextii@}\fi}
\def\readlax#1{\immediate\openin\laxread@=#1.lax\relax
 \ifeof\laxread@\W@{}\W@{File #1.lax not found.}\W@{}\fi
 {\endlinechar\m@ne\noexpands@\catcode`\@=11 \catcode`\~=11
  \loop\ifeof\laxread@\else
   \read\laxread@ to\nextv@
   \ifx\nextv@\empty
   \else
    \expandafter\Finit@\nextv@\Finit@
    \ifx\nextii@ F%
    \else
     \expandafter\getparts@\nextv@
     \expandafter\ifx\csname\nextiv@ @L\endcsname\relax
      \edef\next@{\gdef\csname\nextiv@ @L\endcsname
       {\nextiii@\ifnum\count@=\@ne0\else2\fi}}%
      \next@
     \else
      \Err@{Label \nextiv@\space in #1.lax already used}%
     \fi
    \fi
   \fi
  \repeat}%
 \immediate\closein\laxread@}
\catcode`\~=\active

\def\FNSSP@{\FNSS@\pretendspace@}
\everydisplay{\csname displaymath \endcsname}
\expandafter\def\csname displaymath \endcsname#1$${#1$$\FNSSP@}
\def\locallabel@{\let\thelabel@\Thelabel@\let\thelabel@@\Thelabel@@
 \let\thelabel@@@\Thelabel@@@\let\thelabel@@@@\Thelabel@@@@}
\newcount\tag@C
\tag@C\z@
\let\tag@P\empty
\let\tag@Q\empty
\def\tag@S#1{{\rm(}{#1\/}{\rm)}}
\let\tag@N\arabic
\def\tag@F{\rm}
\def\maketag@{\FN@\maketag@@}
\def\maketag@@{\ifx\next\relax\DN@\relax{\FN@\maketag@@}\else
 \ifx\next"\let\next@\maketag@@@\else
 \let\next@\maketag@@@@\fi\fi\next@}
\def\xdefThelabel@#1{\xdef\Thelabel@{#1{\Thelabel@@@}}}
\def\xdefThelabel@@#1{\xdef\Thelabel@@{#1{\Thelabel@@@@}}}
\def\maketag@@@@#1\maketag@{\global\advance\tag@C\@ne
 {\noexpands@
  \xdef\Thelabel@@@{\number\tag@C}%
  \xdefThelabel@\tag@N
  \xdef\Thelabel@@@@{\ifmathtags@$\tag@P\Thelabel@\tag@Q$\else
   \tag@P\Thelabel@\tag@Q\fi}%
  \xdefThelabel@@\tag@S
  }%
 \locallabel@
 \hbox{\tag@F\thelabel@@}%
 #1}
\def\Qlabel@#1{{\noexpands@\xdef\Thelabel@@{#1}%
 \let\style\empty\xdef\Thelabel@@@@{#1}%
 \let\pre\empty\let\post\empty\xdef\Thelabel@{#1}%
 \let\numstyle\empty\xdef\Thelabel@@@{#1}}}
\def\maketag@@@"#1"#2\maketag@{%
 {\let\pre\tag@P\let\post\tag@Q\let\style\tag@S\let\numstyle\tag@N
  \hbox{\tag@F#1}%
  \noexpands@
  \Qlabel@{#1}%
  }%
 \locallabel@
 #2}
\def\align@{\inalign@true\inany@true
 \vspace@\allowdisplaybreak@\displaybreak@\intertext@
 \def\tag{\global\tag@true\ifnum\and@=\z@
  \DN@{&\omit\global\rwidth@\z@&\relax}\else
  \DN@{&\relax}\fi\next@}%
 \iftagsleft@\DN@{\csname align \endcsname}\else
  \DN@{\csname align \space\endcsname}\fi\next@}
\def\noset@{\def\Offset##1##2{\prevanish@\postvanish@}%
 \def\Reset##1##2{\prevanish@\postvanish@}}
\def\measure@#1\endalign{\global\lwidth@\z@\global\rwidth@\z@
 \global\maxlwidth@\z@\global\maxrwidth@\z@
 \global\and@\z@
 \setbox\z@\vbox
  {\noset@\everycr{\noalign{\global\tag@false\global\and@\z@}}\Let@
  \halign{\setboxz@h{$\m@th\displaystyle{\@lign##}$}%
   \global\lwidth@\wdz@
   \ifdim\lwidth@>\maxlwidth@\global\maxlwidth@\lwidth@\fi
   \global\advance\and@\@ne
   &\setboxz@h{$\m@th\displaystyle{{}\@lign##}$}\global\rwidth@\wdz@
   \ifdim\rwidth@>\maxrwidth@\global\maxrwidth@\rwidth@\fi
   \global\advance\and@\@ne
   &\Tag@\eat@{##}\crcr#1\crcr}}%
 \totwidth@\maxlwidth@\advance\totwidth@\maxrwidth@}
\def\prepost@{\global\let\tag@P@\tag@P\global\let\tag@Q@\tag@Q}
\def\reprepost@{\let\tag@P\tag@P@\let\tag@Q\tag@Q@}
\expandafter\def\csname align \space\endcsname#1\endalign
 {\measure@#1\endalign\global\and@\z@
 \ifingather@\everycr{\noalign{\global\and@\z@}}\else\displ@y@\fi
 \Let@\tabskip\centering@
 \halign to\displaywidth
  {\hfil\strut@\setboxz@h{$\m@th\displaystyle{\@lign##\prepost@}$}%
  \boxz@\global\advance\and@\@ne
  \tabskip\z@skip
  &\setboxz@h{$\m@th\displaystyle{{}\@lign##\prepost@}$}%
  \global\rwidth@\wdz@\boxz@\hfil\global\advance\and@\@ne
  \tabskip\centering@
  &\setboxz@h{\@lign\strut@\reprepost@\maketag@##\maketag@}%
  \dimen@\displaywidth\advance\dimen@-\totwidth@
  \divide\dimen@\tw@\advance\dimen@\maxrwidth@\advance\dimen@-\rwidth@
  \ifdim\dimen@<\tw@\wdz@\llap{\vtop{\normalbaselines\null\boxz@}}%
  \else\llap{\boxz@}\fi
  \tabskip\z@skip
  \crcr#1\crcr
  \black@\totwidth@}}
\expandafter\def\csname align \endcsname#1\endalign{\measure@#1\endalign
 \global\and@\z@
 \ifdim\totwidth@>\displaywidth\let\displaywidth@\totwidth@\else
  \let\displaywidth@\displaywidth\fi
 \ifingather@\everycr{\noalign{\global\and@\z@}}\else\displ@y@\fi
 \Let@\tabskip\centering@\halign to\displaywidth
  {\hfil\strut@\setboxz@h{$\m@th\displaystyle{\@lign##\prepost@}$}%
  \global\lwidth@\wdz@\global\lineht@\ht\z@
  \boxz@\global\advance\and@\@ne
  \tabskip\z@skip&\setboxz@h{$\m@th\displaystyle{{}\@lign##\prepost@}$}%
  \ifdim\ht\z@>\lineht@\global\lineht@\ht\z@\fi
  \boxz@\hfil\global\advance\and@\@ne
  \tabskip\centering@&\kern-\displaywidth@
  \setboxz@h{\@lign\strut@\reprepost@\maketag@##\maketag@}%
  \dimen@\displaywidth\advance\dimen@-\totwidth@
  \divide\dimen@\tw@\advance\dimen@\maxlwidth@\advance\dimen@-\lwidth@
  \ifdim\dimen@<\tw@\wdz@
   \rlap{\vbox{\normalbaselines\boxz@\vbox to\lineht@{}}}\else
   \rlap{\boxz@}\fi
  \tabskip\displaywidth@\crcr#1\crcr\black@\totwidth@}}
\def\attag@#1{\let\Maketag@\maketag@\let\TAG@\Tag@
 \let\Prepost@\prepost@\let\Reprepost@\reprepost@
 \let\Tag@\relax\let\maketag@\relax
 \let\prepost@\relax\let\reprepost@\relax
 \ifmeasuring@
  \def\llap@##1{\setboxz@h{##1}\hbox to\tw@\wdz@{}}%
  \def\rlap@##1{\setboxz@h{##1}\hbox to\tw@\wdz@{}}%
 \else\let\llap@\llap\let\rlap@\rlap\fi
 \toks@{\hfil\strut@
  $\m@th\displaystyle{\@lign\the\hashtoks@\prepost@}$%
  \tabskip\z@skip\global\advance\and@\@ne&
  $\m@th\displaystyle{{}\@lign\the\hashtoks@\prepost@}$\hfil
  \ifxat@\tabskip\centering@\fi\global\advance\and@\@ne}%
 \iftagsleft@
  \toks@@{\tabskip\centering@&\Tag@\kern-\displaywidth
   \rlap@{\@lign\reprepost@\maketag@\the\hashtoks@\maketag@}%
   \global\advance\and@\@ne\tabskip\displaywidth}\else
  \toks@@{\tabskip\centering@&\Tag@\llap@{\@lign\reprepost@\maketag@
   \the\hashtoks@\maketag@}\global\advance\and@\@ne\tabskip\z@skip}\fi
 \atcount@#1\relax\advance\atcount@\m@ne
 \loop\ifnum\atcount@>\z@
  \toks@\expandafter{\the\toks@&\hfil$\m@th\displaystyle{\@lign
  \the\hashtoks@\prepost@}$\global\advance\and@\@ne
  \tabskip\z@skip
  &$\m@th\displaystyle{{}\@lign\the\hashtoks@\prepost@}$\hfil\ifxat@
  \tabskip\centering@\fi\global\advance\and@\@ne}\advance\atcount@\m@ne
 \repeat
 \edef\preamble@{\the\toks@\the\toks@@}%
 \edef\preamble@@{\preamble@}%
 \let\maketag@\Maketag@\let\Tag@\TAG@
 \let\prepost@\Prepost@\let\reprepost@\Reprepost@}
\def\unlabel@{\def\label##1{\prevanish@\postvanish@}%
 \def\pagelabel##1{\prevanish@\postvanish@}}
\newcount\tag@CC
\expandafter\def\csname alignat \endcsname#1#2\endalignat
 {\inany@true\xat@false
 \def\tag{\global\tag@true
  \count@#1\relax\multiply\count@\tw@\advance\count@\m@ne
  \gdef\tag@{&}%
  \loop\ifnum\count@>\and@\xdef\tag@{&\omit\tag@}%
  \advance\count@\m@ne\repeat
  \tag@\relax}%
 \vspace@\allowdisplaybreak@\displaybreak@\intertext@
 \displ@y@\measuring@true\tag@CC\tag@C
 \setbox\savealignat@\hbox{\noset@\unlabel@$\m@th\displaystyle\Let@
  \attag@{#1}\vbox{\halign{\span\preamble@@\crcr#2\crcr}}$}%
 \measuring@false
 \Let@\attag@{#1}\tag@C\tag@CC
 \tabskip\centering@\halign to\displaywidth
  {\span\preamble@@\crcr#2\crcr\black@{\wd\savealignat@}}}
\expandafter\def\csname xalignat \endcsname#1#2\endxalignat
 {\inany@true\xat@true
 \def\tag{\global\tag@true
  \count@#1\relax\multiply\count@\tw@\advance\count@\m@ne
  \gdef\tag@{&}%
  \loop\ifnum\count@>\and@\xdef\tag@{&\omit\tag@}%
  \advance\count@\m@ne\repeat
  \tag@\relax}%
 \vspace@\allowdisplaybreak@\displaybreak@\intertext@
 \displ@y@\measuring@true\tag@CC\tag@C
 \setbox\savealignat@\hbox{\noset@\unlabel@$\m@th\displaystyle\Let@
  \attag@{#1}\vbox{\halign{\span\preamble@@\crcr#2\crcr}}$}%
 \measuring@false\Let@\attag@{#1}\tag@C\tag@CC
 \tabskip\centering@\halign to\displaywidth
 {\span\preamble@@\crcr#2\crcr\black@{\wd\savealignat@}}}
\def\gather{\RIfMIfI@\DN@{\onlydmatherr@\gather}\else
 \ingather@true\inany@true\def\tag{&\relax}%
 \vspace@\allowdisplaybreak@\displaybreak@\intertext@
 \displ@y\Let@
 \iftagsleft@\DN@{\csname gather \endcsname}\else
  \DN@{\csname gather \space\endcsname}\fi\fi
 \else\DN@{\onlydmatherr@\gather}\fi\next@}
\def\exstring@{\expandafter\eat@\string}
\def\newcounter#1{\define#1{}%
 \edef\next@{\def\noexpand#1{\futurelet\noexpand\next
  \csname\exstring@#1@Z\endcsname}}\next@
 \edef\next@{\def\csname\exstring@#1@Z\endcsname
  {\global\advance\csname\exstring@#1@C\endcsname\@ne
  {\csname\exstring@#1@F\endcsname\csname\exstring@#1@S\endcsname
   {\csname\exstring@#1@P\endcsname\csname\exstring@#1@N\endcsname
   {\noexpand\number\csname\exstring@#1@C\endcsname}%
   \csname\exstring@#1@Q\endcsname}}%
  \noexpand\ifx\noexpand\next\noexpand\label
   \def\noexpand\next@\noexpand\label########1{{\noexpand\noexpands@
    \xdef\noexpand\Thelabel@{\csname\exstring@#1@N\endcsname
     {\noexpand\number\csname\exstring@#1@C\endcsname}}%
    \xdef\noexpand\Thelabel@@@{\noexpand\number
     \csname\exstring@#1@C\endcsname}%
    \xdef\noexpand\Thelabel@@{\csname\exstring@#1@S\endcsname
     {\csname\exstring@#1@P\endcsname
     \csname\exstring@#1@N\endcsname
     {\noexpand\number\csname\exstring@#1@C\endcsname}%
     \csname\exstring@#1@Q\endcsname}}%
    \xdef\noexpand\Thelabel@@@@{\csname\exstring@#1@P\endcsname
     \csname\exstring@#1@N\endcsname
     {\noexpand\number\csname\exstring@#1@C\endcsname}%
     \csname\exstring@#1@Q\endcsname}}%
    {\noexpand\locallabel@\noexpand\label{########1}}}%
   \noexpand\else\let\noexpand\next@\relax\noexpand\fi\noexpand\next@}}\next@
 \expandafter\newcount@\csname\exstring@#1@C\endcsname
 \expandafter\let\csname\exstring@#1@N\endcsname\arabic
 \expandafter\def\csname\exstring@#1@S\endcsname##1{##1\/}%
 \expandafter\let\csname\exstring@#1@P\endcsname\empty
 \expandafter\let\csname\exstring@#1@Q\endcsname\empty
 \expandafter\def\csname\exstring@#1@F\endcsname{\rm}%
 }
\def\HASH@#1#2{\ifnum#2=\z@\else
 \edef\next@{\toks@{\the\toks@\the\hashtoks@#2}%
 \toks@@{\the\toks@@{\the\hashtoks@#2}}}\next@\expandafter\HASH@\fi}
\def\HASH@@{\toks@{}\toks@@{}\expandafter\HASH@\macpar@00}
\def\usecounter#1#2{\expandafter\ifx\csname\exstring@#1@Z\endcsname
 \relax\Err@{\noexpand#1not created with \string\newcounter}\fi
 \expandafter\let\csname\exstring@#1@@Z\endcsname\relax
 \expandafter\let\csname\exstring@#1@@Z@\endcsname\relax
 \expandafter\let\csname\exstring@#1@@Z@@\endcsname\relax
 \edef\next@{\def\noexpand#2{\futurelet\noexpand\next
  \csname\exstring@#1@@Z\endcsname}}\next@
 \edef\next@{\def\csname\exstring@#1@@Z\endcsname{\noexpand\ifx
  \noexpand\next\noexpand\label\def\noexpand\next@\noexpand\label
   ########1{\csname\exstring@#1@@Z@\endcsname
   {\noexpand#1\noexpand\label{########1}}}%
   \noexpand\else\noexpand\ifx\noexpand\next
   \noexpand"\def\noexpand\next@\noexpand"########1\noexpand"%
   {\csname\exstring@#1@@Z@\endcsname{{\expandafter\noexpand
   \csname\exstring@#1@F\endcsname
   \let\noexpand\pre\expandafter\noexpand\csname\exstring@#1@P\endcsname
   \let\noexpand\post\expandafter\noexpand\csname\exstring@#1@Q\endcsname
   \let\noexpand\style\expandafter\noexpand\csname\exstring@#1@S\endcsname
   \let\noexpand\numstyle\expandafter\noexpand\csname\exstring@#1@N\endcsname
   ########1}}}\noexpand\else
   \def\noexpand\next@{\csname\exstring@#1@@Z@\endcsname{\noexpand#1}}%
   \noexpand\fi\noexpand\fi\noexpand\next@}}\next@
 \def\next@{\expandafter\expandafter\expandafter\unmacro@\expandafter
  \meaning\csname\exstring@#1@@Z@@\endcsname\unmacro@
  \HASH@@
  \edef\next@{\def\csname\exstring@#1@@Z@\endcsname\the\toks@{%
   \expandafter\noexpand\csname\exstring@#1@@Z@@\endcsname\the\toks@@
   \noexpand\FNSSP@}}\next@}%
 \afterassignment\next@
 \expandafter\def\csname\exstring@#1@@Z@@\endcsname}
\def\listbi@{\penalty50 \medskip}
\def\listbii@{\penalty100 \smallskip}
\let\listbiii@\relax
\let\listbiv@\relax
\let\listbv@\relax
\def\listmi@{\advance\leftskip30\p@\relax}
\let\listmii@\listmi@
\let\listmiii@\listmi@
\let\listmiv@\listmi@
\let\listmv@\listmi@
\def\itemi@#1{\noindent@@\llap{#1\hskip5\p@}}
\let\itemii@\itemi@
\let\itemiii@\itemi@
\let\itemiv@\itemi@
\let\itemv@\itemi@
\def\liste@{\penalty-50 \medskip}
\def\listei@{\penalty-100 \smallskip}
\let\listeii@\relax
\let\listeiii@\relax
\let\listeiv@\relax
\expandafter\newcount\csname list@C1\endcsname
\csname list@C1\endcsname\z@
\expandafter\newcount\csname list@C2\endcsname
\csname list@C2\endcsname\z@
\expandafter\newcount\csname list@C3\endcsname
\csname list@C3\endcsname\z@
\expandafter\newcount\csname list@C4\endcsname
\csname list@C4\endcsname\z@
\expandafter\newcount\csname list@C5\endcsname
\csname list@C5\endcsname\z@
\expandafter\let\csname list@P1\endcsname\empty
\expandafter\let\csname list@P2\endcsname\empty
\expandafter\let\csname list@P3\endcsname\empty
\expandafter\let\csname list@P4\endcsname\empty
\expandafter\let\csname list@P5\endcsname\empty
\expandafter\let\csname list@Q1\endcsname\empty
\expandafter\let\csname list@Q2\endcsname\empty
\expandafter\let\csname list@Q3\endcsname\empty
\expandafter\let\csname list@Q4\endcsname\empty
\expandafter\let\csname list@Q5\endcsname\empty
\expandafter\def\csname list@S1\endcsname#1{{\rm(}{#1\/}{\rm)}}
\expandafter\def\csname list@S2\endcsname#1{{\rm(}{#1\/}{\rm)}}
\expandafter\def\csname list@S3\endcsname#1{{\rm(}{#1\/}{\rm)}}
\expandafter\def\csname list@S4\endcsname#1{{\rm(}{#1\/}{\rm)}}
\expandafter\def\csname list@S5\endcsname#1{{\rm(}{#1\/}{\rm)}}
\expandafter\let\csname list@N1\endcsname\arabic
\expandafter\let\csname list@N2\endcsname\arabic
\expandafter\let\csname list@N3\endcsname\arabic
\expandafter\let\csname list@N4\endcsname\arabic
\expandafter\let\csname list@N5\endcsname\arabic
\expandafter\def\csname list@F1\endcsname{\rm}
\expandafter\def\csname list@F2\endcsname{\rm}
\expandafter\def\csname list@F3\endcsname{\rm}
\expandafter\def\csname list@F4\endcsname{\rm}
\expandafter\def\csname list@F5\endcsname{\rm}
\newcount\listlevel@
\listlevel@\z@
\def\list@@C{\csname list@C\number\listlevel@\endcsname}
\def\list@@P{\csname list@P\number\listlevel@\endcsname}
\def\list@@Q{\csname list@Q\number\listlevel@\endcsname}
\def\list@@S{\csname list@S\number\listlevel@\endcsname}
\def\list@@N{\csname list@N\number\listlevel@\endcsname}
\def\list@@F{\csname list@F\number\listlevel@\endcsname}
\newif\iffirstitemi@
\newif\iffirstitemii@
\newif\iffirstitemiii@
\newif\iffirstitemiv@
\newif\iffirstitemv@
\def\Firstitem@true{\csname firstitem\romannumeral\listlevel@
 @true\endcsname}
\def\Firstitem@false{\csname firstitem\romannumeral\listlevel@
 @false\endcsname}
\def\Listm@{\csname listm\romannumeral\listlevel@ @\endcsname}
\def\Item@{\csname item\romannumeral\listlevel@ @\endcsname}
\def\Liste@{\csname liste\romannumeral\listlevel@ @\endcsname}
\newif\iflistcontinue@
\def\keepitem{\listcontinue@true}
\newcount\list@C@
\def\list{%
 \iflistcontinue@\csname list@C1\endcsname\csname list@C@\endcsname\fi
 \global\csname list@C2\endcsname\z@
 \global\csname list@C3\endcsname\z@
 \global\csname list@C4\endcsname\z@
 \global\csname list@C5\endcsname\z@
 \begingroup
 \firstitemi@true
 \listlevel@\@ne
 \def\item{\FN@\item@}%
 \FN@\list@}
\Invalid@\runinitem
\def\list@{\ifx\next\par
 \DN@\par{\FN@\list@}\else
 \ifx\next\runinitem
  \DN@\runinitem{\FN@\runinitem@}\else
  \DN@{\par\dimen@\parskip\parskip\dimen@}\fi\fi\next@}
\newif\ifoutlevel@
\newif\ifrunin@
\def\item@{%
 \ifoutlevel@\Liste@\outlevel@false\fi
 \ifrunin@\runin@false\par
  \dimen@\parskip\parskip\dimen@
  \Listm@\fi
 \iffirstitemi@\listbi@\listmi@\firstitemi@false\else\par\fi
 \iffirstitemii@\listbii@\listmii@\firstitemii@false\else\par\fi
 \iffirstitemiii@\listbiii@\listmiii@\firstitemiii@false\else\par\fi
 \iffirstitemiv@\listbiv@\listmiv@\firstitemiv@false\else\par\fi
 \iffirstitemv@\listbv@\listmv@\firstitemv@false\else\par\fi
 \DN@"##1"{{\let\pre\list@@P\let\post\list@@Q
  \let\style\list@@S\let\numstyle\list@@N
  \vskip-\parskip
  \Item@{\list@@F##1}%
  \noexpands@
  \Qlabel@{##1}}%
  \locallabel@
  \FNSSP@}%
 \DNii@{\global\advance\list@@C\@ne
  {\noexpands@
   \xdef\Thelabel@@@{\number\list@@C}%
   \xdefThelabel@\list@@N
   \xdef\Thelabel@@@@{\list@@P\Thelabel@\list@@Q}%
   \xdefThelabel@@\list@@S
  }%
  \locallabel@
  \vskip-\parskip
  \Item@{\list@@F\thelabel@@}%
  \FN@\pretendspace@}%
 \ifx\next"\expandafter\next@\else\expandafter\nextii@\fi}
\def\runinitem@{%
  \runin@true
  \Firstitem@false
  \DN@"##1"{{\let\pre\list@@P\let\post\list@@Q
   \let\style\list@@S\let\numstyle\list@@N
   \unskip\space{\list@@F##1} %
   \noexpands@
   \Qlabel@{##1}}%
   \locallabel@
   \ignorespaces}%
  \DNii@{\global\advance\list@@C\@ne
   {\noexpands@
    \xdef\Thelabel@@@{\number\list@@C}%
    \xdefThelabel@\list@@N
    \xdef\Thelabel@@@@{\list@@P\Thelabel@\list@@Q}%
    \xdefThelabel@@\list@@S
   }%
   \locallabel@
   \unskip\space{\list@@F\thelabel@@} }%
  \ifx\next"\expandafter\next@\else\expandafter\nextii@\fi}
\def\inlevel{\ifnum\listlevel@=5
 \DN@{\Err@{Already 5 levels down}}\else
 \DN@{\begingroup\advance\listlevel@\@ne
 \Firstitem@true\FN@\inlevel@}\fi\next@}
\def\inlevel@{\ifx\next\par
 \DN@\par{\FN@\inlevel@}\else
 \ifx\next\runinitem
  \DN@\runinitem{\FN@\runinitem@}\else
  \let\next@\relax\fi\fi\next@}
\def\outlevel{\ifnum\listlevel@=\@ne
 \Err@{At top level}\else
 \par\global\list@@C\z@\endgroup\outlevel@true\fi}
\def\endlist{%
 \expandafter\global\csname list@C@\endcsname\csname list@C1\endcsname
 \par
 \global\toks\@ne{}\count@\listlevel@
 {\loop
  \ifnum\count@>\z@\global\toks\@ne\expandafter{\the\toks\@ne\endgroup}%
  \advance\count@\m@ne
  \repeat}%
 \the\toks\@ne
 \liste@
 \listcontinue@false\global\csname list@C1\endcsname\z@
 \vskip-\parskip
 \noindent@@
 \FN@\pretendspace@}
\newif\iffirstdescribe@
\def\describe{\par
 \begingroup\firstdescribe@true
 \def\item##1{%
  \iffirstdescribe@\penalty50 \medskip\vskip-\parskip
  \firstdescribe@false\else\par\fi
  \noindent@@\hangindent2pc\hangafter\@ne
  {\bf##1}\hskip.5em}}

\Invalid@\pullin
\Invalid@\pullinmore
\newif\iffirstpull@
\def\margins{\par\begingroup\firstpull@true
 \def\pullin##1##2{\par
  \iffirstpull@\firstpull@false\else\endgroup\fi
  \begingroup\DN@{##1}%
  \ifx\next@\empty\leftskip\z@\else\ifx\next@\space\leftskip\z@
  \else\leftskip##1\fi\fi
  \DN@{##2}\ifx\next@\empty\rightskip\z@\else\ifx\next@\space
  \rightskip\z@\else\rightskip##2\fi\fi\ignorespaces}%
 \def\pullinmore##1##2{\par
  \xdef\Next@{\leftskip\the\leftskip\relax\rightskip\the\rightskip\relax}%
  \iffirstpull@\firstpull@false\else\endgroup\fi
  \begingroup\Next@
  \DN@{##1}%
  \ifx\next@\empty\else\ifx\next@\space\else\advance\leftskip##1\fi\fi
  \DN@{##2}\ifx\next@\empty\else\ifx\next@\space\else
  \advance\rightskip##2\fi\fi\ignorespaces}}

\newif\ifnopunct@
\newif\ifnospace@
\newif\ifoverlong@
\let\nofrillslist@\empty
\let\overlonglist@\empty
\def\nopunct{\nopunct@true\FN@\nopunct@}
\def\nospace{\nospace@true\FN@\nospace@}
\def\overlong{\overlong@true\FN@\overlong@}
\def\nopunct@{\ifx\next\nospace
 \DN@\nospace{\nospace@true\FN@\nopnos@}\else\ifx\next\overlong
 \DN@\overlong{\overlong@true\FN@\nopol@}\else
 \let\next@\nopunct@@\fi\fi\next@}
\def\nopunct@@#1{\ismember@\nofrillslist@#1%
 \iftest@\let\next@#1\else
 \DN@{\nopunct@false\Err@{\noexpand\nopunct can't be used with
 \string#1}#1}\fi\next@}
\def\nospace@{\ifx\next\nopunct
 \DN@\nopunct{\nopunct@true\FN@\nopnos@}\else\ifx\next\overlong
 \DN@\overlong{\overlong@true\FN@\nosol@}\else
 \let\next@\nospace@@\fi\fi\next@}
\def\nospace@@#1{\ismember@\nofrillslist@#1%
 \iftest@\let\next@#1\else
 \DN@{\nospace@false\Err@{\noexpand\nospace can't be used with
 \string#1}#1}\fi\next@}
\def\overlong@{\ifx\next\nopunct
 \DN@\nopunct{\nopunct@true\FN@\nopol@}\else\ifx\next\nospace
 \DN@\nospace{\nospace@true\FN@\nosol@}\else
 \let\next@\overlong@@\fi\fi\next@}
\def\overlong@@#1{\ismember@\overlonglist@#1%
 \iftest@\let\next@#1\else
 \DN@{\overlong@false\Err@{\noexpand\overlong can't be used with
 \string#1}#1}\fi\next@}
\def\nopnos@{\ifx\next\overlong
 \DN@\overlong{\overlong@true\nopnosol@}\else
 \let\next@\nopnos@@\fi\next@}
\def\nopol@{\ifx\next\nospace
 \DN@\nospace{\nospace@true\nopnosol@}\else
 \let\next@\nopol@@\fi\next@}
\def\nosol@{\ifx\next\nopunct
 \DN@\nopunct{\nopunct@true\nopnosol@}\else
 \let\next@\nosol@@\fi\next@}
\def\nopnos@@#1{\ismember@\nofrillslist@#1%
 \iftest@\let\next@#1\else
 \DN@{\nopunct@false\nospace@false
  \Err@{\noexpand\nopunct\noexpand\nospace
   can't be used with \string#1}#1}\fi\next@}
\def\testii@#1{\ismember@\nofrillslist@#1%
 \iftest@\let\nextiii@ T\else\let\nextiii@ F\fi
 \ismember@\overlonglist@#1%
 \iftest@\let\nextiv@ T\else\let\nextiv@ F\fi
 \test@false\if\nextiii@ T\if\nextiv@ T\test@true\fi\fi}
\def\nopol@@#1{\testii@{#1}%
 \iftest@\let\next@#1%
 \else\DN@{\if\nextiii@ T\else\nopunct@false\fi
  \if\nextiv@ T\else\overlong@false\fi
  \Err@{\if\nextiii@ T\else\noexpand\nopunct\fi
  \if\nextiv@ T\else\noexpand\overlong\fi can't be used
  with \string#1}#1}\fi\next@}
\def\nosol@@#1{\testii@{#1}%
 \iftest@\let\next@#1%
 \else\DN@{\if\nextiii@ T\else\nospace@false\fi
  \if\nextiv@ T\else\overlong@false\fi
  \Err@{\if\nextiii@ T\else\noexpand\nospace\fi
  \if\nextiv@ T\else\noexpand\overlong\fi can't be used
  with \string#1}#1}\fi\next@}
\def\nopnosol@#1{\testii@{#1}%
 \iftest@\let\next@#1%
 \else\DN@{\if\nextiii@ T\else\nopunct@false\nospace@false\fi
  \if\nextiv@ T\else\overlong@false\fi
  \Err@{\if\nextiii@ T\else\noexpand\nopunct\noexpand\nospace\fi
  \if\nextiv@ T\else\noexpand\overlong\fi can't be used
  with \string#1}#1}\fi\next@}
\def\punct@#1{\ifnopunct@\else#1\fi}
\def\addspace@#1{\ifnospace@\else#1\fi}
\def\hss@{\ifoverlong@\z@ plus\@m\p@ minus\@m\p@
 \else \z@ plus\@m\p@\fi}
\rightadd@\demo\to\nofrillslist@
\newif\ifclaim@
\def\exxx@{\expandafter\expandafter\expandafter\eat@\expandafter\string}
\let\colon@:
\def\demo#1{\ifclaim@
 \Err@{Previous \expandafter\noexpand\claimtype@ has
  no matching \string\end\exxx@\claimtype@}%
 \let\next@\relax
 \else
  \par
  \ifdim\lastskip<\smallskipamount\removelastskip\smallskip\fi
  \begingroup
  \noindent@@{\smc\ignorespaces#1\unskip
   \punct@{\null\colon@}\addspace@\enspace}%
  \nopunct@false\nospace@false
  \rm
  \DN@{\FNSSP@}%
 \fi
 \next@}
\def\enddemo{\par\endgroup\nopunct@false\nospace@false\smallskip}
\rightadd@\claim\to\nofrillslist@
\def\claim@F{\smc}
\def\claim@@@F{\csname\exxx@\claimtype@ @F\endcsname}
\def\claimformat@#1#2#3{%
 \medbreak\noindent@@{\smc#1 {\claim@@@F#2} #3%
 \punct@{\null.}\addspace@\enspace}\sl}
\def\claimformat@@#1#2{\claimformat@{\ignorespaces#1\unskip}%
 {\ifx\thelabel@@\empty\unskip\else\thelabel@@\fi}%
 {\ignorespaces#2\unskip}%
 \let\Claimformat@@\claimformat@@\FNSSP@}
\let\Claimformat@@\claimformat@@
\def\claim@@@P{\csname\exxx@\claimtype@ @P\endcsname}
\def\claim@@@Q{\csname\exxx@\claimtype@ @Q\endcsname}
\def\claim@@@S{\csname\exxx@\claimtype@ @S\endcsname}
\def\claim@@@N{\csname\exxx@\claimtype@ @N\endcsname}
\def\claim@@@C{\csname claim@C\claimclass@\endcsname}
\newcount\claim@C
\claim@C\z@
\let\claim@P\empty
\let\claim@Q\empty
\def\claim@S#1{#1\/}
\let\claim@N\arabic
\def\claim{\claim@true\let\claimclass@\empty
 \def\claimtype@{\claim}\FN@\claim@}
\def\claim@{%
 \ifx\next\c
  \let\next@\claim@c
 \else
  \ifx\next"%
   \let\next@\claim@q
  \else
   \begingroup\global\advance\claim@C\@ne
   {\noexpands@
    \xdef\Thelabel@@@{\number\claim@C}%
    \xdefThelabel@\claim@N
    \xdef\Thelabel@@@@{\claim@P\Thelabel@\claim@Q}%
    \xdefThelabel@@\claim@S
   }%
   \locallabel@
   \let\next@\Claimformat@@
  \fi
 \fi
 \next@}
\def\claim@c\c#1{\claim@true\begingroup
 \expandafter
 \ifx\csname claim@C#1\endcsname\relax
  \expandafter\newcount@\csname claim@C#1\endcsname
  \global\csname claim@C#1\endcsname\@ne
 \else
  \global\advance\csname claim@C#1\endcsname\@ne
 \fi
 \def\claimclass@{#1}%
 {\noexpands@
  \xdef\Thelabel@@@{\number\claim@@@C}%
  \xdefThelabel@\claim@@@N
  \xdef\Thelabel@@@@{\claim@@@P\Thelabel@\claim@@@Q}%
  \xdefThelabel@@\claim@@@S
 }%
 \locallabel@
 \FNSS@\claim@c@}
\def\claim@q"#1"{\begingroup
 {\let\pre\claim@@@P\let\post\claim@@@Q
  \let\style\claim@@@S\let\numstyle\claim@@@N
  \noexpands@
  \Qlabel@{#1}}%
 \locallabel@
 \FNSS@\claim@q@}
\def\claim@c@{\ifx\next"%
 \global\advance\claim@@@C\m@ne\let\next@\claim@cq
 \else\let\next@\Claimformat@@\fi\next@}
\def\claim@cq"#1"{{\let\pre\claim@@@P\let\post\claim@@@Q
 \let\style\claim@@@S\let\numstyle\claim@@@N
 \noexpands@
 \Qlabel@{#1}}%
 \locallabel@
 \FNSS@\Claimformat@@}
\def\claim@q@{\ifx\next\c\expandafter\claim@qc
 \else\expandafter\Claimformat@@\fi}
\def\claim@qc\c#1{\expandafter\ifx\csname claim@C#1\endcsname\relax
 \expandafter\newcount@\csname claim@C#1\endcsname
 \global\csname claim@C#1\endcsname\z@\fi
 \FNSS@\Claimformat@@}
\def\endclaim{\endgroup\claim@false\nopunct@false\nospace@false
 \let\Claimformat@@\claimformat@@\medbreak}
\Invalid@\claimclause
\def\newclaim{\FN@\newclaim@}
\def\newclaim@{\ifx\next\claimclause
 \DN@\claimclause##1{\newclaim@@{##1}}\else
 \DN@{\newclaim@@\relax}\fi\next@}
\def\claimlist@{\\\claim}
\newtoks\claim@i
\newtoks\claim@v
\let\noclaimclause@=F
\def\newclaim@@#1#2#3\c#4#5{\define#2{}%
 \rightadd@#2\to\claimlist@\rightadd@#2\to\nofrillslist@%
 \expandafter\def\csname\exstring@#2@P\endcsname{\claim@P}%
 \expandafter\def\csname\exstring@#2@Q\endcsname{\claim@Q}%
 \expandafter\def\csname\exstring@#2@S\endcsname{\claim@S}%
 \expandafter\def\csname\exstring@#2@N\endcsname{\claim@N}%
 \expandafter\def\csname\exstring@#2@F\endcsname{\claim@F}%
 \expandafter\def\csname end\exstring@#2\endcsname{\endclaim}%
 \expandafter\ifx\csname claim@C#4\endcsname\relax
  \expandafter\newcount@\csname claim@C#4\endcsname
  \global\csname claim@C#4\endcsname\z@\fi
 \edef\next@{\let\csname\exstring@#2@C\endcsname
   \csname claim@C#4\endcsname}\next@
 \def#2{\ifx\noclaimclause@ T\else#1\fi
  \global\claim@i{#1}\gdef\claim@iv{#4}\global\claim@v{#5}%
  \def\claimtype@{#2}\def\Claimformat@@{\claimformat@@{#5}}\claim@c\c{#4}}}
\def\shortenclaim#1#2{\define#2{}%
 \ismember@\claimlist@#1%
 \iftest@
  \rightadd@#2\to\nofrillslist@%
  \expandafter\def\csname\exstring@#2@P\endcsname
   {\csname\exstring@#1@P\endcsname}%
  \expandafter\def\csname\exstring@#2@Q\endcsname
   {\csname\exstring@#1@Q\endcsname}%
  \expandafter\def\csname\exstring@#2@S\endcsname
   {\csname\exstring@#1@S\endcsname}%
  \expandafter\def\csname\exstring@#2@N\endcsname
   {\csname\exstring@#1@N\endcsname}%
  \expandafter\def\csname\exstring@#2@F\endcsname
   {\csname\exstring@#1@F\endcsname}%
  \expandafter\def\csname end\exstring@#2\endcsname{\endclaim}%
  \edef\next@{\let\csname\exstring@#2@C\endcsname
    \csname claim\exstring@#1C\endcsname}\next@
  \setbox\z@\vbox{\let\noclaimclause@ T#1""\relax\endgroup}%
  \edef#2{\the\claim@i
   \def\noexpand\claimtype@{\noexpand#2}%
   \def\noexpand\Claimformat@@{\noexpand\claimformat@@{\the\claim@v}\relax}%
   \noexpand\claim@c\noexpand\c{\claim@iv}}%
 \else
  \Err@{\noexpand#1not yet created by \string\newclaim}%
 \fi}
\def\classtest@#1{\DN@{#1}\ifx\next@\claimclass@
 \test@true\else\test@false\fi}
\def\typetest@#1{\DN@{#1}\ifx\next@\claimtype@\test@true\else
  \test@false\fi}
\newif\iftoc@
\def\tocfile{\iftoc@\else\alloc@@7\write\chardef\sixt@@n\toc@
 \immediate\openout\toc@=\jobname.toc
 \alloc@@7\write\chardef\sixt@@n\tic@
 \immediate\openout\tic@=\jobname.tic
 \global\toc@true\fi}
\rightadd@\hl\to\nofrillslist@
\rightadd@\HL\to\overlonglist@
\def\HL@@C{\csname HL@C\HLlevel@\endcsname}
\def\HL@@P{\csname HL@P\HLlevel@\endcsname}
\def\HL@@Q{\csname HL@Q\HLlevel@\endcsname}
\def\HL@@S{\csname HL@S\HLlevel@\endcsname}
\def\HL@@N{\csname HL@N\HLlevel@\endcsname}
\def\HL@@F{\csname HL@F\HLlevel@\endcsname}
\def\HL@@@C{\csname\exxx@\HLtype@ @C\endcsname}
\def\HL@@@P{\csname\exxx@\HLtype@ @P\endcsname}
\def\HL@@@Q{\csname\exxx@\HLtype@ @Q\endcsname}
\def\HL@@@S{\csname\exxx@\HLtype@ @S\endcsname}
\def\HL@@@N{\csname\exxx@\HLtype@ @N\endcsname}
\def\HL#1{\expandafter
 \ifx\csname HL@C#1\endcsname\relax
  \DN@{\Err@{\string\HL#1 not defined in this style}}%
 \else
  \DN@{\gdef\HLlevel@{#1}\def\HLname@{\HL{#1}}\let\HLtype@\relax\FNSS@\HL@}%
 \fi
 \next@}%
\newif\ifquoted@
\let\aftertoc@\relax
\def\HL@{%
 \DN@"##1"##2\endHL{\def\entry@{##2}\quoted@true
  {\noexpands@
  \ifx\HLtype@\relax
   \let\pre\HL@@P\let\post\HL@@Q\let\style\HL@@S\let\numstyle\HL@@N
  \else
   \let\pre\HL@@@P\let\post\HL@@@Q\let\style\HL@@@S\let\numstyle\HL@@@N
  \fi
  \Qlabel@{##1}\let\style\relax\xdef\Qlabel@@@@{##1}%
  \xdef\Thepref@{\Thelabel@@@@}}%
  \csname HL@\HLlevel@\endcsname##2\endHL
  \let\pref\Thepref@
  \csname HL@I\HLlevel@\endcsname
  \csname HL@J\HLlevel@\endcsname
  \let\pref\pref@
  \HLtoc@
  \aftertoc@
  \let\aftertoc@\relax\overlong@false}%
 \DNii@##1\endHL{\def\entry@{##1}\quoted@false
  {\noexpands@
  \ifx\HLtype@\relax
   \global\advance\HL@@C\@ne
   \xdef\Thelabel@@@{\number\HL@@C}%
   \xdefThelabel@{\HL@@N}%
   \xdef\Thelabel@@@@{\HL@@P\Thelabel@\HL@@Q}%
   \xdefThelabel@@{\HL@@S}%
  \else
   \global\advance\HL@@@C\@ne
   \xdef\Thelabel@@@{\number\HL@@@C}%
   \xdefThelabel@{\HL@@@N}%
   \xdef\Thelabel@@@@{\HL@@@P\Thelabel@\HL@@@Q}%
   \xdefThelabel@@{\HL@@@S}%
  \fi
  \xdef\Thepref@{\Thelabel@@@@}}%
  \csname HL@\HLlevel@\endcsname##1\endHL
  \let\pref\Thepref@
  \csname HL@I\HLlevel@\endcsname
  \csname HL@J\HLlevel@\endcsname
  \let\pref\pref@
  \HLtoc@
  \aftertoc@
  \let\aftertoc@\relax\overlong@false}%
 \ifx\next"\expandafter\next@\else\expandafter\nextii@\fi}%
\Invalid@\endHL
\def\hl@@C{\csname hl@C\hllevel@\endcsname}
\def\hl@@P{\csname hl@P\hllevel@\endcsname}
\def\hl@@Q{\csname hl@Q\hllevel@\endcsname}
\def\hl@@S{\csname hl@S\hllevel@\endcsname}
\def\hl@@N{\csname hl@N\hllevel@\endcsname}
\def\hl@@F{\csname hl@F\hllevel@\endcsname}
\def\hl@@@C{\csname\exxx@\hltype@ @C\endcsname}
\def\hl@@@P{\csname\exxx@\hltype@ @P\endcsname}
\def\hl@@@Q{\csname\exxx@\hltype@ @Q\endcsname}
\def\hl@@@S{\csname\exxx@\hltype@ @S\endcsname}
\def\hl@@@N{\csname\exxx@\hltype@ @N\endcsname}
\def\hl#1{\expandafter
 \ifx\csname hl@C#1\endcsname\relax
  \DN@{\Err@{\string\hl#1 not defined in this style}}%
 \else
  \DN@{\gdef\hllevel@{#1}\def\hlname@{\hl{#1}}\let\hltype@\relax\FNSS@\hl@}%
 \fi
 \next@}
\def\hl@{%
 \DN@"##1"##2{\def\entry@{##2}\quoted@true
  {\noexpands@
  \ifx\hltype@\relax
   \let\pre\hl@@P\let\post\hl@@Q\let\style\hl@@S\let\numstyle\hl@@N
  \else
   \let\pre\hl@@@P\let\post\hl@@@Q\let\style\hl@@@S\let\numstyle\hl@@@N
  \fi
  \Qlabel@{##1}\let\style\relax\xdef\Qlabel@@@@{##1}%
  \xdef\Thepref@{\Thelabel@@@@}}%
  \csname hl@\hllevel@\endcsname{##2}%
  \let\pref\Thepref@
  \csname hl@I\hllevel@\endcsname
  \csname hl@J\hllevel@\endcsname
  \let\pref\pref@
  \hltoc@
  \aftertoc@
  \let\aftertoc@\relax\nopunct@false\nospace@false\FNSSP@}%
 \DNii@##1{\def\entry@{##1}\quoted@false
  {\noexpands@
  \ifx\hltype@\relax
   \global\advance\hl@@C\@ne
   \xdef\Thelabel@@@{\number\hl@@C}%
   \xdefThelabel@{\hl@@N}%
   \xdef\Thelabel@@@@{\hl@@P\Thelabel@\hl@@Q}%
   \xdefThelabel@@{\hl@@S}%
  \else
   \global\advance\hl@@@C\@ne
   \xdef\Thelabel@@@{\number\hl@@@C}%
   \xdefThelabel@{\hl@@@N}%
   \xdef\Thelabel@@@@{\hl@@@P\Thelabel@\hl@@@Q}%
   \xdefThelabel@@{\hl@@@S}%
  \fi
  \xdef\Thepref@{\Thelabel@@@@}}%
  \csname hl@\hllevel@\endcsname{##1}%
  \let\pref\Thepref@
  \csname hl@I\hllevel@\endcsname
  \csname hl@J\hllevel@\endcsname
  \let\pref\pref@
  \hltoc@
  \aftertoc@
  \let\aftertoc@\relax\nopunct@false\nospace@false\FNSSP@}%
 \ifx\next"\expandafter\next@\else\expandafter\nextii@\fi}%
\def\six@#1#2 #3 #4 #5 #6 #7 {\DN@{#2}\ifx\next@\empty
 \DN@##1\six@{}\else
 \write#1{ #2 #3 #4 #5 #6 #7}\DN@{\six@#1}\fi
 \next@}
\def\Sixtoc@{\ifx\macdef@\empty\else
 \DN@##1##2\next@{\def\macdef@{##1##2}}%
 \expandafter\next@\macdef@\next@
 \edef\next@
  {\noexpand\six@\toc@\macdef@
  \space\space\space\space\space\space\space\space\space\space\space\space
  \noexpand\six@}%
 \next@\let\macdef@\relax\fi}
\def\QorThelabel@@@@{\ifquoted@
 \noexpand\noexpand\noexpand"\Qlabel@@@@\noexpand\noexpand\noexpand"\else
 \Thelabel@@@@\fi}
\def\HLtoc@{%
 \iftoc@
 \expandafter\expandafter\expandafter\unmacro@
  \expandafter\meaning\csname HL@W\HLlevel@\endcsname\unmacro@
  {\noexpands@\let\style\relax
   \edef\next@{\write\toc@{\noexpand\noexpand\expandafter\noexpand\HLname@
   {\macdef@}{\QorThelabel@@@@}}}%
  \next@}%
  \expandafter\unmacro@\meaning\entry@\unmacro@
  \Sixtoc@
  \write\toc@{\noexpand\Page{\number\pageno}{\page@N}%
   {\page@P}{\page@Q}^^J}%
 \fi}
\def\hltoc@{%
 \iftoc@
 \expandafter\expandafter\expandafter\unmacro@
  \expandafter\meaning\csname hl@W\hllevel@\endcsname\unmacro@
  {\noexpands@\let\style\relax
  \edef\next@{\write\toc@{%
   \ifnopunct@\noexpand\noexpand\noexpand\nopunct\fi
   \ifnospace@\noexpand\noexpand\noexpand\nospace\fi
   \noexpand\noexpand\expandafter\noexpand\hlname@
   {\macdef@}{\QorThelabel@@@@}}}%
  \next@}%
  \expandafter\unmacro@\meaning\entry@\unmacro@
  \Sixtoc@
  \write\toc@{\noexpand\Page{\number\pageno}{\page@N}%
   {\page@P}{\page@Q}^^J}%
 \fi}
\def\mainfile#1{\def\mainfile@{#1}}
\def\checkmainfile@{\ifx\mainfile@\undefined
 \Err@{No \noexpand\mainfile specified}\fi}
\expandafter\newcount@\csname HL@C1\endcsname
\csname HL@C1\endcsname\z@
\expandafter\def\csname HL@S1\endcsname#1{#1\null.}
\expandafter\let\csname HL@N1\endcsname\arabic
\expandafter\let\csname HL@P1\endcsname\empty
\expandafter\let\csname HL@Q1\endcsname\empty
\expandafter\def\csname HL@F1\endcsname{\bf}
\expandafter\let\csname HL@W1\endcsname\empty
\expandafter\newcount@\csname hl@C1\endcsname
\csname hl@C1\endcsname\z@
\expandafter\def\csname hl@S1\endcsname#1{#1\/}
\expandafter\let\csname hl@N1\endcsname\arabic
\expandafter\let\csname hl@P1\endcsname\empty
\expandafter\let\csname hl@Q1\endcsname\empty
\expandafter\def\csname hl@F1\endcsname{\bf}
\expandafter\let\csname hl@W1\endcsname\empty
\expandafter\def\csname HL@1\endcsname#1\endHL{\bigbreak
 {\locallabel@
  \global\setbox\@ne\vbox{\Let@\tabskip\hss@
  \halign to\hsize{\bf\hfil\ignorespaces##\unskip\hfil\cr
  \expandafter\ifx\csname HL@W1\endcsname\empty\else
   \csname HL@W1\endcsname\space\fi
  {\HL@@F\ifx\thelabel@@\empty\else\thelabel@@\space\fi}%
  \ignorespaces#1\crcr}}%
  }%
 \unvbox\@ne\nobreak\medskip}
\expandafter\def\csname hl@1\endcsname#1{\medbreak\noindent@@
 {\locallabel@
 \bf{\hl@@F\ifx\thelabel@@\empty\else\thelabel@@\space\fi}%
 \ignorespaces#1\unskip\punct@{\null.}\addspace@\enspace}}
\expandafter\def\csname HL@I1\endcsname{\Reset\hl1{1}%
 \ifx\pref\empty\newpre\hl1{}\else\newpre\hl1{\pref.}\fi}
\def\NameHL#1#2{\define#2{}%
 \expandafter\ifx\csname HL@R#1\endcsname\relax
 \else
  \def\nextiv@{\let\nextiii@}%
  \expandafter\nextiv@\csname HL@R#1\endcsname
  \expandafter\let\nextiii@\undefined
  \expandafter\let\csname\exxx@\nextiii@ @C\endcsname\relax
  \expandafter\let\csname\exxx@\nextiii@ @P\endcsname\relax
  \expandafter\let\csname\exxx@\nextiii@ @Q\endcsname\relax
  \expandafter\let\csname\exxx@\nextiii@ @S\endcsname\relax
  \expandafter\let\csname\exxx@\nextiii@ @N\endcsname\relax
  \expandafter\let\csname\exxx@\nextiii@ @F\endcsname\relax
  \expandafter\let\csname\exxx@\nextiii@ @W\endcsname\relax
  \expandafter\let\csname end\exxx@\nextiii@\endcsname\undefined
 \fi
 \expandafter\gdef\csname HL@R#1\endcsname{#2}%
 \expandafter\gdef\csname\exstring@#2@R\endcsname{{HL}{#1}}%
 \iftoc@\write\toc@{\noexpand\NameHL#1\noexpand#2^^J}\fi
 \rightadd@#2\to\overlonglist@
 \edef\next@{\let\csname\exstring@#2@C\endcsname\expandafter\noexpand
  \csname HL@C#1\endcsname}\next@
 \edef\next@{\let\csname\exstring@#2@P\endcsname\expandafter\noexpand
  \csname HL@P#1\endcsname}\next@
 \edef\next@{\let\csname\exstring@#2@Q\endcsname\expandafter\noexpand
  \csname HL@Q#1\endcsname}\next@
 \edef\next@{\let\csname\exstring@#2@S\endcsname\expandafter\noexpand
  \csname HL@S#1\endcsname}\next@
 \edef\next@{\let\csname\exstring@#2@N\endcsname\expandafter\noexpand
  \csname HL@N#1\endcsname}\next@
 \edef\next@{\let\csname\exstring@#2@F\endcsname\expandafter\noexpand
  \csname HL@F#1\endcsname}\next@
 \edef\next@{\let\csname\exstring@#2@W\endcsname\expandafter\noexpand
  \csname HL@W#1\endcsname}\next@
 \edef\next@{\def\noexpand#2####1\expandafter\noexpand
  \csname end\exstring@#2\endcsname
  {\def\noexpand\HLtype@{\noexpand#2}%
   \def\noexpand\HLname@{\noexpand#2}%
   \gdef\noexpand\HLlevel@{#1}%
   \noexpand\FNSS@\noexpand\HL@####1\noexpand\endHL}}%
  \next@
 \edef\next@{\noexpand\Invalid@\expandafter\noexpand
  \csname end\exstring@#2\endcsname}%
 \next@}
\def\Namehl#1#2{\define#2{}%
 \expandafter\ifx\csname hl@R#1\endcsname\relax
 \else
  \def\nextiv@{\let\nextiii@}%
  \expandafter\nextiv@\csname hl@R#1\endcsname
  \expandafter\let\nextiii@\undefined
  \expandafter\let\csname\exxx@\nextiii@ @C\endcsname\relax
  \expandafter\let\csname\exxx@\nextiii@ @P\endcsname\relax
  \expandafter\let\csname\exxx@\nextiii@ @Q\endcsname\relax
  \expandafter\let\csname\exxx@\nextiii@ @S\endcsname\relax
  \expandafter\let\csname\exxx@\nextiii@ @N\endcsname\relax
  \expandafter\let\csname\exxx@\nextiii@ @F\endcsname\relax
  \expandafter\let\csname\exxx@\nextiii@ @W\endcsname\relax
 \fi
 \expandafter\gdef\csname hl@R#1\endcsname{#2}%
 \expandafter\gdef\csname\exstring@#2@R\endcsname{{hl}{#1}}%
 \iftoc@\write\toc@{\noexpand\Namehl#1\noexpand#2^^J}\fi
 \rightadd@#2\to\nofrillslist@%
 \edef\next@{\let\csname\exstring@#2@C\endcsname\expandafter\noexpand
  \csname hl@C#1\endcsname}\next@
 \edef\next@{\let\csname\exstring@#2@P\endcsname\expandafter\noexpand
  \csname hl@P#1\endcsname}\next@
 \edef\next@{\let\csname\exstring@#2@Q\endcsname\expandafter\noexpand
  \csname hl@Q#1\endcsname}\next@
 \edef\next@{\let\csname\exstring@#2@S\endcsname\expandafter\noexpand
  \csname hl@S#1\endcsname}\next@
 \edef\next@{\let\csname\exstring@#2@N\endcsname\expandafter\noexpand
  \csname hl@N#1\endcsname}\next@
 \edef\next@{\let\csname\exstring@#2@F\endcsname\expandafter\noexpand
  \csname hl@F#1\endcsname}\next@
 \edef\next@{\let\csname\exstring@#2@W\endcsname\expandafter\noexpand
  \csname hl@W#1\endcsname}\next@
 \edef\next@{\def\noexpand#2{%
  \def\noexpand\hltype@{\noexpand#2}%
  \def\noexpand\hlname@{\noexpand#2}%
  \gdef\noexpand\hllevel@{#1}%
  \noexpand\FNSS@\noexpand\hl@}}%
 \next@}%
\def\Initialize{\FN@\Init@}
\def\Init@{\ifx\next\HL\let\next@\InitH@\else\ifx\next\hl\let\next@\InitH@
  \else\let\next@\InitS@\fi\fi\next@}
\def\InitH@#1#2{\expandafter\ifx\csname\exstring@#1@C#2\endcsname\relax
 \DN@{\Err@{\noexpand#1level #2 not defined in this style}}\else
 \DN@{\expandafter\gdef\csname\exstring@#1@J#2\endcsname}\fi\next@}
\def\InitC@#1#2{\edef\nextii@{\expandafter\noexpand\csname#1\endcsname{#2}}}
\def\InitS@#1{\expandafter\ifx\csname\exstring@#1@R\endcsname\relax
 \Err@{\noexpand#1not defined in this style}\let\next@\relax\else
 \DN@{\let\next@}\expandafter\next@\csname\exstring@#1@R\endcsname
 \expandafter\InitC@\next@
 \DN@{\expandafter\InitH@\nextii@}\fi\next@}
\def\value#1{\expandafter
 \ifx\csname\exstring@#1@C\endcsname\relax
  \expandafter\ifx\csname\exstring@#1@C1\endcsname\relax
   \DN@{\Err@{\noexpand\value can't be used with \string#1}}%
  \else
   \DN@{\value@#1}%
  \fi
 \else
  \DN@{\number\csname\exstring@#1@C\endcsname\relax}%
 \fi
 \next@}
\def\value@#1#2{\expandafter
 \ifx\csname\exstring@#1@C#2\endcsname\relax
  \DN@{\Err@{\string\value\string#1 can't be followed by \string#2}}%
 \else
  \DN@{\number\csname\exstring@#1@C#2\endcsname\relax}%
 \fi
 \next@}
\newcount\Value
\def\Evaluate#1{\expandafter
 \ifx\csname\exstring@#1@C\endcsname\relax
  \expandafter\ifx\csname\exstring@#1@C1\endcsname\relax
   \DN@{\Err@{\noexpand\Evaluate can't be used with \string#1}}%
  \else
   \DN@{\Evaluate@#1}%
  \fi
 \else
  \DN@{\global\Value\csname\exstring@#1@C\endcsname}%
 \fi
 \next@}
\def\Evaluate@#1#2{\expandafter
 \ifx\csname\exstring@#1@C#2\endcsname\relax
  \DN@{\Err@{\string\Evaluate\string#1 can't be followed by \string#2}}%
 \else
  \DN@{\global\Value\csname\exstring@#1@C#2\endcsname}%
 \fi\next@}
\def\pre#1{\expandafter
 \ifx\csname\exstring@#1@P\endcsname\relax
  \expandafter\ifx\csname\exstring@#1@P1\endcsname\relax
   \DN@{\Err@{\noexpand\pre can't be used with \string#1}}%
  \else
   \DN@{\pre@#1}%
  \fi
 \else
  \DN@{{\csname\exstring@#1@P\endcsname}}%
 \fi
 \next@}
\def\pre@#1#2{\expandafter
 \ifx\csname\exstring@#1@P#2\endcsname\relax
  \DN@{\Err@{\string\pre\string#1 can't be followed by \string#2}}%
 \else
  \DN@{{\csname\exstring@#1@P#2\endcsname}}%
 \fi
 \next@}
\def\post#1{\expandafter
 \ifx\csname\exstring@#1@Q\endcsname\relax
  \expandafter\ifx\csname\exstring@#1@Q1\endcsname\relax
   \DN@{\Err@{\noexpand\post can't be used with \string#1}}%
  \else
   \DN@{\post@#1}%
  \fi
 \else
  \DN@{{\csname\exstring@#1@Q\endcsname}}%
 \fi
 \next@}
\def\post@#1#2{\expandafter
 \ifx\csname\exstring@#1@Q#2\endcsname\relax
  \DN@{\Err@{\string\post\string#1 can't be followed by \string#2}}%
 \else
  \DN@{{\csname\exstring@#1@Q#2\endcsname}}%
 \fi
 \next@}
\def\style#1{\expandafter
 \ifx\csname\exstring@#1@S\endcsname\relax
  \expandafter\ifx\csname\exstring@#1@S1\endcsname\relax
   \DN@{\Err@{\noexpand\style can't be used with \string#1}}%
  \else
   \DN@{\style@#1}%
  \fi
 \else
  \DN@{\csname\exstring@#1@S\endcsname}%
 \fi
 \next@}
\def\style@#1#2{\expandafter
 \ifx\csname\exstring@#1@S#2\endcsname\relax
  \DN@{\Err@{\string\style\string#1 can't be followed by \string#2}}%
 \else
  \DN@{\csname\exstring@#1@S#2\endcsname}%
 \fi
 \next@}
\def\fontstyle#1{\expandafter
 \ifx\csname\exstring@#1@F\endcsname\relax
  \expandafter\ifx\csname\exstring@#1@F1\endcsname\relax
   \DN@{\Err@{\noexpand\fontstyle can't be used with \string#1}}%
  \else
   \DN@{\fontstyle@#1}%
  \fi
 \else
  \DN@##1{{\csname\exstring@#1@F\endcsname##1}}%
 \fi
 \next@}
\def\fontstyle@#1#2{\expandafter
 \ifx\csname\exstring@#1@F#2\endcsname\relax
  \DN@{\Err@{\string\fontstyle\string#1 can't be followed by \string#2}}%
 \else
  \DN@##1{{\csname\exstring@#1@F#2\endcsname##1}}%
 \fi
 \next@}
\def\Reset#1{\expandafter
 \ifx\csname\exstring@#1@C\endcsname\relax
  \expandafter\ifx\csname\exstring@#1@C1\endcsname\relax
   \DN@{\Err@{\noexpand\Reset can't be used with \string#1}}%
  \else
   \DN@{\Reset@#1}%
  \fi
 \else
  \DN@##1{\count@##1\relax\ifx#1\page\else\advance\count@\m@ne\fi
   \global\csname\exstring@#1@C\endcsname\count@}%
 \fi
 \next@}
\def\Reset@#1#2{\expandafter
 \ifx\csname\exstring@#1@C#2\endcsname\relax
  \DN@{\Err@{\string\Reset\string#1 can't be followed by \string#2}}%
 \else
  \DN@##1{\count@##1\relax\advance\count@\m@ne
   \global\csname\exstring@#1@C#2\endcsname\count@}%
 \fi
 \next@}
\def\Offset#1{\expandafter
 \ifx\csname\exstring@#1@C\endcsname\relax
  \expandafter\ifx\csname\exstring@#1@C1\endcsname\relax
   \DN@{\Err@{\noexpand\Offset can't be used with \string#1}}%
  \else
   \DN@{\Offset@#1}%
  \fi
 \else
  \DN@##1{\count@##1\relax\advance\count@\m@ne\global\advance
   \csname\exstring@#1@C\endcsname\count@}%
 \fi
 \next@}
\def\Offset@#1#2{\expandafter
 \ifx\csname\exstring@#1@C#2\endcsname\relax
  \DN@{\Err@{\string\Offset\string#1 can't be followed by \string#2}}%
 \else
  \DN@##1{\count@##1\relax\advance\count@\m@ne
   \global\advance\csname\exstring@#1@C#2\endcsname\count@}%
 \fi
 \next@}
\def\getR@#1#2{\def\nextiv@{\let\nextiii@}\expandafter\nextiv@
 \csname\exstring@#1@R#2\endcsname}
\def\letR@#1#2#3{\expandafter\let\csname#1@#3#2\endcsname\Next@}
\def\letR@@#1#2{\expandafter\let\csname\exstring@#1@#2\endcsname\Next@}
\def\newpre#1{\expandafter
 \ifx\csname\exstring@#1@P\endcsname\relax
  \expandafter\ifx\csname\exstring@#1@P1\endcsname\relax
   \DN@{\Err@{\noexpand\newpre can't be used with \string#1}}%
  \else
   \DN@{\newpre@#1}%
  \fi
 \else
  \DN@{%
   \DNii@{%
    \endgroup
    \expandafter\let\csname\exstring@#1@P\endcsname\Next@
    \expandafter\ifx\csname\exstring@#1@R\endcsname\relax\else
    \getR@#1{}\expandafter\letR@\nextiii@ P\fi
    }%
   \begingroup\noexpands@\afterassignment\nextii@\xdef\Next@}%
 \fi
 \next@}
\def\newpre@#1#2{\expandafter
 \ifx\csname\exstring@#1@P#2\endcsname\relax
  \DN@{\Err@{\string\newpre\string#1 can't be followed by \string#2}}%
 \else
  \DN@{%
   \DNii@{%
    \endgroup
    \expandafter\let\csname\exstring@#1@P#2\endcsname\Next@
    \expandafter\ifx\csname\exstring@#1@R#2\endcsname\relax\else
    \getR@#1{#2}\expandafter\letR@@\nextiii@ P\fi
    }%
   \begingroup\noexpands@\afterassignment\nextii@\xdef\Next@}%
 \fi
 \next@}
\def\newpost#1{\expandafter
 \ifx\csname\exstring@#1@Q\endcsname\relax
  \expandafter\ifx\csname\exstring@#1@Q1\endcsname\relax
   \DN@{\Err@{\noexpand\newpost can't be used with \string#1}}%
  \else
   \DN@{\newpost@#1}%
  \fi
 \else
  \DN@{%
   \DNii@{%
    \endgroup
    \expandafter\let\csname\exstring@#1@Q\endcsname\Next@
    \expandafter\ifx\csname\exstring@#1@R\endcsname\relax\else
    \getR@#1{}\expandafter\letR@\nextiii@ Q\fi
    }%
   \begingroup\noexpands@\afterassignment\nextii@\xdef\Next@}%
 \fi
 \next@}
\def\newpost@#1#2{\expandafter
 \ifx\csname\exstring@#1@Q#2\endcsname\relax
  \DN@{\Err@{\string\newpost\string#1 can't be followed by \string#2}}%
 \else
  \DN@{%
   \DNii@{%
    \endgroup
    \expandafter\let\csname\exstring@#1@Q#2\endcsname\Next@
    \expandafter\ifx\csname\exstring@#1@R#2\endcsname\relax\else
    \getR@#1{#2}\expandafter\letR@@\nextiii@ Q\fi
    }%
   \begingroup\noexpands@\afterassignment\nextii@\xdef\Next@}%
 \fi
 \next@}
\def\newstyle#1{\expandafter
 \ifx\csname\exstring@#1@S\endcsname\relax
  \expandafter\ifx\csname\exstring@#1@S1\endcsname\relax
   \DN@{\Err@{\noexpand\newstyle can't be used
    with \string#1}}%
  \else
   \DN@{\newstyle@#1}%
  \fi
 \else
  \DN@{%
   \DNii@{%
    \expandafter\let\csname\exstring@#1@S\endcsname\Next@
    \expandafter\ifx\csname\exstring@#1@R\endcsname\relax\else
    \getR@#1{}\expandafter\letR@\nextiii@ S\fi
    }%
   \afterassignment\nextii@\gdef\Next@}%
 \fi
 \next@}
\def\newstyle@#1#2{\expandafter
 \ifx\csname\exstring@#1@S#2\endcsname\relax
  \DN@{\Err@{\string\newstyle\string#1 can't be followed by
   \string#2}}%
 \else
  \DN@{%
   \DNii@{%
    \expandafter\let\csname\exstring@#1@S#2\endcsname\Next@
    \expandafter\ifx\csname\exstring@#1@R#2\endcsname\relax\else
    \getR@#1{#2}\expandafter\letR@@\nextiii@ S\fi
    }%
   \afterassignment\nextii@\gdef\Next@}%
 \fi
 \next@}
\def\newnumstyle#1{\expandafter
 \ifx\csname\exstring@#1@N\endcsname\relax
  \expandafter\ifx\csname\exstring@#1@N1\endcsname\relax
   \DN@{\Err@{\noexpand\newnumstyle can't be used with
    \string#1}}%
  \else
   \DN@{\newnumstyle@#1}%
  \fi
 \else
  \DN@##1{%
   \gdef\Next@{##1}%
    \expandafter\let\csname\exstring@#1@N\endcsname\Next@
    \expandafter\ifx\csname\exstring@#1@R\endcsname\relax\else
    \getR@#1{}\expandafter\letR@\nextiii@ N\fi
    }%
 \fi
 \next@}
\def\newnumstyle@#1#2{\expandafter
 \ifx\csname\exstring@#1@N#2\endcsname\relax
  \DN@{\Err@{\string\newnumstyle\string#1 can't be followed by
   \string#2}}%
 \else
  \DN@##1{%
   \gdef\Next@{##1}%
    \expandafter\let\csname\exstring@#1@N#2\endcsname\Next@
    \expandafter\ifx\csname\exstring@#1@R#2\endcsname\relax\else
    \getR@#1{#2}\expandafter\letR@@\nextiii@ N\fi
    }%
  \fi
 \next@}
\def\newfontstyle#1{\expandafter
 \ifx\csname\exstring@#1@F\endcsname\relax
  \expandafter\ifx\csname\exstring@#1@F1\endcsname\relax
   \DN@{\Err@{\noexpand\newfontstyle can't be used with
    \string#1}}%
  \else
   \DN@{\newfontstyle@#1}%
  \fi
 \else
  \DN@##1{%
   \gdef\Next@{##1}%
    \expandafter\let\csname\exstring@#1@F\endcsname\Next@
    \expandafter\ifx\csname\exstring@#1@R\endcsname\relax\else
    \getR@#1{}\expandafter\letR@\nextiii@ F\fi
    }%
 \fi
 \next@}
\def\newfontstyle@#1#2{\expandafter
 \ifx\csname\exstring@#1@F#2\endcsname\relax
  \DN@{\Err@{\string\newfontstyle\string#1 can't be followed by
   \string#2}}%
 \else
  \DN@##1{%
   \gdef\Next@{##1}%
    \expandafter\let\csname\exstring@#1@F#2\endcsname\Next@
    \expandafter\ifx\csname\exstring@#1@R#2\endcsname\relax\else
    \getR@#1{#2}\expandafter\letR@@\nextiii@ F\fi
    }%
 \fi
 \next@}
\def\word#1{\expandafter
 \ifx\csname\exstring@#1@W\endcsname\relax
  \expandafter\ifx\csname\exstring@#1@W1\endcsname\relax
   \DN@{\Err@{\noexpand\word can't be used with \string#1}}%
  \else
   \DN@{\word@#1}%
  \fi
 \else
  \DN@{{\csname\exstring@#1@W\endcsname}}%
 \fi
 \next@}
\def\word@#1#2{\expandafter
 \ifx\csname\exstring@#1@W#2\endcsname\relax
  \DN@{\Err@{\string\word\noexpand#1can't be followed by \string#2}}%
 \else
  \DN@{{\csname\exstring@#1@W#2\endcsname}}%
 \fi
 \next@}
\def\newword#1{\expandafter
 \ifx\csname\exstring@#1@W\endcsname\relax
  \expandafter\ifx\csname\exstring@#1@W1\endcsname\relax
   \DN@{\Err@{\noexpand\newword can't be used  with \string#1}}%
  \else
   \DN@{\newword@#1}%
  \fi
 \else
  \DN@{%
   \DNii@{%
    \expandafter\let\csname\exstring@#1@W\endcsname\Next@
    \expandafter\ifx\csname\exstring@#1@R\endcsname\relax\else
     \getR@#1{}\expandafter\letR@\nextiii@ W\fi
    }%
   \afterassignment\nextii@\gdef\Next@}%
 \fi
 \next@}
\def\newword@#1#2{\expandafter
 \ifx\csname\exstring@#1@W#2\endcsname\relax
  \DN@{\Err@{\string\newword\noexpand#1can't be followed by \string#2}}%
 \else
  \DN@{%
   \DNii@{%
    \expandafter\let\csname\exstring@#1@W#2\endcsname\Next@
    \expandafter\ifx\csname\exstring@#1@R#2\endcsname\relax\else
     \getR@#1{#2}\expandafter\letR@@\nextiii@ W\fi
    }%
   \afterassignment\nextii@\gdef\Next@}%
 \fi
 \next@}
\newif\iffn@
\newcount\footmark@C
\footmark@C\z@
\def\footmark@S#1{$^{#1}$}
\let\footmark@N\arabic
\def\footmark@F{\rm}
\def\foottext@S#1{$^{#1}$}
\def\foottext@F{\rm}
\let\modifyfootnote@\relax
\def\modifyfootnote#1{\def\modifyfootnote@{#1}}
\def\vfootnote@#1{\insert\footins
 \bgroup
 \floatingpenalty\@MM\interlinepenalty\interfootnotelinepenalty
 \leftskip\z@\rightskip\z@\spaceskip\z@\xspaceskip\z@
 \rm\splittopskip\ht\strutbox\splitmaxdepth\dp\strutbox
 \locallabel@\noindent@@{\foottext@F#1}\modifyfootnote@
 \footstrut\FN@\fo@t}
\def\fo@t{\ifcat\bgroup\noexpand\next\expandafter\f@@t\else
 \expandafter\f@t\fi}
\def\f@t#1{#1\@foot}
\def\f@@t{\bgroup\aftergroup\@foot\afterassignment\FNSSP@\let\next@}
\def\@foot{\unskip\lower\dp\strutbox\vbox to\dp\strutbox{}\egroup
 \iffn@\expandafter\fn@false\else
 \expandafter\postvanish@\fi}
\newif\ifplainfn@
\plainfn@true
\def\fancyfootnotes{\plainfn@false}
\newcount\fancyfootmarkcount@
\fancyfootmarkcount@\z@
\newcount\lastfnpage@
\lastfnpage@-\@M
\let\justfootmarklist@\empty
\def\footmark{\let\@sf\empty
 \ifhmode\edef\@sf{\spacefactor\the\spacefactor}\/\fi
 \DN@{\ifx"\next\expandafter\nextii@\else\expandafter\footmark@\fi}%
 \DNii@"##1"{%
  \iffirstchoice@
   {\let\style\footmark@S\let\numstyle\footmark@N
   \footmark@F##1%
   \noexpands@
   \let\style\foottext@S
   \Qlabel@{##1}%
   }%
   \iffn@\else
    {\noexpands@
    \xdef\Next@{{\Thelabel@}{\Thelabel@@}{\Thelabel@@@}{\Thelabel@@@@}}%
    }%
    \expandafter\rightappend@\Next@\to\justfootmarklist@
   \fi
  \fi
  \@sf\relax}%
 \FN@\next@}
\def\footmark@{%
 \iffirstchoice@
  \global\advance\footmark@C\@ne
  \ifplainfn@
   \xdef\adjustedfootmark@{\number\footmark@C}%
  \else
   {\let\\\or\xdef\Next@{\ifcase\number\footmark@C\fnpages@\else
     -\@M\fi}}%
   \ifnum\Next@=-\@M
    \xdef\adjustedfootmark@{\number\footmark@C}%
   \else
    \ifnum\Next@=\lastfnpage@
     \global\advance\fancyfootmarkcount@\@ne
    \else
     \global\fancyfootmarkcount@\@ne
     \global\lastfnpage@\Next@
    \fi
    \xdef\adjustedfootmark@{\number\fancyfootmarkcount@}%
   \fi
  \fi
  {\noexpands@
  \xdef\Thelabel@@@{\adjustedfootmark@}%
  \xdefThelabel@\footmark@N
  \xdef\Thelabel@@@@{\Thelabel@}%
  \xdefThelabel@@\foottext@S
  }%
  \iffn@\else
   {\noexpands@
   \xdef\Next@{{\Thelabel@}{\Thelabel@@}{\Thelabel@@@}{\Thelabel@@@@}}%
   }%
   \expandafter\rightappend@\Next@\to\justfootmarklist@
  \fi
  \ifplainfn@
  \else
   \edef\next@{\write\laxwrite@{F\noexpand\the\pageno}}\next@
  \fi
 \fi
 \footmark@S{\footmark@N{\adjustedfootmark@}}%
 \@sf\relax}
\def\foottext{\prevanish@
 \ifx\justfootmarklist@\empty
  \Err@{There is no \noexpand\footmark for this \string\foottext}\fi
 \DN@\\##1##2\next@{\DN@{##1}\gdef\justfootmarklist@{##2}}%
 \expandafter\next@\justfootmarklist@\next@
 \expandafter\foottext@\next@}
\def\foottext@#1#2#3#4{{\noexpands@
  \xdef\Thelabel@{#1}\xdef\Thelabel@@{#2}%
  \xdef\Thelabel@@@{#3}\xdef\Thelabel@@@@{#4}}%
  \vfootnote@{\thelabel@@}}
\rightadd@\foottext\to\vanishlist@
\def\footnote{\fn@true
 \let\@sf\empty
 \ifhmode\edef\@sf{\spacefactor\the\spacefactor}\/\fi
 \DN@{\ifx"\next\expandafter\nextii@\else\expandafter\nextiii@\fi}%
 \DNii@"##1"{\footmark"##1"\vfootnote@{\let\style\foottext@S
  \let\numstyle\footmark@N##1}}%
 \def\nextiii@{\footmark\vfootnote@{\foottext@S{\footmark@N
  {\adjustedfootmark@}}}}%
 \FN@\next@}
\newdimen\litindent
\litindent20\p@
\newbox\litbox@
\newbox\Litbox@
\newcount\interlitpenalty@
\interlitpenalty@\@M
\newcount\litlines@
{\obeyspaces\gdef\defspace@{\def {\allowbreak\hskip.5emminus.15em}}}
{\obeylines\gdef\letM@{\let^^M\CtrlM@}}
\def\CtrlM@{\egroup
 \ifcase\litlines@\advance\litlines@\@ne\or
 \box\litbox@\advance\litlines@\@ne\else
 \penalty\interlitpenalty@\box\litbox@\fi
 \Lit@}
\def\Lit@{\setbox\litbox@\hbox\bgroup\litdefs@\hskip\litindent}
\newcount\littab@
\littab@8
\def\littab#1{\littab@#1\relax}
{\catcode`\^^I=\active\gdef\letTAB@{\let^^I\TAB@}}
\def\TAB@{\egroup
 \dimen@\wd\litbox@
 \advance\dimen@-\litindent
 \setboxz@h{\tt0}%
 \dimen@ii\littab@\wdz@
 \divide\dimen@\dimen@ii
 \multiply\dimen@\dimen@ii
 \advance\dimen@\littab@\wdz@
 \advance\dimen@\litindent
 \setbox\litbox@\hbox\bgroup\litdefs@\hbox to\dimen@{\unhbox\litbox@\hfil}}
{\catcode`\`=\active\gdef`{\relax\lq}}
\let\litbs@\relax
\let\litbs@@\relax
\def\litbackslash#1{%
 \edef\litbs@{\catcode`\string#1=\z@
 \def\noexpand\litbs@@{\def\expandafter\noexpand\csname\string#1\endcsname
  {\char`\string#1}}}}
\def\litcodes@{\catcode`\\=12
 \catcode`\{=12 \catcode`\}=12
 \catcode`\$=12 \catcode`\&=12
 \catcode`\#=12
 \catcode`\^=12 \catcode`\_=12
 \catcode`\@=12 \catcode`\~=12 \catcode`\"=12
 \catcode`\;=12 \catcode`\:=12 \catcode`\!=12 \catcode`\?=12
 \catcode`\%=12 \litbs@\catcode`\`=\active\obeyspaces\defspace@}
\def\activate@#1#2{{\lccode`\~=`#2%
 \lowercase{%
  \if0#1%
  \gdef\Next@{\def~{\egroup\endgroup\bigskip\vskip-\parskip
   \def\next@{\noindent@@\FN@\pretendspace@}\FNSS@\next@}}\else
  \gdef\Next@{\def~{\egroup\egroup\endgroup}}\fi
  }%
 }}
\def\litdefs@{\let\0\empty\let\1\litdelim@\def\ {\char32 }\litbs@@}%
\def\litdelimiter#1{%
 \edef\litdelim@{\char`#1}%
 \def\lit#1{\leavevmode\begingroup\litcodes@\litdefs@
  \tt\hyphenchar\tentt\m@ne\lit@}%
 \def\lit@##1#1{##1\endgroup\null}%
 \def\Lit#1{\ifhmode$$\abovedisplayskip\bigskipamount
  \abovedisplayshortskip\bigskipamount
  \belowdisplayskip\z@\belowdisplayshortskip\z@
  \postdisplaypenalty\@M
  $$\vskip-\baselineskip\else\bigskip\fi
  \begingroup\litlines@\z@
  \catcode`#1=\active\activate@0#1\Next@
  \def\displaybreak{\egroup\break\litlines@\z@\Lit@}%
  \def\allowdisplaybreak{\egroup\allowbreak\litlines@\z@\Lit@}%
  \def\allowdisplaybreaks{\egroup\allowbreak\interlitpenalty@\z@
   \litlines@\z@\Lit@}%
  \litcodes@\tt\catcode`\^^I=\active\letTAB@
  \obeylines\letM@\Lit@}%
 \def\Litbox##1=#1{\begingroup\ifodd##1\relax\aftergroup\global\fi
  \aftergroup\setbox\aftergroup##1\aftergroup\box\aftergroup\Litbox@
  \def\allowdisplaybreak{\egroup\allowbreak\litlines@\z@\Lit@}%
  \def\allowdisplaybreaks{\egroup\allowbreak\interlitpenalty@\z@
   \litlines@\z@\Lit@}%
  \catcode`#1=\active\activate@1#1\Next@
  \litcodes@\tt\catcode`\^^I=\active\letTAB@
  \obeylines\letM@\global\setbox\Litbox@\vbox\bgroup\litindent\z@%
  \litlines@\z@\Lit@}%
}
\newbox\titlebox@
\setbox\titlebox@\vbox{}
\rightadd@\title\to\overlonglist@
\def\title{\begingroup\Let@
 \global\setbox\titlebox@\vbox\bgroup\tabskip\hss@
 \halign to\hsize\bgroup\bf\hfil\ignorespaces##\unskip\hfil\cr}
\def\endtitle{\crcr\egroup\egroup\endgroup\overlong@false}
\newbox\authorbox@
\rightadd@\author\to\overlonglist@
\def\author{\begingroup\Let@
 \global\setbox\authorbox@\vbox\bgroup\tabskip\hss@
 \halign to\hsize\bgroup\rm\hfil\ignorespaces##\unskip\hfil\cr}
\def\endauthor{\crcr\egroup\egroup\endgroup\overlong@false}
\newbox\affilbox@
\def\affil{\begingroup\Let@
 \global\setbox\affilbox@\vbox\bgroup\tabskip\hss@
 \halign to\hsize\bgroup\rm\hfil\ignorespaces##\unskip\hfil\cr}%
\def\endaffil{\crcr\egroup\egroup\endgroup\overlong@false}
\let\date@\relax
\def\date#1{\gdef\date@{\ignorespaces#1\unskip}}
\def\today{\ifcase\month\or January\or February\or March\or April\or May\or
 June\or July\or August\or September\or October\or November\or December\fi
 \space\number\day, \number\year}
\def\maketitle{\hrule\height\z@\vskip-\topskip
 \vskip24\p@ plus12\p@ minus12\p@
 \unvbox\titlebox@
 \ifvoid\authorbox@\else\vskip12\p@ plus6\p@ minus3\p@\unvbox\authorbox@\fi
 \ifvoid\affilbox@\else\vskip10\p@ plus5\p@ minus2\p@\unvbox\affilbox@\fi
 \ifx\date@\relax\else\vskip6\p@ plus2\p@ minus\p@\centerline{\rm\date@}\fi
 \vskip18\p@ plus12\p@ minus6\p@}
\def\cite{%
 \DNii@(##1)##2{{\rm[}{##2}, {##1\/}{\rm]}}%
 \def\nextiii@##1{{\rm[}{##1\/}{\rm]}}%
 \DN@{\ifx\next(\expandafter\nextii@\else\expandafter\nextiii@\fi}%
 \FN@\next@}
\def\makebib@W{Bibliography}
\def\makebibp@W{References}
\def\makebib{\begingroup\rm\bigbreak\centerline{\smc\makebib@W}%
 \nobreak\medskip
 \sfcode`\.=\@m\everypar{}\parindent\z@
 \def\nopunct{\nopunct@true}\def\nospace{\nospace@true}%
 \nopunct@false\nospace@false
 \def\lkerns@{\null\kern\m@ne sp\kern\@ne sp}%
 \def\nkerns@{\null\kern-\tw@ sp\kern\tw@ sp}%
}

\newif\ifnoprepunct@
\newif\ifnoprespace@
\newif\ifnoquotes@
\def\noprepunct{\noprepunct@true}
\def\noprespace{\noprespace@true}
\def\noquotes{\noquotes@true}
\newbox\nobox@
\newbox\keybox@
\newbox\bybox@
\newbox\paperbox@
\newbox\paperinfobox@
\newbox\jourbox@
\newbox\volbox@
\newbox\issuebox@
\newbox\yrbox@
\newbox\pgbox@
\newbox\ppbox@
\newbox\bookbox@
\newbox\inbookbox@
\newbox\bookinfobox@
\newbox\publbox@
\newbox\publaddrbox@
\newbox\edbox@
\newbox\edsbox@
\newbox\langbox@
\newbox\translbox@
\newbox\finalinfobox@
\def\setbibinfo@#1{\edef\next@{\ifnopunct@1\else0\fi
 \ifnospace@1\else0\fi\ifnoprepunct@1\else0\fi\ifnoprespace@1\else0\fi
 \ifnoquotes@1\else0\fi}%
 \DNii@{00000}%
 \ifx\next@\nextii@\else\xdef\bibinfo@{\bibinfo@\the#1,\next@}%
 \fi}
\def\getbibinfo@#1{\ifx\bibinfo@\empty
 \let\next@0\let\nextii@0\let\nextiii@0\let\nextiv@0\let\nextv@0\else
 \edef\next@{\def
  \noexpand\next@####1\the#1,####2####3####4####5####6####7\noexpand\next@
  {\let\noexpand\next@####2\let\noexpand\nextii@####3%
  \let\noexpand\nextiii@####4\let\noexpand\nextiv@####5%
  \let\noexpand\nextv@####6}%
  \noexpand\next@\bibinfo@\the#1,00000\noexpand\next@}\next@
 \fi}
\newif\ifbookinquotes@
\def\bookinquotes{\bookinquotes@true}
\newif\ifpaperinquotes@
\def\paperinquotes{\paperinquotes@true}
\newif\ifininbook@
\def\ininbook{\ininbook@true}
\newif\ifopenquotes@
\def\closequotes@{\ifopenquotes@''\openquotes@false\fi}
\newif\ifbeginbib@
\newif\ifendbib@
\newif\ifprevjour@
\newif\ifprevbook@
\newdimen\bibindent@
\bibindent@20\p@
\def\bib{\global\let\bibinfo@\empty\global\let\translinfo@\relax\beginbib@true
 \begingroup\noindent@
 \hangindent\bibindent@\hangafter\@ne\bib@}
\def\v@id#1{\setbox#1\box\voidb@x}
\def\bib@{\v@id\nobox@\v@id\keybox@\v@id\bybox@\v@id\paperbox@
 \v@id\paperinfobox@\v@id\jourbox@\v@id\volbox@\v@id\issuebox@
 \v@id\yrbox@\v@id\pgbox@\v@id\ppbox@\v@id\bookbox@\v@id\inbookbox@
 \v@id\bookinfobox@\v@id\publbox@\v@id\publaddrbox@\v@id\edbox@
 \v@id\edsbox@\v@id\langbox@\v@id\translbox@\v@id\finalinfobox@
 \bgroup}
\def\Setnonemptybox@#1#2{\unskip\setbibinfo@#1\egroup#2%
 \def\aftergroup@{\ifdim\wd#1=\z@\setbox#1\box\voidb@x\fi}%
 \setbox#1\vbox\bgroup\aftergroup\aftergroup@\hsize\maxdimen\leftskip\z@
 \rightskip\z@\hbadness\@M\hfuzz\maxdimen\noindent}
\def\setnonemptybox@#1{\Setnonemptybox@#1\relax}
\def\no{\setnonemptybox@\nobox@}
\def\key{\setnonemptybox@\keybox@\bf}
\def\by{\setnonemptybox@\bybox@}
\def\bysame{\setnonemptybox@\bybox@\leaders\hrule\hskip3em\null}
\def\paper{\setnonemptybox@\paperbox@
 \ifpaperinquotes@\getbibinfo@\paperbox@
 \if\nextv@1\else``\fi\else\it\fi}
\def\paperinfo{\setnonemptybox@\paperinfobox@}
\def\jour{\Setnonemptybox@\jourbox@\prevjour@true}
\def\vol{\setnonemptybox@\volbox@\bf}
\def\issue{\setnonemptybox@\issuebox@}
\def\yr{\setnonemptybox@\yrbox@}

\def\pg{\setnonemptybox@\pgbox@}
\def\pp{\setnonemptybox@\ppbox@}
\def\book{\Setnonemptybox@\bookbox@\prevbook@true
 \ifbookinquotes@\getbibinfo@\bookbox@
 \if\nextv@1\else``\fi\else\it\fi}
\def\inbook{\Setnonemptybox@\inbookbox@\prevbook@true
 \ifininbook@ in \fi\ifbookinquotes@\getbibinfo@\inbookbox@
 \if\nextv@1\else``\fi\fi}
\def\bookinfo{\setnonemptybox@\bookinfobox@}
\def\publ{\setnonemptybox@\publbox@}
\def\publaddr{\setnonemptybox@\publaddrbox@}
\def\ed{\setnonemptybox@\edbox@}
\def\eds{\setnonemptybox@\edsbox@}
\def\lang{\setnonemptybox@\langbox@}
\def\finalinfo{\setnonemptybox@\finalinfobox@}
\def\setboxzl@{\setbox\z@\lastbox}
\def\getbox@#1{\setbox\z@\vbox{\vskip-\@M\p@
 \unvbox#1%
 \setboxzl@
 \global\setbox\@ne\hbox{\unhbox\z@\unskip\unskip\unpenalty}%
 \ifdim\lastskip=-\@M\p@\else
 \loop\ifdim\lastskip=-\@M\p@
 \else\unskip\unpenalty\setboxzl@
 \global\setbox\@ne\hbox{\unhbox\z@\unhbox\@ne}%
 \repeat\fi}%
 \unhbox\@ne}
\def\adjustpunct@#1{\count@\lastkern
 \ifnum\count@=\z@#1\closequotes@\else
 \ifnum\count@>\tw@#1\closequotes@\else
 \ifnum\count@<-\tw@#1\closequotes@\else
  \unkern\unkern\setboxzl@
  \skip@\lastskip\unskip
  \count@@\lastpenalty\unpenalty
  \ifnum\count@=\tw@\unskip\setboxzl@\fi
  \ifdim\skip@=\z@\else\hskip\skip@\fi
  #1\closequotes@
  \ifnum\count@=\tw@\null\hfill\fi
  \penalty\count@@
 \fi\fi\fi}
\def\prepunct@#1#2{\getbibinfo@#2%
 \ifnopunct@
 \else
  \if\nextiii@0\adjustpunct@#1\fi
 \fi
 \closequotes@
 \ifnospace@
 \else
  \if\nextiv@0\space\else\fi
 \fi
 \nopunct@false\nospace@false
 \if\next@1\nopunct@true\fi
 \if\nextii@1\nospace@true\fi}
\def\ppunbox@#1#2{\prepunct@{#1}#2%
 \getbox@#2}
\let\semicolon@;
\def\endbib@{%
 \ifbeginbib@
  \ifvoid\nobox@
   \ifvoid\keybox@\else\hbox to\bibindent@{[\getbox@\keybox@]\hss}\fi
  \else\hbox to\bibindent@{\hss\getbox@\nobox@. }\fi
  \ifvoid\bybox@\else\getbox@\bybox@\fi
 \else
  \nopunct@true
  \ifvoid\bybox@\else\ppunbox@\relax\bybox@\fi
 \fi
 \ifvoid\translbox@\else\ppunbox@,\translbox@\fi
 \ifvoid\paperbox@\else\ppunbox@,\paperbox@\ifpaperinquotes@
  \if\nextv@1\else\openquotes@true\fi\fi
 \fi
 \ifvoid\paperinfobox@\else\ppunbox@,\paperinfobox@\fi
 \test@false
 \ifvoid\jourbox@\else\test@true\ppunbox@,\jourbox@\fi
 \ifprevjour@\test@true\fi
 \iftest@
  \ifvoid\volbox@\else\ppunbox@\relax\volbox@\fi
  \ifvoid\issuebox@
   \else\prepunct@\relax\issuebox@ no.~\getbox@\issuebox@\fi
  \ifvoid\yrbox@\else\prepunct@\relax\yrbox@(\getbox@\yrbox@)\fi
  \ifvoid\ppbox@\else\ppunbox@,\ppbox@\fi
  \ifvoid\pgbox@\else\prepunct@,\pgbox@ p.~\getbox@\pgbox@\fi
 \fi
 \test@false
 \ifvoid\bookbox@\else\test@true\ppunbox@,\bookbox@\ifbookinquotes@
  \if\nextv@1\else\openquotes@true\fi\fi\fi
 \ifvoid\inbookbox@\else\test@true\ppunbox@,\inbookbox@\ifbookinquotes@
  \if\nextv@1\else\openquotes@true\fi\fi\fi
 \ifprevbook@\test@true\fi
 \iftest@
  \ifvoid\edbox@\else\prepunct@\relax\edbox@(\getbox@\edbox@, ed.)\fi
  \ifvoid\edsbox@\else\prepunct@\relax\edsbox@(\getbox@\edsbox@, eds.)\fi
  \ifvoid\bookinfobox@\else\ppunbox@,\bookinfobox@\fi
  \ifvoid\publbox@\else\ppunbox@,\publbox@\fi
  \ifvoid\publaddrbox@\else\ppunbox@,\publaddrbox@\fi
  \ifvoid\yrbox@\else\ppunbox@,\yrbox@\fi
  \ifvoid\ppbox@\else\prepunct@,\ppbox@ pp.~\getbox@\ppbox@\fi
  \ifvoid\pgbox@\else\prepunct@,\pgbox@ p.~\getbox@\pgbox@\fi
 \fi
 \ifvoid\finalinfobox@
  \ifendbib@
   \ifnopunct@\else.\closequotes@\fi
  \else
  \ifvoid\langbox@\else\space(\getbox@\langbox@)\fi
   \/\semicolon@\closequotes@
  \fi
 \else
  \ifendbib@
   \ppunbox@{.\spacefactor3000\relax}\finalinfobox@
    \ifnopunct@\else.\fi
  \else
   \ppunbox@,\finalinfobox@\/\semicolon@\fi
 \fi
 \ifvoid\langbox@\else\space(\getbox@\langbox@)\fi
}
\def\endbib{\unskip\egroup\endbib@true\endbib@\par\endgroup}
\def\morebib{\unskip\egroup
 \endbib@false\endbib@
 \global\let\bibinfo@\empty\beginbib@false
 \bib@}
\def\anotherbib{\unskip\egroup
 \endbib@false\endbib@
 \global\let\bibinfo@\empty\beginbib@false
 \prevjour@false\prevbook@false\bib@}
\def\transl{\unskip
 \xdef\translinfo@{\the\translbox@,\ifnopunct@1\else0\fi
 \ifnospace@1\else0\fi\ifnoprepunct@1\else0\fi\ifnoprespace@1\else0\fi0}%
 \egroup\endbib@false\endbib@
 \global\let\bibinfo@\translinfo@\beginbib@false
 \bib@
 \egroup
 \def\aftergroup@{\ifdim\wd\translbox@=\z@\setbox\translbox@\box\voidb@x\fi}%
 \setbox\translbox@\vbox\bgroup\aftergroup\aftergroup@
 \hsize\maxdimen\leftskip\z@\rightskip\z@\hbadness\@M\hfuzz\maxdimen
 \noindent}
\newwrite\auxwrite@
\newread\bbl@
\def\UseBibTeX{\immediate\openout\auxwrite@=\jobname.aux
 \let\cite\BTcite@
 \def\nocite##1{\immediate\write\auxwrite@{\string\citation{##1}}}%
 \def\bibliographystyle##1{\immediate\write\auxwrite@{\string
  \bibstyle{##1}}}%
 \def\bibliography@W{Bibliography}%
 \def\bibliography##1{\immediate\write\auxwrite@{\string\bibdata{##1}}%
  \immediate\openin\bbl@=\jobname.bbl
  \ifeof\bbl@
   \W@{No .bbl file}%
  \else
   \immediate\closein\bbl@
   \begingroup\input bibtex \input\jobname.bbl \endgroup
  \fi}%
 }
\def\BTcite@{%
 \DNii@(##1)##2{{\rm[}\BTcite@@##2,\BTcite@@{\rm, }{##1\/}{\rm]}%
  \immediate\write\auxwrite@{\string\citation{##2}}}%
 \def\nextiii@##1{{\rm[}\BTcite@@##1,\BTcite@@\/{\rm]}%
  \immediate\write\auxwrite@{\string\citation{##1}}}%
 \DN@{\ifx\next(\expandafter\nextii@\else\expandafter\nextiii@\fi}%
 \FN@\next@}%
\def\BTcite@@#1,{\BTcite@@@{#1}\FN@\BTcite@@@@}
\def\BTcite@@@@{\ifx\next\BTcite@@
 \expandafter\eat@\else{\rm, }\expandafter\BTcite@@\fi}
\catcode`\~=11
\def\BTcite@@@#1{\nolabel@\cite{#1}\relax
 \DNii@##1~##2\nextii@{##1}%
 \csL@{#1}\expandafter\nextii@\Next@\nextii@\fi}
\catcode`\~=\active

\def\beginthebibliography@#1{\rm\setboxz@h{#1\ }\bibindent@\wdz@
 \bigbreak\centerline{\smc\bibliography@W}\nobreak\medskip
 \sfcode`\.=\@m\everypar{}\parindent\z@}

\newif\iffigproofing@
\def\Figureproofing{\figproofing@true}
\def\noFigureproofing{\figproofing@false}
\newif\ifHby@
\def\Hbyw#1{\global\Hby@true\hbyw\vsize{#1}}
\def\hbyw#1#2{%
 \hbox{%
  \ifHby@
  \else
   \iffigproofing@
    \setbox\z@\vbox{\hrule\width5\p@}\ht\z@\z@
    \vbox to#1{\hrule\height5\p@\width.4\p@\vfil\hrule\height5\p@\width.4\p@}%
    \kern-.4\p@\rlap{\copy\z@}\raise#1\hbox{\rlap{\copy\z@}}%
   \fi
  \fi
  \vbox to#1{\hbox to#2{}\vfil}%
  \ifHby@
  \else
   \iffigproofing@
    \vbox to#1{\hrule\height5\p@\width.4\p@\vfil\hrule\height5\p@\width.4\p@}%
    \kern-.4\p@\llap{\copy\z@}\raise#1\hbox{\llap{\boxz@}}%
   \fi
  \fi}}
\newcount\island@C
\let\island@P\empty
\let\island@Q\empty
\def\island@S#1{#1\null.}
\let\island@N\arabic
\def\island@F{\rm}
\def\island@@@P{\csname\exxx@\islandtype@ @P\endcsname}
\def\island@@@Q{\csname\exxx@\islandtype@ @Q\endcsname}
\def\island@@@S{\csname\exxx@\islandtype@ @S\endcsname}
\def\island@@@N{\csname\exxx@\islandtype@ @N\endcsname}
\def\island@@@F{\csname\exxx@\islandtype@ @F\endcsname}
\def\island@@@C{\csname island@C\islandclass@\endcsname}
\newif\ifplace@
\newif\ifisland@
\def\island{%
 \ifplace@
  \DN@{\let\islandclass@\empty\def\islandtype@{\island}\FN@\island@}%
 \else
  \long\DN@##1\endisland{\Err@{\noexpand\island must be used after some
   type of \string\...place}}%
 \fi
 \next@}
\def\island@{\ifx\next\c\let\next@\island@c\else
 \DN@{\FN@\island@@}\fi\next@}
\def\island@@{\ifcat\bgroup\noexpand\next\let\next@\island@@@\else
 \DN@{\Err@{\noexpand\island must be followed by a {prefix} for
 \string\caption's}}\fi\next@}
\newbox\islandbox@
\newcount\captioncount@
\def\island@@@#1{\def\captionprefix@{#1}\captioncount@\z@
 \global\setbox\islandbox@\vbox\bgroup}
\def\island@c\c#1{%
 \ifplace@
 \DN@{\def\islandclass@{#1}%
  \expandafter\ifx\csname island@C#1\endcsname\relax
  \expandafter\newcount@\csname island@C#1\endcsname
   \global\csname island@C#1\endcsname\z@\fi
  \FNSS@\island@c@}%
 \else
 \DN@{\edef\next@{\long\def\noexpand\next@########1\expandafter\noexpand
  \csname end\exxx@\islandtype@\endcsname{\noexpand\Err@{\noexpand\noexpand
  \expandafter\noexpand
  \islandtype@ must be used after some type of \noexpand\string
   \noexpand\...place}}}\next@\next@}%
 \fi
 \next@}
\def\island@c@{%
 \ifcat\bgroup\noexpand\next
  \let\next@\island@c@@
 \else
  \DN@{\Err@{\noexpand\island\string\c{\expandafter\string\islandclass@} must
   be followed by a {prefix} for \string\caption's}}%
 \fi\next@}
\def\island@c@@#1{\def\captionprefix@{#1}%
 \captioncount@\z@\global\setbox\islandbox@\vbox\bgroup}
\rightadd@\caption\to\nofrillslist@
\newbox\captionbox@
\newbox\Captionbox@
\def\caption{%
 \ifnum\captioncount@=\z@
  \ifnopunct@
   \DN@{\egroup\nopunct@true}%
  \else
   \let\next@\egroup
  \fi
 \else
  \let\next@\relax
 \fi
 \next@
 \advance\captioncount@\@ne
 \FN@\caption@}
\def\caption@{\ifx\next"\expandafter\caption@q\else\expandafter\caption@@\fi}
\def\caption@q"#1"{\quoted@true
 {\noexpands@
 \let\pre\island@@@P\let\post\island@@@Q
 \let\style\island@@@S\let\numstyle\island@@@N
 \Qlabel@{#1}\let\style\relax\xdef\Qlabel@@@@{#1}}%
 \finishcaption@}
\def\caption@@{\quoted@false
 \global\advance\island@@@C\@ne
 {\noexpands@
 \xdef\Thelabel@@@{\number\island@@@C}%
 \xdefThelabel@\island@@@N
 \xdef\Thelabel@@@@{\island@@@P\Thelabel@\island@@@Q}%
 \xdefThelabel@@\island@@@S
 \xdef\Thepref@{\Thelabel@@@@}}%
 \finishcaption@}
\long\def\captionformat@#1#2#3{\rm\strut#1 {\island@@@F#2} #3%
 \punct@.\strut}
\long\def\widerthanisland@#1#2#3{\test@true\setbox\z@\vbox{\hsize\maxdimen
 \noindent@@\captionformat@{#1}{#2}{#3}\par\setboxzl@}%
 \ifdim\wdz@=\z@
  \global\setbox\captionbox@\hbox{\noset@\unlabel@
   \captionformat@{#1}{#2}{#3}}%
  \ifdim\wd\captionbox@>\wd\islandbox@\else\test@false\fi
 \fi}
\long\def\captionformat@@#1#2#3{\widerthanisland@{#1}{#2}{#3}%
 \iftest@
  \global\setbox\captionbox@\vbox{\hsize\wd\islandbox@
   \vskip-\parskip\noindent@@\noset@\unlabel@
   \captionformat@{#1}{#2}{#3}\par}%
 \else
  \global\setbox\captionbox@
   \hbox to\wd\islandbox@{\hfil\box\captionbox@\hfil}%
 \fi}
\long\def\finishcaption@#1{\def\entry@{#1}%
 {\locallabel@
 \captionformat@@
  {\expandafter\ignorespaces\captionprefix@\unskip}%
  {\ifx\thelabel@@\empty\unskip\else\thelabel@@\fi}%
  {\ignorespaces#1\unskip}%
 \ifnum\captioncount@=\@ne
  \global\setbox\islandbox@\vbox{\ticwrite@\vbox{\box\islandbox@}}%
  \global\setbox\Captionbox@\vbox{\box\captionbox@}%
 \else
  \global\setbox\islandbox@\vbox{\unvbox\islandbox@\setboxzl@
   \ticwrite@\boxz@}%
  \global\setbox\Captionbox@\vbox{\unvbox\Captionbox@
   \smallskip\box\captionbox@}%
 \fi}%
 \nopunct@false\nospace@false\ignorespaces}
\def\Sixtic@{\ifx\macdef@\empty\else
 \DN@##1##2\next@{\def\macdef@{##1##2}}%
 \expandafter\next@\macdef@\next@
 \edef\next@
  {\noexpand\six@\tic@\macdef@
  \space\space\space\space\space\space\space\space\space\space\space\space
  \noexpand\six@}%
 \next@\let\macdef@\relax\fi}
\def\ticwrite@{%
 \iftoc@
  {\noexpands@\let\style\relax
  \DN@{\island}%
  \edef\next@{\write\tic@{%
   \ifnopunct@\noexpand\noexpand\noexpand\nopunct\fi
   \ifx\islandtype@\next@\noexpand\noexpand\noexpand\island
    \noexpand\string\noexpand\c{\islandclass@}{\captionprefix@}%
     {\QorThelabel@@@@}\else\noexpand\noexpand\expandafter\noexpand
     \islandtype@{\QorThelabel@@@@}}\fi}%
  \next@}%
  \expandafter\unmacro@\meaning\entry@\unmacro@
  \Sixtic@
  \write\tic@{\noexpand\Page{\number\pageno}{\page@N}{\page@P}{\page@Q}^^J}%
 \fi}
\def\Htrim@#1{%
 \ifHby@
  \dimen@\vsize
  \ifnum\captioncount@=\z@
  \else
   \advance\dimen@-\ht\Captionbox@
   \advance\dimen@-#1%
  \fi
  \global\Hby@false
  \dimen@ii\wd\islandbox@
  \global\setbox\islandbox@\vbox
   {\unvbox\islandbox@\setboxzl@
   \vbox to\z@{\vss\boxz@}\nointerlineskip\hbyw\dimen@\dimen@ii}%
  \global\Hby@true
 \fi}
\newif\ifdata@
\def\iclasstest@#1{\DN@{#1}\ifx\next@\islandclass@
 \test@true\else\test@false\fi}
\skipdef\skipi@=1
\def\endisland{\ifnum\captioncount@=\z@\expandafter\egroup\fi
 \ifdata@
 \else
  \iclasstest@{T}%
  \iftest@
   {\rm\global\skipi@-\dp\strutbox}\global\advance\skipi@\bigskipamount
   \Htrim@\skipi@
   \global\setbox\islandbox@\vbox
    {\ifnum\captioncount@=\z@\else
     \box\Captionbox@
     \nointerlineskip
     \vskip\skipi@\fi
     \box\islandbox@}%
  \else
   {\rm\global\skipi@\dp\strutbox}\global\advance\skipi@\medskipamount
   \Htrim@\skipi@
   \global\setbox\islandbox@\vbox
    {\box\islandbox@
     \ifnum\captioncount@=\z@\else
     \nointerlineskip
     \vskip\skipi@
     \box\Captionbox@
     \fi}%
  \fi
  \ifHby@
  \else
   \dimen@\ht\islandbox@\advance\dimen@\dp\islandbox@
   \ifdim\dimen@>\vsize
    \DN@{\island}%
    \Err@{%
     \ifx\islandtype@\next@\noexpand\island\else
      \expandafter\noexpand\islandtype@\fi
     \ifnum\captioncount@=\z@\else
       with \noexpand\caption\fi
      is larger than page}%
     \ht\islandbox@=\vsize
   \fi
  \fi
 \fi
 \global\Hby@false\island@true}
\def\newisland#1\c#2#3{\define#1{}%
 \iftoc@\immediate\write\tic@{\noexpand\newisland\noexpand#1%
  \string\c{#2}{#3}^^J}\fi
 \expandafter\def\csname\exstring@#1@S\endcsname{\island@S}%
 \expandafter\def\csname\exstring@#1@N\endcsname{\island@N}%
 \expandafter\def\csname\exstring@#1@P\endcsname{\island@P}%
 \expandafter\def\csname\exstring@#1@Q\endcsname{\island@Q}%
 \expandafter\def\csname\exstring@#1@F\endcsname{\island@F}%
 \expandafter\def\csname end\exstring@#1\endcsname{\endisland}%
 \expandafter
 \ifx\csname island@C#2\endcsname\relax
  \expandafter\newcount@\csname island@C#2\endcsname
  \global\csname island@C#2\endcsname\z@
 \fi
 \edef\next@{\noexpand\expandafter\noexpand\let\noexpand
  \csname\exstring@#1@C\noexpand\endcsname
  \csname island@C#2\endcsname}%
 \next@
 \def#1{\def\islandtype@{#1}\island@c\c{#2}{#3}}}
\newisland\Figure\c{F}{Figure}
\newisland\Table\c{T}{Table}
\newbox\islandboxi
\newbox\islandboxii
\newbox\islandboxiii
\newbox\captionboxi
\newbox\captionboxii
\newbox\captionboxiii
\long\def\islandpairdata#1#2{{\data@true
 \place@true
 #1%
 \global\setbox\islandboxi\box\islandbox@
 \global\setbox\captionboxi\box\Captionbox@
 #2%
 \global\setbox\islandboxii\box\islandbox@
 \global\setbox\captionboxii\box\Captionbox@
 }}
\long\def\islandpairbox#1#2{\islandpairdata{#1}{#2}%
 \dimen@\ht\captionboxi
 \ifdim\ht\captionboxii>\dimen@\dimen@\ht\captionboxii\fi
 \ifdim\dimen@>\z@
  \ifdim\ht\captionboxi<\dimen@
   \global\setbox\captionboxi\vbox to\dimen@{\unvbox\captionboxi\vfil}\fi
  \ifdim\ht\captionboxii<\dimen@
   \global\setbox\captionboxii\vbox to\dimen@{\unvbox\captionboxii\vfil}\fi
 \fi
 \global\setbox\islandbox@\vbox
 {\hbox to\hsize{\hfil\box\islandboxi\hfil\box\islandboxii\hfil}%
 \ifdim\dimen@>\z@\nointerlineskip
 {\rm\global\skipi@\dp\strutbox}\global\advance\skipi@\medskipamount
  \vskip\skipi@
  \hbox to\hsize{\hfil\box\captionboxi\hfil\box\captionboxii\hfil}\fi}}
\long\def\islandpairboxa#1#2{\islandpairdata{#1}{#2}%
 \dimen@\ht\captionboxi
 \ifdim\ht\captionboxii>\dimen@\dimen@\ht\captionboxii\fi
 \ifdim\dimen@>\z@
  \ifdim\ht\captionboxi<\dimen@
   \global\setbox\captionboxi\vbox to\dimen@{\vfil\unvbox\captionboxi}\fi
  \ifdim\ht\captionboxii<\dimen@
   \global\setbox\captionboxii\vbox to\dimen@{\vfil\unvbox\captionboxii}\fi
 \fi
 \dimen@ii\ht\islandboxi
 \ifdim\ht\islandboxii>\dimen@ii \dimen@ii\ht\islandboxii\fi
 \ifdim\dimen@ii>\z@
  \ifdim\ht\islandboxi<\dimen@ii
   \global\setbox\islandboxi\vbox to\dimen@ii{\box\islandboxi\vfil}\fi
  \ifdim\ht\islandboxii<\dimen@ii
   \global\setbox\islandboxii\vbox to\dimen@ii{\box\islandboxii\vfil}\fi
 \fi
 \global\setbox\islandbox@\vbox{\ifdim\dimen@>\z@
  \hbox to\hsize{\hfil\box\captionboxi\hfil\box\captionboxii\hfil}%
  \nointerlineskip{\rm\global\skipi@-\dp\strutbox}%
  \global\advance\skipi@\bigskipamount\vskip\skipi@\fi
  \hbox to\hsize{\hfil\box\islandboxi\hfil\box\islandboxii\hfil}}}
\long\def\islandtripledata#1#2#3{{\data@true\place@true
 #1%
 \global\setbox\islandboxi\box\islandbox@
 \global\setbox\captionboxi\box\Captionbox@
 #2%
 \global\setbox\islandboxii\box\islandbox@
 \global\setbox\captionboxii\box\Captionbox@
 #3%
 \global\setbox\islandboxiii\box\islandbox@
 \global\setbox\captionboxiii\box\Captionbox@
 }}
\long\def\islandtriplebox#1#2#3{\islandtripledata{#1}{#2}{#3}%
 \dimen@\ht\captionboxi
 \ifdim\ht\captionboxii>\dimen@ \dimen@\ht\captionboxii\fi
 \ifdim\ht\captionboxiii>\dimen@ \dimen@\ht\captionboxiii\fi
 \ifdim\dimen@>\z@
  \ifdim\ht\captionboxi<\dimen@
   \global\setbox\captionboxi\vbox to\dimen@{\unvbox\captionboxi\vfil}\fi
  \ifdim\ht\captionboxii<\dimen@
   \global\setbox\captionboxii\vbox to\dimen@{\unvbox\captionboxii\vfil}\fi
  \ifdim\ht\captionboxiii<\dimen@
   \global\setbox\captionboxiii\vbox to\dimen@{\unvbox\captionboxiii\vfil}\fi
 \fi
 \global\setbox\islandbox@\vbox
  {\hbox to\hsize{\hfil\box\islandboxi\hfil\box\islandboxii\hfil
   \box\islandboxiii\hfil}%
 \ifdim\dimen@>\z@\nointerlineskip
  {\rm\global\skipi@\dp\strutbox}\global\advance\skipi@\medskipamount
  \vskip\skipi@
  \hbox to\hsize{\hfil\box\captionboxi\hfil\box\captionboxii\hfil
   \box\captionboxiii\hfil}\fi}}
\def\islandtripleboxa#1#2#3{\islandtripledata{#1}{#2}{#3}%
 \dimen@\ht\captionboxi
 \ifdim\ht\captionboxii>\dimen@ \dimen@\ht\captionboxii\fi
 \ifdim\ht\captionboxiii>\dimen@ \dimen@\ht\captionboxiii\fi
 \ifdim\dimen@>\z@
  \ifdim\ht\captionboxi<\dimen@
   \global\setbox\captionboxi\vbox to\dimen@{\vfil\unvbox\captionboxi}\fi
  \ifdim\ht\captionboxii<\dimen@
   \global\setbox\captionboxii\vbox to\dimen@{\vfil\unvbox\captionboxii}\fi
  \ifdim\ht\captionboxiii<\dimen@
   \global\setbox\captionboxiii\vbox to\dimen@{\vfil\unvbox\captionboxiii}\fi
 \fi
 \dimen@ii\ht\islandboxi
 \ifdim\ht\islandboxii>\dimen@ii \dimen@ii\ht\islandboxii\fi
 \ifdim\ht\islandboxiii>\dimen@ii \dimen@ii\ht\islandboxiii\fi
 \ifdim\dimen@ii>\z@
  \ifdim\ht\islandboxi<\dimen@ii
   \global\setbox\islandboxi\vbox to\dimen@ii{\box\islandboxi\vfil}\fi
  \ifdim\ht\islandboxii<\dimen@ii
   \global\setbox\islandboxii\vbox to\dimen@ii{\box\islandboxii\vfil}\fi
  \ifdim\ht\islandboxiii<\dimen@ii
   \global\setbox\islandboxiii\vbox to\dimen@ii{\box\islandboxiii\vfil}\fi
 \fi
 \global\setbox\islandbox@\vbox
  {\ifdim\dimen@>\z@
  \hbox to\hsize{\hfil\box\captionboxi\hfil\box\captionboxii\hfil
   \box\captionboxiii\hfil}%
  \nointerlineskip{\rm\global\skipi@-\dp\strutbox}%
  \global\advance\skipi@\bigskipamount\vskip\skipi@\fi
  \hbox to\hsize{\hfil\box\islandboxi\hfil\box\islandboxii\hfil
   \box\islandboxiii\hfil}}}
\def\Figurepair#1\and#2\endFigurepair{\island@true
 \islandpairbox{\Figure#1\endFigure}{\Figure#2\endFigure}}
\def\Figuretriple#1\and#2\and#3\endFiguretriple{\island@true
 \islandtriplebox{\Figure#1\endFigure}{\Figure#2\endFigure}%
  {\Figure#3\endFigure}}
\def\Tablepair#1\and#2\endTablepair{\island@true
 \islandpairboxa{\Table#1\endTable}{\Table#2\endTable}}
\def\Tabletriple#1\and#2\and#3\endTabletriple{\island@true
 \islandtripleboxa{\Table#1\endTable}{\Table#2\endTable}%
 {\Table#3\endTable}}
\def\place#1{\place@true\island@false
 #1%
 \ifisland@
  \box\islandbox@
 \else
  \Err@{Whoa ... there's no \string\Figure, \string\Table,
   etc., here}%
 \fi
 \place@false}
\newskip\belowtopfigskip
\belowtopfigskip 15\p@ plus 5\p@ minus5\p@
\newskip\abovebotfigskip
\abovebotfigskip 18\p@ plus 6\p@ minus6\p@
\newdimen\minpagesize
\minpagesize 5pc
\dimen@\belowtopfigskip
\advance\dimen@-\abovebotfigskip
\skip\topins\dimen@
\dimen\topins\z@
\newcount\topinscount@
\newbox\topinsdims@
\def\storedim@{\global\setbox\topinsdims@
 \vbox{\hbox to\dimen@{}\unvbox\topinsdims@}}
\def\advancedimtopins@{%
 \ifnum\pageno=\@ne
 \else
   \advance\dimen@\dimen\topins
   \global\dimen\topins\dimen@
 \fi}
\newcount\flipcount@
\def\fliptopins@{%
 \global\flipcount@\z@
 \ifvoid\topins\else
 \setbox\z@\vbox
  {\vskip\p@
   \unvbox\topins
   \global\setbox\topins\vbox{}%
   \loop
    \test@false
    \ifdim\lastskip=\z@\unskip
     \ifdim\lastskip=\z@
      \test@true\fi\fi
    \iftest@
    \global\advance\flipcount@\@ne
    \setboxzl@
    \global\setbox\topins\vbox{\unvbox\topins\boxz@}%
    \unpenalty
   \repeat}\fi}
\newif\ifPar@
\newcount\Parcount@
\newbox\Parbox@
\expandafter\newbox\csname Parfigbox1\endcsname
\expandafter\newbox\csname Parfigbox2\endcsname
\expandafter\newbox\csname Parfigbox3\endcsname
\expandafter\newbox\csname Parfigbox4\endcsname
\expandafter\newbox\csname Parfigbox5\endcsname
\expandafter\newdimen\csname Parprev1\endcsname
\expandafter\newdimen\csname Parprev2\endcsname
\expandafter\newdimen\csname Parprev3\endcsname
\expandafter\newdimen\csname Parprev4\endcsname
\expandafter\newdimen\csname Parprev5\endcsname
\expandafter\newdimen\csname Parprev6\endcsname
\def\Par{\par\global\csname Parprev1\endcsname\prevdepth
 \global\Parcount@\@ne
 \global\Par@true\global\let\Parlist@\empty
 \global\setbox\Parbox@\vbox\bgroup\break}
\def\place@#1#2{%
 \ifisland@
  \ifhmode
   \ifPar@
    \ifnum\Parcount@>5
     \Err@{Only 5 \string\place's allowed per
      \string\Par...\noexpand\endPar paragraph}%
    \else
     \expandafter\expandafter\expandafter
      \global\expandafter\setbox
       \csname Parfigbox\number\Parcount@\endcsname\box\islandbox@
     \global\advance\Parcount@\@ne
     \xdef\Parlist@{\Parlist@#1}%
    \fi
   \else
    \vadjust{#2}%
   \fi
  \else
   #2%
  \fi
 \else
  \Err@{Whoa ... there's no \string\Figure,
   \string\Table, etc., here}%
 \fi
 \place@false}
\long\def\Aplace#1{\prevanish@
 \place@true\island@false
 #1%
 \place@ a\Aplace@
 \postvanish@}
\long\def\AAplace#1{\prevanish@\place@true\island@false
 #1%
 \place@ A\AAplace@
 \postvanish@}
\newif\ifAA@
\def\AAplace@{\AA@true\Aplace@\AA@false}
\let\AAlist@\empty
\def\Aplace@{\allowbreak
 \dimen@=\ht\islandbox@
 \advance\dimen@\abovebotfigskip
 \ht\islandbox@\dimen@
 \advance\dimen@\dp\islandbox@
 \storedim@
 \ifAA@
  \xdef\AAlist@{\AAlist@1}%
  \advancedimtopins@
 \else
  \xdef\AAlist@{\AAlist@0}%
  \ifnum\topinscount@>\@ne\else\advancedimtopins@\fi
 \fi
 \insert\topins{\penalty\z@\splittopskip\z@\floatingpenalty\z@
  \box\islandbox@}%
 \global\advance\topinscount@\@ne}
\long\def\Bplace#1{\prevanish@\place@true\island@false
 #1%
 \place@ b\Bplace@
 \postvanish@}
\def\Bplace@{\allowbreak
 \ifnum\topinscount@=\z@
  \setbox\z@\vbox{\vbox to-\belowtopfigskip{}}%
  \dimen@-\skip\topins
  \ht\z@\dimen@
  \storedim@
  \advancedimtopins@
  \insert\topins{\boxz@}%
  \global\advance\topinscount@\@ne
  \xdef\AAlist@{\AAlist@0}%
 \fi
 \dimen@\ht\islandbox@
 \advance\dimen@\abovebotfigskip
 \ht\islandbox@\dimen@
 \advance\dimen@\dp\islandbox@
 \storedim@
 \xdef\AAlist@{\AAlist@0}%
 \ifnum\topinscount@>\@ne\else\advancedimtopins@\fi
 \insert\topins{\penalty\z@\splittopskip\z@
  \floatingpenalty\z@
  \box\islandbox@}%
 \global\advance\topinscount@\@ne}
\def\breakisland@{\global\setbox\@ne\lastbox\global\skipi@\lastskip\unskip
 \global\setbox\thr@@\lastbox}%
\def\printisland@{\centerline{\box\thr@@}\nobreak\nointerlineskip
 \vskip\skipi@
 \ifdim\ht\@ne<\z@\box\@ne\else\centerline{\box\@ne}\fi}
\def\bottomfigs@{%
 \count@\@ne
 \loop
  \ifnum\count@<\flipcount@
  \nointerlineskip
  \vskip\abovebotfigskip
  \global\setbox\topins\vbox{\unvbox\topins\setboxzl@
   \unvbox\z@
   \breakisland@}%
  \printisland@
  \advance\count@\@ne
  \repeat}
\def\resetdimtopins@{%
 \global\advance\topinscount@-\flipcount@
 \global\setbox\topinsdims@\vbox
  {\unvbox\topinsdims@
   \count@\z@
   \DN@##1##2\next@{\gdef\AAlist@{##2}}%
   \loop
    \ifnum\count@<\flipcount@\setboxzl@
    \expandafter\next@\AAlist@\next@
    \advance\count@\@ne
    \repeat
   \dimen@\z@
   \count@\z@
   \setbox\tw@\vbox{}%
   \edef\nextiii@{\AAlist@}%
   \DN@##1##2\next@{\DNii@{##1}\def\nextiii@{##2}}%
   \loop
    \test@false
    \ifnum\count@<\topinscount@
    \expandafter\next@\nextiii@\next@
     \ifnum\count@<\tw@
      \test@true
     \else
      \if\nextii@ 1\test@true\fi
     \fi
    \fi
    \iftest@
     \setboxzl@
     \advance\dimen@\wdz@
     \setbox\tw@\vbox{\boxz@\unvbox\tw@}%
     \advance\count@\@ne
    \repeat
    \unvbox\tw@
    \global\dimen\topins\dimen@}}
\def\Place@#1#2{%
 \ifisland@
  \ifhmode
   \ifPar@
    \ifnum\Parcount@>5
     \Err@{Only 5 \string\place's allowed per
       \string\Par...\noexpand\endPar paragraph}%
    \else
     \expandafter\expandafter\expandafter\global\expandafter\setbox
      \csname Parfigbox\number\Parcount@\endcsname\box\islandbox@
     \global\advance\Parcount@\@ne
     \xdef\Parlist@{\Parlist@#1}%
     \vadjust{\break}%
    \fi
   \else
    \Err@{\noexpand#2allowed only in a \string\Par...\noexpand\endPar
     paragraph}%
   \fi
  \else
   #2%
  \fi
 \else
  \Err@{Who ... there's no \string\Figure, \string\Table,
   etc., here}%
 \fi
 \place@false}
\newif\ifC@
\newdimen\Cdim@
\long\def\Cplace#1{\prevanish@\place@true\island@false
 #1%
 \Place@ c\Cplace@
 \postvanish@}
\def\Cplace@{\allowbreak
 \ifnum\topinscount@>\z@\else
  \global\C@true\global\Cdim@\pagetotal\fi
 \Aplace@}
\long\def\Mplace#1{\prevanish@\place@true\island@false
 #1%
 \Place@ m\Mplace@
 \postvanish@}
\long\def\MXplace#1{\prevanish@\place@true\island@false
 #1%
 \Place@ M\MXplace@
 \postvanish@}
\newif\ifMX@
\def\MXplace@{\MX@true\Mplace@\MX@false}
\def\Mplace@{\allowbreak
 \dimen@\ht\islandbox@\advance\dimen@\dp\islandbox@
 \ifdim\pagetotal=\z@\else
  \ifdim\lastskip<\abovebotfigskip\advance\dimen@\abovebotfigskip
  \advance\dimen@-\lastskip\fi
 \fi
 \advance\dimen@\pagetotal
 \ifdim\dimen@>\pagegoal
  \Aplace@
 \else
  \nointerlineskip
  \ifdim\lastskip<\abovebotfigskip\removelastskip\vskip\abovebotfigskip\fi
  \setbox\z@\vbox{\unvbox\islandbox@
   \breakisland@}%
  \printisland@
  \ifnum\topinscount@=\z@
   \setbox\z@\vbox{\vbox to-\belowtopfigskip{}}%
   \dimen@-\skip\topins
   \ht\z@\dimen@
   \storedim@
   \advancedimtopins@
   \insert\topins{\boxz@}%
   \global\advance\topinscount@\@ne
   \xdef\AAlist@{\AAlist@0}%
  \fi
  \ifMX@
   \ifnum\topinscount@=\@ne
    \setbox\z@\vbox{\vbox to-\abovebotfigskip{}}%
    \ht\z@\z@
    \dimen@\z@
    \storedim@
    \advancedimtopins@
    \insert\topins{\boxz@}%
    \global\advance\topinscount@\@ne
    \xdef\AAlist@{\AAlist@0}%
   \fi
  \fi
  \nointerlineskip
  \vskip\belowtopfigskip
 \fi}
\expandafter\newbox\csname Parbox1\endcsname
\expandafter\newbox\csname Parbox2\endcsname
\expandafter\newbox\csname Parbox3\endcsname
\expandafter\newbox\csname Parbox4\endcsname
\expandafter\newbox\csname Parbox5\endcsname
\def\endPar{\egroup
 \count@\@ne
 {\vbadness\@M\vfuzz\maxdimen\splitmaxdepth\maxdimen\splittopskip\ht\strutbox
 \setbox\z@\vsplit\Parbox@ to\ht\Parbox@
 \loop
  \ifnum\count@<\Parcount@
  \expandafter\expandafter\expandafter\global\expandafter\setbox
   \csname Parbox\number\count@\endcsname\vsplit\Parbox@ to\ht\Parbox@
  \count@@\count@\advance\count@@\@ne
  \global\csname Parprev\number\count@@\endcsname
   \dp\csname Parbox\number\count@\endcsname
  \advance\count@\@ne
  \repeat}%
 \vskip\parskip
 \count@\@ne
 \def\nextv@##1##2\nextv@{\DN@{##1}\gdef\Parlist@{##2}}%
 \loop
  \ifnum\count@<\Parcount@
   \dimen@\csname Parprev\number\count@\endcsname
   \advance\dimen@\ht\strutbox
   \ifdim\dimen@<\baselineskip
    \advance\dimen@-\baselineskip\vskip-\dimen@
   \else
    \vskip\lineskip
   \fi
   \unvbox\csname Parbox\number\count@\endcsname
   \global\setbox\islandbox@\box\csname Parfigbox\number\count@\endcsname
   \expandafter\nextv@\Parlist@\nextv@
   \if a\next@\Aplace@\else
   \if A\next@\AAplace@\else
   \if b\next@\Bplace@\else
   \if c\next@\Cplace@\else
   \if m\next@\Mplace@\else
   \if M\next@\MXplace@\fi\fi\fi\fi\fi\fi
  \advance\count@\@ne
  \repeat
 \global\Par@false
 \ifvoid\Parbox@
  \prevdepth\csname Parprev\number\count@\endcsname
 \else
  \dimen@\csname Parprev\number\count@\endcsname\advance\dimen@\ht\strutbox
  \ifdim\dimen@<\baselineskip
    \advance\dimen@-\baselineskip\vskip-\dimen@
  \else
    \vskip\lineskip
  \fi
  \dimen@\dp\Parbox@
  \unvbox\Parbox@
  \prevdepth\dimen@
 \fi}
\def\folio{{\page@F\page@S{\page@P\page@N{\number\page@C}\page@Q}}}
\def\advancepageno{\global\advance\pageno\@ne}
\newif\ifspecialsplit@
\newbox\outbox@
\let\shipout@\shipout
\def\plainoutput{\specialsplit@false\ifvoid\topins\else\ifdim\ht\topins=\z@
 \specialsplit@true\advance\minpagesize-\skip\topins\fi\fi
 \fliptopins@
 \setbox\outbox@\vbox{\makeheadline\pagebody\makefootline}%
 {\noexpands@\let\style\relax
 \shipout@\box\outbox@}%
 \advancepageno
 \resetdimtopins@
 \ifvoid\@cclv\else\unvbox\@cclv\penalty\outputpenalty\fi
 \ifnum\outputpenalty>-\@MM\else\dosupereject\fi}
\def\pagebody{\vbox to\vsize{\boxmaxdepth\maxdepth
 \ifvoid\margin@\else
 \rlap{\kern\hsize\vbox to\z@{\kern4\p@\box\margin@\vss}}\fi
 \pagecontents}}
\newif\ifonlytop@
\def\pagecontents{%
 \onlytop@false
 \ifdim\ht\@cclv<\minpagesize\ifnum\flipcount@<\tw@\ifvoid\footins
  \onlytop@true\fi\fi\fi
 \test@false
 \ifC@
  \ifnum\flipcount@=\@ne
   \global\multiply\Cdim@\tw@
   \ifdim\Cdim@>\ht\@cclv
    \test@true
   \fi
  \fi
 \fi
 \global\C@false
 \iftest@
  \dimen@\ht\@cclv
  \advance\dimen@\skip\topins
  {\vfuzz\maxdimen\vbadness\@M
  \splitmaxdepth\maxdepth\splittopskip\topskip
  \setbox\z@\vsplit\@cclv to\dimen@
  \unvbox\z@}%
  \global\setbox\topins\vbox{\unvbox\topins
   \global\setbox\@ne\lastbox}%
  \setbox\z@\vbox{\unvbox\@ne
   \breakisland@}%
  \nointerlineskip
  \vskip\abovebotfigskip
  \printisland@
 \else
  \ifnum\flipcount@>\z@
   \global\setbox\topins\vbox{\unvbox\topins\global\setbox\@ne\lastbox}%
   \setbox\z@\vbox{\unvbox\@ne
    \breakisland@}%
   \printisland@
   \ifonlytop@\kern-\prevdepth\vfill\else\vskip\belowtopfigskip\fi
  \fi
 \fi
 \ifdim\ht\@cclv<\minpagesize
  \ifonlytop@\else\vfill\fi
 \else
  \ifspecialsplit@
   {\vfuzz\maxdimen\vbadness\@M
   \splitmaxdepth\maxdepth\splittopskip\topskip
   \dimen@ii\ht\@cclv \advance\dimen@ii\skip\topins
   \setbox\z@\vsplit\@cclv to\dimen@ii
   \unvbox\z@}%
  \else
   \unvbox\@cclv
  \fi
 \fi
 \bottomfigs@
 \ifvoid\footins\else\vskip\skip\footins\footnoterule\unvbox\footins\fi}
\newread\readdata@
\def\readthedata@#1{\expandafter
 \ifx\csname#1@D\endcsname\relax
  \immediate\openin\readdata@=#1.dat
  \ifeof\readdata@
   \Err@{No file #1.dat}%
  \else
   {\endlinechar\m@ne\gdef\Next@{}%
   \DNii@##1 ##2 ##3pt{\global\data@ht##1\global\data@dp##2%
    \global\data@wd##3pt}%
   \loop
    \ifeof\readdata@
    \else
    \read\readdata@ to\next@
    \ifx\next@\empty\else
     \edef\next@{\expandafter\nextii@\next@}%
     \expandafter\rightadd@\next@\to\Next@
    \fi
    \repeat}%
   \immediate\closein\readdata@
   \expandafter\expandafter\expandafter\global\expandafter
    \let\csname#1@D\endcsname\Next@\global\let\Next@\relax
  \fi
 \fi}
\newdimen\data@ht
\newdimen\data@dp
\newdimen\data@wd
\newif\ifgetdata@
\def\getdata@#1#2{\global\getdata@true\count@#2\relax
 {\let\\\or\xdef\Next@{\ifcase\number\count@#1\else
 \global\noexpand\getdata@false\fi}}\Next@}
\def\paste#1#2{\readthedata@{#1}%
 \getdata@{\csname#1@D\endcsname}{#2}%
 \ifgetdata@
 \dimen@\data@ht \advance\dimen@\data@dp
  \hbox{\special{dvipaste: #1 #2}%
   \lower\data@dp\vbox to\dimen@{\hbox to\data@wd{}\vfil}}%
 \else
  {\lccode`\Z=`\#\lccode`\N=`\N\lccode`\F=`\F%
   \lowercase{\Err@{No data for File [#1], Z#2}}}%
 \fi}
\newdimen\httable
\newdimen\dptable
\newdimen\wdtable
\def\measuretable#1#2{\readthedata@{#1}%
 \getdata@{\csname#1@D\endcsname}{#2}%
 \ifgetdata@
  \httable\data@ht \dptable\data@dp \wdtable\data@wd
 \else
  {\lccode`\Z=`\#\lccode`\N=`\N\lccode`\F=`\F%
  \lowercase{\Err@{No data for File [#1], Z#2}}}%
 \fi}
\def\East#1#2{\setboxz@h{$\m@th\ssize\;{#1}\;\;$}%
 \setbox\tw@\hbox{$\m@th\ssize\;{#2}\;\;$}\setbox4=\hbox{$\m@th#2$}%
 \dimen@\minaw@
 \ifdim\wdz@>\dimen@\dimen@\wdz@\fi\ifdim\wd\tw@>\dimen@\dimen@\wd\tw@\fi
 \ifdim\wd4 >\z@
  \mathrel{\mathop{\hbox to\dimen@{\rightarrowfill}}\limits^{#1}_{#2}}%
 \else
  \mathrel{\mathop{\hbox to\dimen@{\rightarrowfill}}\limits^{#1}}%
 \fi}
\def\West#1#2{\setboxz@h{$\m@th\ssize\;\;{#1}\;$}%
 \setbox\tw@\hbox{$\m@th\ssize\;\;{#2}\;$}\setbox4=\hbox{$\m@th#2$}%
 \dimen@\minaw@
 \ifdim\wdz@>\dimen@\dimen@\wdz@\fi\ifdim\wd\tw@>\dimen@\dimen@\wd\tw@\fi
 \ifdim\wd4 >\z@
  \mathrel{\mathop{\hbox to\dimen@{\leftarrowfill}}\limits^{#1}_{#2}}%
 \else
  \mathrel{\mathop{\hbox to\dimen@{\leftarrowfill}}\limits^{#1}}%
 \fi}
\font\arrow@i=lams1
\font\arrow@ii=lams2
\font\arrow@iii=lams3
\font\arrow@iv=lams4
\font\arrow@v=lams5
\newdimen\standardcgap
\standardcgap40\p@
\newdimen\hunit
\hunit\tw@\p@
\newdimen\standardrgap
\standardrgap32\p@
\newdimen\vunit
\vunit1.6\p@
\def\Cgaps#1{\RIfM@
 \standardcgap#1\standardcgap\relax\hunit#1\hunit\relax
 \else\nonmatherr@\Cgaps\fi}
\def\Rgaps#1{\RIfM@
 \standardrgap#1\standardrgap\relax\vunit#1\vunit\relax
 \else\nonmatherr@\Rgaps\fi}
\newdimen\getdim@
\def\getcgap@#1{\ifcase#1\or\getdim@\z@\else\getdim@\standardcgap\fi}
\def\getrgap@#1{\ifcase#1\getdim@\z@\else\getdim@\standardrgap\fi}
\def\cgaps{\RIfM@\expandafter\cgaps@\else\expandafter\nonmatherr@
 \expandafter\cgaps\fi}
\def\cgaps@{\ifnum\catcode`\;=\active\expandafter\cgapsA@\else
 \expandafter\cgapsO@\fi}
\def\cgapsO@#1{\toks@{\ifcase\i@\or\getdim@=\z@}%
 \gaps@@\standardcgap#1;\gaps@@\gaps@@
 \edef\next@{\the\toks@\noexpand\else\noexpand\getdim@\noexpand\standardcgap
  \noexpand\fi}%
 \toks@=\expandafter{\next@}%
 \edef\getcgap@##1{\i@##1\relax\the\toks@}\toks@{}}
{\catcode`\;=\active
 \gdef\cgapsA@#1{\toks@{\ifcase\i@\or\getdim@=\z@}%
 \gaps@@\standardcgap#1;\gaps@@\gaps@@
 \edef\next@{\the\toks@\noexpand\else\noexpand\getdim@\noexpand\standardcgap
  \noexpand\fi}%
 \toks@=\expandafter{\next@}%
 \edef\getcgap@##1{\i@##1\relax\the\toks@}\toks@{}}
}
\def\Gaps@@{\gaps@@}
\def\gaps@@#1#2;#3{\mgaps@#1#2\mgaps@
 \edef\next@{\the\toks@\noexpand\or\noexpand\getdim@
  \noexpand#1\the\mgapstoks@@}%
 \toks@\expandafter{\next@}%
 \DN@{#3}%
 \ifx\next@\Gaps@@\def\next@##1\gaps@@{}\else
  \def\next@{\gaps@@#1#3}\fi\next@}
{\catcode`\;=\active
 \gdef\rgaps#1{\RIfM@{\ifnum\catcode`\;=\active\def;{\string;}\fi
   \xdef\Next@{\noexpand\rgaps@{#1}}}%
  \Next@\edef\getrgap@##1{\i@##1\relax\the\toks@}\toks@{}\else
  \nonmatherr@\rgaps\fi}
}
\def\rgaps@#1{\toks@{\ifcase\i@\getdim@=\z@}%
 \gaps@@\standardrgap#1;\gaps@@\gaps@@
 \edef\next@{\the\toks@\noexpand\else\noexpand\getdim@\noexpand\standardrgap
  \noexpand\fi}%
 \toks@=\expandafter{\next@}}
\newbox\ZER@
\def\mgaps@#1{\let\mgapsnext@#1\FNSS@\mgaps@@}
\def\mgaps@@{\ifx\next\w\expandafter\mgaps@@@\else
 \expandafter\mgaps@@@@\fi}
\newtoks\mgapstoks@@
\def\mgaps@@@@#1\mgaps@{\getdim@\mgapsnext@\getdim@#1\getdim@
 \edef\next@{\noexpand\getdim@\the\getdim@}%
 \mgapstoks@@\expandafter{\next@}}
\def\mgaps@@@\w#1#2\mgaps@{\mgaps@@@@#2\mgaps@
 \setbox\ZER@\hbox{$\m@th\hskip15\p@\tsize@#1$}%
 \dimen@\wd\ZER@
 \ifdim\dimen@>\getdim@\getdim@\dimen@\fi
 \edef\next@{\noexpand\getdim@\the\getdim@}%
 \mgapstoks@@\expandafter{\next@}}
\def\changewidth#1#2{\setbox\ZER@{$\m@th#2}%
 \hbox to\wd\ZER@{\hss$\m@th#1$\hss}}
\atdef@({\FN@\ARROW@}
\def\ARROW@{\ifx\next)\let\next@\OPTIONS@\else
 \DN@{\csname\string @(\endcsname}\fi\next@}
\newif\ifoptions@
\def\OPTIONS@){\ifoptions@\let\next@\relax\else
 \DN@{\global\options@true\begingroup\optioncodes@}\fi\next@}
\newif\ifN@
\newif\ifE@
\newif\ifNESW@
\newif\ifH@
\newif\ifV@
\newif\ifHshort@
\expandafter\def\csname\string @(\endcsname #1,#2){%
 \ifoptions@\expandafter\endgroup\fi
 \N@false\E@false\H@false\V@false\Hshort@false
 \ifnum#1>\z@\E@true\fi
 \ifnum#1=\z@\V@true\global\tX@false\global\tY@false\global\a@false\fi
 \ifnum#2>\z@\N@true\fi
 \ifnum#2=\z@\H@true\global\tX@false\global\tY@false\global\a@false
  \ifshort@\Hshort@true\fi\fi
 \NESW@false
 \ifN@\ifE@\NESW@true\fi\else\ifE@\else\NESW@true\fi\fi
 \arrow@{#1}{#2}%
 \global\options@false
 \global\scount@\z@\global\tcount@\z@\global\arrcount@\z@
 \global\s@false\global\sxdimen@\z@\global\sydimen@\z@
 \global\tX@false\global\tXdimen@i\z@\global\tXdimen@ii\z@
 \global\tY@false\global\tYdimen@i\z@\global\tYdimen@ii\z@
 \global\a@false\global\exacount@\z@
 \global\x@false\global\xdimen@\z@
 \global\X@false\global\Xdimen@\z@
 \global\y@false\global\ydimen@\z@
 \global\Y@false\global\Ydimen@\z@
 \global\p@false\global\pdimen@\z@
 \global\label@ifalse\global\label@iifalse
 \global\dl@ifalse\global\ldimen@i\z@
 \global\dl@iifalse\global\ldimen@ii\z@
 \global\short@false\global\unshort@false}
\newif\iflabel@i
\newif\iflabel@ii
\newcount\scount@
\newcount\tcount@
\newcount\arrcount@
\newif\ifs@
\newdimen\sxdimen@
\newdimen\sydimen@
\newif\iftX@
\newdimen\tXdimen@i
\newdimen\tXdimen@ii
\newif\iftY@
\newdimen\tYdimen@i
\newdimen\tYdimen@ii
\newif\ifa@
\newcount\exacount@
\newif\ifx@
\newdimen\xdimen@
\newif\ifX@
\newdimen\Xdimen@
\newif\ify@
\newdimen\ydimen@
\newif\ifY@
\newdimen\Ydimen@
\newif\ifp@
\newdimen\pdimen@
\newif\ifdl@i
\newif\ifdl@ii
\newdimen\ldimen@i
\newdimen\ldimen@ii
\newif\ifshort@
\newif\ifunshort@
\def\zero@#1{\ifnum\scount@=\z@
 \if#1e\global\scount@\m@ne\else
 \if#1t\global\scount@\tw@\else
 \if#1h\global\scount@\thr@@\else
 \if#1'\global\scount@6 \else
 \if#1`\global\scount@7 \else
 \if#1(\global\scount@8 \else
 \if#1)\global\scount@9 \else
 \if#1s\global\scount@12 \else
 \if#1H\global\scount@13 \else
 \Err@{\Invalid@@ option \string\0}\fi\fi\fi\fi\fi\fi\fi\fi\fi
 \fi}
\def\one@#1{\ifnum\tcount@=\z@
 \if#1e\global\tcount@\m@ne\else
 \if#1h\global\tcount@\tw@\else
 \if#1t\global\tcount@\thr@@\else
 \if#1'\global\tcount@4 \else
 \if#1`\global\tcount@5 \else
 \if#1(\global\tcount@\ten@ \else
 \if#1)\global\tcount@11 \else
 \if#1s\global\tcount@12 \else
 \if#1H\global\tcount@13 \else
 \Err@{\Invalid@@ option \string\1}\fi\fi\fi\fi\fi\fi\fi\fi\fi
 \fi}
\def\a@#1{\ifnum\arrcount@=\z@
 \if#10\global\arrcount@\m@ne\else
 \if#1+\global\arrcount@\@ne\else
 \if#1-\global\arrcount@\tw@\else
 \if#1=\global\arrcount@\thr@@\else
 \Err@{\Invalid@@ option \string\a}\fi\fi\fi\fi
 \fi}
\def\ds@{\ifnum\catcode`\;=\active\expandafter\dsA@\else
 \expandafter\dsO@\fi}
\def\dsO@(#1;#2){\ds@@{#1}{#2}}
\def\ds@@#1#2{\ifs@\else
 \global\s@true
 \global\sxdimen@\hunit\global\sxdimen@#1\sxdimen@\relax
 \global\sydimen@\vunit\global\sydimen@#2\sydimen@\relax
 \fi}
\def\dtX@{\ifnum\catcode`\;=\active\expandafter\dtXA@\else
 \expandafter\dtXO@\fi}
\def\dtXO@(#1;#2){\dtX@@{#1}{#2}}
\def\dtX@@#1#2{\iftX@\else
 \global\tX@true
 \global\tXdimen@i\hunit\global\tXdimen@i#1\tXdimen@i\relax
 \global\tXdimen@ii\vunit\global\tXdimen@ii#2\tXdimen@ii\relax
 \fi}
\def\dtY@{\ifnum\catcode`\;=\active\expandafter\dtYA@\else
 \expandafter\dtYO@\fi}
\def\dtYO@(#1;#2){\dtY@@{#1}{#2}}
\def\dtY@@#1#2{\iftY@\else
 \global\tY@true
 \global\tYdimen@i\hunit\global\tYdimen@i#1\tYdimen@i\relax
 \global\tYdimen@ii\vunit\global\tYdimen@ii#2\tYdimen@ii\relax
 \fi}
{\catcode`\;=\active
 \gdef\dsA@(#1;#2){\ds@@{#1}{#2}}
 \gdef\dtXA@(#1;#2){\dtX@@{#1}{#2}}
 \gdef\dtYA@(#1;#2){\dtY@@{#1}{#2}}
}
\def\da@#1{\ifa@\else\global\a@true\global\exacount@#1\relax\fi}
\def\dx@#1{\ifx@\else
 \global\x@true
 \global\xdimen@\hunit\global\xdimen@#1\xdimen@\relax
 \fi}
\def\dX@#1{\ifX@\else
 \global\X@true
 \global\Xdimen@\hunit\global\Xdimen@#1\Xdimen@\relax
 \fi}
\def\dy@#1{\ify@\else
 \global\y@true
 \global\ydimen@\vunit\global\ydimen@#1\ydimen@\relax
 \fi}
\def\dY@#1{\ifY@\else
 \global\Y@true
 \global\Ydimen@\vunit\global\Ydimen@#1\Ydimen@\relax
 \fi}
\def\p@@#1{\ifp@\else
 \global\p@true
 \global\pdimen@\hunit\global\divide\pdimen@\tw@
 \global\pdimen@#1\pdimen@\relax
 \fi}
\def\L@#1{\iflabel@i\else
 \global\label@itrue\gdef\label@i{#1}%
 \fi}
\def\l@#1{\iflabel@ii\else
 \global\label@iitrue\gdef\label@ii{#1}%
 \fi}
\def\dL@#1{\ifdl@i\else
 \global\dl@itrue\global\ldimen@i\hunit\global\ldimen@i#1\ldimen@i\relax
 \fi}
\def\dl@#1{\ifdl@ii\else
 \global\dl@iitrue\global\ldimen@ii\hunit\global\ldimen@ii#1\ldimen@ii\relax
 \fi}
\def\s@{\ifunshort@\else\global\short@true\fi}
\def\uns@{\ifshort@\else\global\unshort@true\global\short@false\fi}
\def\optioncodes@{\let\0\zero@\let\1\one@\let\a\a@\let\ds\ds@\let\dtX\dtX@
 \let\dtY\dtY@\let\da\da@\let\dx\dx@\let\dX\dX@\let\dY\dY@\let\dy\dy@
 \let\p\p@@\let\L\L@\let\l\l@\let\dL\dL@\let\dl\dl@\let\s\s@\let\uns\uns@}
\def\slopes@{\\161\\152\\143\\134\\255\\126\\357\\238\\349\\45{10}\\56{11}%
 \\11{12}\\65{13}\\54{14}\\43{15}\\32{16}\\53{17}\\21{18}\\52{19}\\31{20}%
 \\41{21}\\51{22}\\61{23}}
\newcount\tan@i
\newcount\tan@ip
\newcount\tan@ii
\newcount\tan@iip
\newdimen\slope@i
\newdimen\slope@ip
\newdimen\slope@ii
\newdimen\slope@iip
\newcount\angcount@
\newcount\extracount@
\def\slope@{{\slope@i\secondy@\advance\slope@i-\firsty@
 \ifN@\else\multiply\slope@i\m@ne\fi
 \slope@ii\secondx@\advance\slope@ii-\firstx@
 \ifE@\else\multiply\slope@ii\m@ne\fi
 \ifdim\slope@ii<\z@
  \global\tan@i6 \global\tan@ii\@ne\global\angcount@23
 \else
  \dimen@\slope@i\multiply\dimen@6
  \ifdim\dimen@<\slope@ii
   \global\tan@i\@ne\global\tan@ii6 \global\angcount@\@ne
  \else
   \dimen@\slope@ii\multiply\dimen@6
   \ifdim\dimen@<\slope@i
    \global\tan@i6 \global\tan@ii\@ne\global\angcount@23
   \else
    \global\tan@ip\z@\global\tan@iip\@ne
    \def\\##1##2##3{\global\angcount@##3\relax
     \slope@ip\slope@i\slope@iip\slope@ii
     \multiply\slope@iip##1\relax\multiply\slope@ip##2\relax
     \ifdim\slope@iip<\slope@ip
      \global\tan@ip##1\relax\global\tan@iip##2\relax
     \else
      \global\tan@i##1\relax\global\tan@ii##2\relax
      \def\\####1####2####3{}%
     \fi}%
    \slopes@
    \slope@i\secondy@\advance\slope@i-\firsty@
    \ifN@\else\multiply\slope@i\m@ne\fi
    \multiply\slope@i\tan@ii\multiply\slope@i\tan@iip\multiply\slope@i\tw@
    \count@\tan@i\multiply\count@\tan@iip
    \extracount@\tan@ip\multiply\extracount@\tan@ii
    \advance\count@\extracount@
    \slope@ii\secondx@\advance\slope@ii-\firstx@
    \ifE@\else\multiply\slope@ii\m@ne\fi
    \multiply\slope@ii\count@
    \ifdim\slope@i<\slope@ii
     \global\tan@i\tan@ip\global\tan@ii\tan@iip
     \global\advance\angcount@\m@ne
    \fi
   \fi
  \fi
 \fi}%
}
\def\slope@a#1{{\def\\##1##2##3{\ifnum##3=#1\global\tan@i##1\relax
 \global\tan@ii##2\relax\fi}\slopes@}}
\newcount\i@
\newcount\j@
\newcount\colcount@
\newcount\Colcount@
\newcount\tcolcount@
\newdimen\rowht@
\newdimen\rowdp@
\newcount\rowcount@
\newcount\Rowcount@
\newcount\maxcolrow@
\newtoks\colwidthtoks@
\newtoks\Rowheighttoks@
\newtoks\Rowdepthtoks@
\newtoks\widthtoks@
\newtoks\Widthtoks@
\newtoks\heighttoks@
\newtoks\Heighttoks@
\newtoks\depthtoks@
\newtoks\Depthtoks@
\newif\iffirstCDcr@
\def\dotoks@i{%
 \global\widthtoks@\expandafter{\the\widthtoks@\else\getdim@\z@\fi}%
 \global\heighttoks@\expandafter{\the\heighttoks@\else\getdim@\z@\fi}%
 \global\depthtoks@\expandafter{\the\depthtoks@\else\getdim@\z@\fi}}
\def\dotoks@ii{%
 \global\widthtoks@{\ifcase\j@}%
 \global\heighttoks@{\ifcase\j@}%
 \global\depthtoks@{\ifcase\j@}}
\def\preCD@#1\endCD{\setbox\ZER@
 \vbox{%
  \def\arrow@##1##2{{}}%
  \global\rowcount@\m@ne\global\colcount@\z@\global\Colcount@\z@
  \global\firstCDcr@true\toks@{}%
  \global\widthtoks@{\ifcase\j@}%
  \global\Widthtoks@{\ifcase\i@}%
  \global\heighttoks@{\ifcase\j@}%
  \global\Heighttoks@{\ifcase\i@}%
  \global\depthtoks@{\ifcase\j@}%
  \global\Depthtoks@{\ifcase\i@}%
  \global\Rowheighttoks@{\ifcase\i@}%
  \global\Rowdepthtoks@{\ifcase\i@}%
  \Let@
  \everycr{%
   \noalign{%
    \global\advance\rowcount@\@ne
    \ifnum\colcount@<\Colcount@
    \else
     \global\Colcount@\colcount@\global\maxcolrow@\rowcount@
    \fi
    \global\colcount@\z@
    \iffirstCDcr@
     \global\firstCDcr@false
    \else
     \edef\next@{\the\Rowheighttoks@\noexpand\or\noexpand\getdim@\the\rowht@}%
      \global\Rowheighttoks@\expandafter{\next@}%
     \edef\next@{\the\Rowdepthtoks@\noexpand\or\noexpand\getdim@\the\rowdp@}%
      \global\Rowdepthtoks@\expandafter{\next@}%
     \global\rowht@\z@\global\rowdp@\z@
     \dotoks@i
     \edef\next@{\the\Widthtoks@\noexpand\or\the\widthtoks@}%
      \global\Widthtoks@\expandafter{\next@}%
     \edef\next@{\the\Heighttoks@\noexpand\or\the\heighttoks@}%
      \global\Heighttoks@\expandafter{\next@}%
     \edef\next@{\the\Depthtoks@\noexpand\or\the\depthtoks@}%
      \global\Depthtoks@\expandafter{\next@}%
     \dotoks@ii
    \fi}}%
  \tabskip\z@
  \halign{&\setbox\ZER@\hbox{\vrule\height\ten@\p@\width\z@\depth\z@     %1
   $\m@th\displaystyle{##}$}\copy\ZER@
   \ifdim\ht\ZER@>\rowht@\global\rowht@\ht\ZER@\fi
   \ifdim\dp\ZER@>\rowdp@\global\rowdp@\dp\ZER@\fi
   \global\advance\colcount@\@ne
   \edef\next@{\the\widthtoks@\noexpand\or\noexpand\getdim@\the\wd\ZER@}%
    \global\widthtoks@\expandafter{\next@}%
   \edef\next@{\the\heighttoks@\noexpand\or\noexpand\getdim@\the\ht\ZER@}%
    \global\heighttoks@\expandafter{\next@}%
   \edef\next@{\the\depthtoks@\noexpand\or\noexpand\getdim@\the\dp\ZER@}%
    \global\depthtoks@\expandafter{\next@}%
   \cr#1\crcr}}%
 \Rowcount@\rowcount@
 \global\Widthtoks@\expandafter{\the\Widthtoks@\fi\relax}%
 \edef\Width@##1##2{\i@##1\relax\j@##2\relax\the\Widthtoks@}%
 \global\Heighttoks@\expandafter{\the\Heighttoks@\fi\relax}%
 \edef\Height@##1##2{\i@##1\relax\j@##2\relax\the\Heighttoks@}%
 \global\Depthtoks@\expandafter{\the\Depthtoks@\fi\relax}%
 \edef\Depth@##1##2{\i@##1\relax\j@##2\relax\the\Depthtoks@}%
 \edef\next@{\the\Rowheighttoks@\noexpand\fi\relax}%
 \global\Rowheighttoks@\expandafter{\next@}%
 \edef\Rowheight@##1{\i@##1\relax\the\Rowheighttoks@}%
 \edef\next@{\the\Rowdepthtoks@\noexpand\fi\relax}%
 \global\Rowdepthtoks@\expandafter{\next@}%
 \edef\Rowdepth@##1{\i@##1\relax\the\Rowdepthtoks@}%
 \global\colwidthtoks@{\fi}%
 \setbox\ZER@\vbox{%
  \unvbox\ZER@
  \count@\rowcount@
  \loop
   \unskip\unpenalty
   \setbox\ZER@\lastbox
   \ifnum\count@>\maxcolrow@\advance\count@\m@ne
   \repeat
  \hbox{%
   \unhbox\ZER@
   \count@\z@
   \loop
    \unskip
    \setbox\ZER@\lastbox
    \edef\next@{\noexpand\or\noexpand\getdim@\the\wd\ZER@\the\colwidthtoks@}%
     \global\colwidthtoks@\expandafter{\next@}%
    \advance\count@\@ne
    \ifnum\count@<\Colcount@
    \repeat}}%
 \edef\next@{\noexpand\ifcase\noexpand\i@\the\colwidthtoks@}%
  \global\colwidthtoks@\expandafter{\next@}%
 \edef\Colwidth@##1{\i@##1\relax\the\colwidthtoks@}%
 \global\colwidthtoks@{}\global\Rowheighttoks@{}\global\Rowdepthtoks@{}%
 \global\widthtoks@{}\global\Widthtoks@{}\global\heighttoks@{}%
 \global\Heighttoks@{}\global\depthtoks@{}\global\Depthtoks@{}%
}
\newcount\xoff@
\newcount\yoff@
\newcount\endcount@
\newcount\rcount@
\newdimen\firstx@
\newdimen\firsty@
\newdimen\secondx@
\newdimen\secondy@
\newdimen\tocenter@
\newdimen\charht@
\newdimen\charwd@
\def\outside@{\Err@{This arrow points outside the \string\CD}}
\newif\ifsvertex@
\newif\iftvertex@
\def\arrow@#1#2{\global\xoff@#1\relax\global\yoff@#2\relax
 \count@\rowcount@\advance\count@-\yoff@
 \ifnum\count@<\@ne\outside@\else\ifnum\count@>\Rowcount@\outside@\fi\fi
 \count@\colcount@\advance\count@\xoff@
 \ifnum\count@<\@ne\outside@\else\ifnum\count@>\Colcount@\outside@\fi\fi
 \tcolcount@\colcount@\advance\tcolcount@\xoff@
 \Width@\rowcount@\colcount@\divide\getdim@\tw@\tocenter@-\getdim@
 \ifdim\getdim@=\z@
  \firstx@\z@\firsty@\mathaxis@\svertex@true
 \else
  \svertex@false
  \ifHshort@
   \Colwidth@\colcount@\divide\getdim@\tw@
   \ifE@ \firstx@\getdim@ \else \firstx@-\getdim@ \fi
  \else
   \ifE@ \firstx@\getdim@ \else \firstx@-\getdim@ \fi
  \fi
  \ifE@
   \ifH@ \advance\firstx@\thr@@\p@ \else \advance\firstx@-\thr@@\p@ \fi  %2
  \else
   \ifH@ \advance\firstx@-\thr@@\p@ \else \advance\firstx@\thr@@\p@ \fi  %3
  \fi
  \ifN@
   \Height@\rowcount@\colcount@ \firsty@\getdim@                         %4
   \ifV@ \advance\firsty@\thr@@\p@ \fi                                   %5
  \else
   \ifV@
    \Depth@\rowcount@\colcount@ \firsty@-\getdim@                        %6
    \advance\firsty@-\thr@@\p@                                           %7
   \else
    \firsty@\z@                                                          %8
   \fi
  \fi
 \fi
 \ifV@
 \else
  \Colwidth@\colcount@\divide\getdim@\tw@
  \ifE@\secondx@\getdim@\else\secondx@-\getdim@\fi
  \ifE@\else\getcgap@\colcount@\advance\secondx@-\getdim@\fi
  \endcount@\colcount@\advance\endcount@\xoff@
  \count@\colcount@
  \ifE@
   \advance\count@\@ne
   \loop
    \ifnum\count@<\endcount@
    \Colwidth@\count@\advance\secondx@\getdim@
    \getcgap@\count@\advance\secondx@\getdim@
    \advance\count@\@ne
    \repeat
  \else
   \advance\count@\m@ne
   \loop
    \ifnum\count@>\endcount@
    \Colwidth@\count@\advance\secondx@-\getdim@
    \getcgap@\count@\advance\secondx@-\getdim@
    \advance\count@\m@ne
    \repeat
  \fi
  \Colwidth@\count@\divide\getdim@\tw@
  \ifHshort@
  \else
   \ifE@\advance\secondx@\getdim@\else\advance\secondx@-\getdim@\fi
  \fi
  \ifE@\getcgap@\count@\advance\secondx@\getdim@\fi
  \rcount@\rowcount@\advance\rcount@-\yoff@
  \Width@\rcount@\count@\divide\getdim@\tw@
  \tvertex@false
  \ifH@\ifdim\getdim@=\z@\tvertex@true\Hshort@false\fi\fi
  \ifHshort@
  \else
   \ifE@\advance\secondx@-\getdim@\else\advance\secondx@\getdim@\fi
  \fi
  \iftvertex@
   \advance\secondx@.4\p@
  \else
   \ifE@\advance\secondx@-\thr@@\p@\else\advance\secondx@\thr@@\p@\fi    %9
  \fi
 \fi
 \ifH@
 \else
  \ifN@
   \Rowheight@\rowcount@\secondy@\getdim@
  \else
   \Rowdepth@\rowcount@\secondy@-\getdim@
   \getrgap@\rowcount@\advance\secondy@-\getdim@
  \fi
  \endcount@\rowcount@\advance\endcount@-\yoff@
  \count@\rowcount@
  \ifN@
   \advance\count@\m@ne
   \loop
    \ifnum\count@>\endcount@
    \Rowheight@\count@\advance\secondy@\getdim@
    \Rowdepth@\count@\advance\secondy@\getdim@
    \getrgap@\count@\advance\secondy@\getdim@
    \advance\count@\m@ne
    \repeat
  \else
   \advance\count@\@ne
   \loop
    \ifnum\count@<\endcount@
    \Rowheight@\count@\advance\secondy@-\getdim@
    \Rowdepth@\count@\advance\secondy@-\getdim@
    \getrgap@\count@\advance\secondy@-\getdim@
    \advance\count@\@ne
    \repeat
  \fi
  \tvertex@false
  \ifV@\Width@\count@\colcount@\ifdim\getdim@=\z@\tvertex@true\fi\fi
  \ifN@
   \getrgap@\count@\advance\secondy@\getdim@
   \Rowdepth@\count@\advance\secondy@\getdim@
   \iftvertex@
    \advance\secondy@\mathaxis@
   \else
    \Depth@\count@\tcolcount@\advance\secondy@-\getdim@
    \advance\secondy@-\thr@@\p@                                          %10
   \fi
  \else
   \Rowheight@\count@\advance\secondy@-\getdim@
   \iftvertex@
    \advance\secondy@\mathaxis@
   \else
    \Height@\count@\tcolcount@\advance\secondy@\getdim@
    \advance\secondy@\thr@@\p@                                           %11
   \fi
  \fi
 \fi
 \ifV@\else\advance\firstx@\sxdimen@\fi
 \ifH@\else\advance\firsty@\sydimen@\fi
 \iftX@
  \advance\secondy@\tXdimen@ii
  \advance\secondx@\tXdimen@i
  \slope@
 \else
  \iftY@
   \advance\secondy@\tYdimen@ii
   \advance\secondx@\tYdimen@i
   \slope@
   \secondy@\secondx@\advance\secondy@-\firstx@
   \ifNESW@\else\multiply\secondy@\m@ne\fi
   \multiply\secondy@\tan@i\divide\secondy@\tan@ii\advance\secondy@\firsty@
  \else
   \ifa@
    \slope@
    \ifNESW@\global\advance\angcount@\exacount@\else
     \global\advance\angcount@-\exacount@\fi
    \ifnum\angcount@>23 \global\angcount@23 \fi
    \ifnum\angcount@<\@ne\global\angcount@\@ne\fi
    \slope@a\angcount@
    \ifY@
     \advance\secondy@\Ydimen@
    \else
     \ifX@
      \advance\secondx@\Xdimen@
      \dimen@\secondx@\advance\dimen@-\firstx@
      \ifNESW@\else\multiply\dimen@\m@ne\fi
      \multiply\dimen@\tan@i\divide\dimen@\tan@ii
      \advance\dimen@\firsty@\secondy@\dimen@
     \fi
    \fi
   \else
    \ifH@\else\ifV@\else\slope@\fi\fi
   \fi
  \fi
 \fi
 \ifH@\else\ifV@\else\ifsvertex@\else
  \dimen@6\p@\multiply\dimen@\tan@ii
  \count@\tan@i\advance\count@\tan@ii\divide\dimen@\count@
  \ifE@\advance\firstx@\dimen@\else\advance\firstx@-\dimen@\fi
  \multiply\dimen@\tan@i\divide\dimen@\tan@ii
  \ifN@\advance\firsty@\dimen@\else\advance\firsty@-\dimen@\fi
 \fi\fi\fi
 \ifp@
  \ifH@\else\ifV@\else
   \getcos@\pdimen@\advance\firsty@\dimen@\advance\secondy@\dimen@
   \ifNESW@\advance\firstx@-\dimen@ii\else\advance\firstx@\dimen@ii\fi
  \fi\fi
 \fi
 \ifH@\else\ifV@\else
  \ifnum\tan@i>\tan@ii
   \charht@\ten@\p@\charwd@\ten@\p@
   \multiply\charwd@\tan@ii\divide\charwd@\tan@i
  \else
   \charwd@\ten@\p@\charht@\ten@\p@
   \divide\charht@\tan@ii\multiply\charht@\tan@i
  \fi
  \ifnum\tcount@=\thr@@
   \ifN@\advance\secondy@-.3\charht@\else\advance\secondy@.3\charht@\fi
  \fi
  \ifnum\scount@=\tw@
   \ifE@\advance\firstx@.3\charht@\else\advance\firstx@-.3\charht@\fi
  \fi
  \ifnum\tcount@=12
   \ifN@\advance\secondy@-\charht@\else\advance\secondy@\charht@\fi
  \fi
  \iftY@
  \else
   \ifa@
    \ifX@
    \else
     \secondx@\secondy@\advance\secondx@-\firsty@
     \ifNESW@\else\multiply\secondx@\m@ne\fi
     \multiply\secondx@\tan@ii\divide\secondx@\tan@i
     \advance\secondx@\firstx@
    \fi
   \fi
  \fi
 \fi\fi
 \ifH@\harrow@\else\ifV@\varrow@\else\arrow@@\fi\fi}
\newdimen\mathaxis@
\mathaxis@90\p@\divide\mathaxis@36
\def\harrow@b{\ifE@\hskip\tocenter@\hskip\firstx@\fi}
\def\harrow@bb{\ifE@\hskip\xdimen@\else\hskip\Xdimen@\fi}
\def\harrow@e{\ifE@\else\hskip-\firstx@\hskip-\tocenter@\fi}
\def\harrow@ee{\ifE@\hskip-\Xdimen@\else\hskip-\xdimen@\fi}
\def\harrow@{\dimen@\secondx@\advance\dimen@-\firstx@
 \ifE@\let\next@\rlap\else\multiply\dimen@\m@ne\let\next@\llap\fi
 \next@{%
  \harrow@b
  \smash{\raise\pdimen@\hbox to\dimen@
   {\harrow@bb\arrow@ii
    \ifnum\arrcount@=\m@ne\else\ifnum\arrcount@=\thr@@\else
     \ifE@
      \ifnum\scount@=\m@ne
      \else
       \ifcase\scount@\or\or\char118 \or\char117 \or\or\or\char119 \or
       \char120 \or\char121 \or\char122 \or\or\or\arrow@i\char125 \or
       \char117 \hskip\thr@@\p@\char117 \hskip-\thr@@\p@\fi
      \fi
     \else
      \ifnum\tcount@=\m@ne
      \else
       \ifcase\tcount@\char117 \or\or\char117 \or\char118 \or\char119 \or
       \char120 \or\or\or\or\or\char121 \or\char122 \or\arrow@i\char125
       \or\char117 \hskip\thr@@\p@\char117 \hskip-\thr@@\p@\fi
      \fi
     \fi
    \fi\fi
    \dimen@\mathaxis@\advance\dimen@.2\p@
    \dimen@ii\mathaxis@\advance\dimen@ii-.2\p@
    \ifnum\arrcount@=\m@ne
     \let\leads@\null
    \else
     \ifcase\arrcount@
      \def\leads@{\hrule\height\dimen@\depth-\dimen@ii}\or
      \def\leads@{\hrule\height\dimen@\depth-\dimen@ii}\or
      \def\leads@{\hbox to\ten@\p@{%
       \leaders\hrule\height\dimen@\depth-\dimen@ii\hfil
       \hfil
      \leaders\hrule\height\dimen@\depth-\dimen@ii\hskip\z@ plus2fil\relax
       \hfil
       \leaders\hrule\height\dimen@\depth-\dimen@ii\hfil}}\or
     \def\leads@{\hbox{\hbox to\ten@\p@{\dimen@\mathaxis@\advance\dimen@1.2\p@
       \dimen@ii\dimen@\advance\dimen@ii-.4\p@
       \leaders\hrule\height\dimen@\depth-\dimen@ii\hfil}%
       \kern-\ten@\p@
       \hbox to\ten@\p@{\dimen@\mathaxis@\advance\dimen@-1.2\p@
       \dimen@ii\dimen@\advance\dimen@ii-.4\p@
       \leaders\hrule\height\dimen@\depth-\dimen@ii\hfil}}}\fi
    \fi
    \cleaders\leads@\hfil
    \ifnum\arrcount@=\m@ne\else\ifnum\arrcount@=\thr@@\else
     \arrow@i
     \ifE@
      \ifnum\tcount@=\m@ne
      \else
       \ifcase\tcount@\char119 \or\or\char119 \or\char120 \or\char121 \or
       \char122 \or \or\or\or\or\char123 \or\char124 \or
       \char125 \or\char119 \hskip-\thr@@\p@\char119 \hskip\thr@@\p@\fi
      \fi
     \else
      \ifcase\scount@\or\or\char120 \or\char119 \or\or\or\char121 \or\char122
      \or\char123 \or\char124 \or\or\or\char125 \or
      \char119 \hskip-\thr@@\p@\char119 \hskip\thr@@\p@\fi
     \fi
    \fi\fi
    \harrow@ee}}%
  \harrow@e}%
 \iflabel@i
  \dimen@ii\z@\setbox\ZER@\hbox{$\m@th\tsize@@\label@i$}%
  \ifnum\arrcount@=\m@ne
  \else
   \advance\dimen@ii\mathaxis@
   \advance\dimen@ii\dp\ZER@\advance\dimen@ii\tw@\p@
   \ifnum\arrcount@=\thr@@\advance\dimen@ii\tw@\p@\fi
  \fi
  \advance\dimen@ii\pdimen@
  \next@{\harrow@b\smash{\raise\dimen@ii\hbox to\dimen@
   {\harrow@bb\hskip\tw@\ldimen@i\hfil\box\ZER@\hfil\harrow@ee}}\harrow@e}%
 \fi
 \iflabel@ii
  \ifnum\arrcount@=\m@ne
  \else
   \setbox\ZER@\hbox{$\m@th\tsize@\label@ii$}%
   \dimen@ii-\ht\ZER@\advance\dimen@ii-\tw@\p@
   \ifnum\arrcount@=\thr@@\advance\dimen@ii-\tw@\p@\fi
   \advance\dimen@ii\mathaxis@\advance\dimen@ii\pdimen@
   \next@{\harrow@b\smash{\raise\dimen@ii\hbox to\dimen@
    {\harrow@bb\hskip\tw@\ldimen@ii\hfil\box\ZER@\hfil\harrow@ee}}\harrow@e}%
  \fi
 \fi}
\let\tsize@\tsize
\def\tsizeCDlabels{\let\tsize@\tsize}
\def\ssizeCDlabels{\let\tsize@\ssize}
\def\tsize@@{\ifnum\arrcount@=\m@ne\else\tsize@\fi}
\def\varrow@{\dimen@\secondy@\advance\dimen@-\firsty@
 \ifN@\else\multiply\dimen@\m@ne\fi
 \setbox\ZER@\vbox to\dimen@
  {\ifN@\vskip-\Ydimen@\else\vskip\ydimen@\fi
   \ifnum\arrcount@=\m@ne\else\ifnum\arrcount@=\thr@@\else
    \hbox{\arrow@iii
     \ifN@
      \ifnum\tcount@=\m@ne
      \else
       \ifcase\tcount@\char117 \or\or\char117 \or\char118 \or\char119 \or
       \char120 \or\or\or\or\or\char121 \or\char122 \or\char123 \or
       \vbox{\hbox{\char117}\nointerlineskip\vskip\thr@@\p@
       \hbox{\char117}\vskip-\thr@@\p@}\fi
      \fi
     \else
      \ifcase\scount@\or\or\char118 \or\char117 \or\or\or\char119 \or
      \char120 \or\char121 \or\char122 \or\or\or\char123 \or
      \vbox{\hbox{\char117}\nointerlineskip\vskip\thr@@\p@
      \hbox{\char117}\vskip-\thr@@\p@}\fi
     \fi}%
    \nointerlineskip
   \fi\fi
   \ifnum\arrcount@=\m@ne
    \let\leads@\null
   \else
    \ifcase\arrcount@\let\leads@\vrule\or\let\leads@\vrule\or
    \def\leads@{\vbox to\ten@\p@{%
     \hrule\height1.67\p@\depth\z@\width.4\p@
     \vfil
     \hrule\height3.33\p@\depth\z@\width.4\p@
     \vfil
     \hrule\height1.67\p@\depth\z@\width.4\p@}}\or
    \def\leads@{\hbox{\vrule\height\p@\hskip\tw@\p@\vrule}}\fi
   \fi
  \cleaders\leads@\vfill\nointerlineskip
   \ifnum\arrcount@=\m@ne\else\ifnum\arrcount@=\thr@@\else
    \hbox{\arrow@iv
     \ifN@
      \ifcase\scount@\or\or\char118 \or\char117 \or\or\or\char119 \or
      \char120 \or\char121 \or\char122 \or\or\or\arrow@iii\char123 \or
      \vbox{\hbox{\char117}\nointerlineskip\vskip-\thr@@\p@
      \hbox{\char117}\vskip\thr@@\p@}\fi
     \else
      \ifnum\tcount@=\m@ne
      \else
       \ifcase\tcount@\char117 \or\or\char117 \or\char118 \or\char119 \or
       \char120 \or\or\or\or\or\char121 \or\char122 \or\arrow@iii\char123 \or
       \vbox{\hbox{\char117}\nointerlineskip\vskip-\thr@@\p@
       \hbox{\char117}\vskip\thr@@\p@}\fi
      \fi
     \fi}%
   \fi\fi
   \ifN@\vskip\ydimen@\else\vskip-\Ydimen@\fi}%
 \ifN@
  \dimen@ii\firsty@
 \else
  \dimen@ii-\firsty@\advance\dimen@ii\ht\ZER@\multiply\dimen@ii\m@ne
 \fi
 \rlap{\smash{\hskip\tocenter@\hskip\pdimen@\raise\dimen@ii\box\ZER@}}%
 \iflabel@i
  \setbox\ZER@\vbox to\dimen@{\vfil
   \hbox{$\m@th\tsize@@\label@i$}\vskip\tw@\ldimen@i\vfil}%
  \rlap{\smash{\hskip\tocenter@\hskip\pdimen@
  \ifnum\arrcount@=\m@ne\let\next@\relax\else\let\next@\llap\fi
  \next@{\raise\dimen@ii\hbox{\ifnum\arrcount@=\m@ne\hskip-.5\wd\ZER@\fi
   \box\ZER@\ifnum\arrcount@=\m@ne\else\hskip\tw@\p@\fi}}}}%
 \fi
 \iflabel@ii
  \ifnum\arrcount@=\m@ne
  \else
   \setbox\ZER@\vbox to\dimen@{\vfil
    \hbox{$\m@th\tsize@\label@ii$}\vskip\tw@\ldimen@ii\vfil}%
   \rlap{\smash{\hskip\tocenter@\hskip\pdimen@
   \rlap{\raise\dimen@ii\hbox{\ifnum\arrcount@=\thr@@\hskip4.5\p@\else
    \hskip2.5\p@\fi\box\ZER@}}}}%
  \fi
 \fi
}
\newdimen\goal@
\newdimen\shifted@
\newcount\Tcount@
\newcount\Scount@
\newbox\shaft@
\newcount\slcount@
\def\getcos@#1{%
 \ifnum\tan@i<\tan@ii
  \dimen@#1%
  \ifnum\slcount@<8 \count@9 \else \ifnum\slcount@<12 \count@8 \else
   \count@7 \fi\fi
  \multiply\dimen@\count@\divide\dimen@\ten@
  \dimen@ii\dimen@\multiply\dimen@ii\tan@i\divide\dimen@ii\tan@ii
 \else
  \dimen@ii#1%
  \count@-\slcount@\advance\count@24
  \ifnum\count@<8 \count@9 \else \ifnum\count@<12 \count@8
   \else\count@7 \fi\fi
  \multiply\dimen@ii\count@\divide\dimen@ii\ten@
  \dimen@\dimen@ii\multiply\dimen@\tan@ii\divide\dimen@\tan@i
 \fi}
\newdimen\adjust@
\def\Nnext@{\ifN@\let\next@\raise\else\let\next@\lower\fi}
\def\arrow@@{\slcount@\angcount@
 \ifNESW@
  \ifnum\angcount@<\ten@
   \let\arrowfont@\arrow@i\global\advance\angcount@\m@ne
   \global\multiply\angcount@13
  \else
   \ifnum\angcount@<19
    \let\arrowfont@\arrow@ii\global\advance\angcount@-\ten@
    \global\multiply\angcount@13
   \else
    \let\arrowfont@\arrow@iii\global\advance\angcount@-19
    \global\multiply\angcount@13
  \fi\fi
  \Tcount@\angcount@
 \else
  \ifnum\angcount@<5
   \let\arrowfont@\arrow@iii\global\advance\angcount@\m@ne
   \global\multiply\angcount@13 \global\advance\angcount@65
  \else
   \ifnum\angcount@<14
    \let\arrowfont@\arrow@iv\global\advance\angcount@-5
    \global\multiply\angcount@13
   \else
    \ifnum\angcount@<23
     \let\arrowfont@\arrow@v\global\advance\angcount@-14
     \global\multiply\angcount@13
    \else
     \let\arrowfont@\arrow@i\global\angcount@117
  \fi\fi\fi
  \ifnum\angcount@=117 \Tcount@115 \else\Tcount@\angcount@\fi
 \fi
 \Scount@\Tcount@
 \ifE@
  \ifnum\tcount@=\z@\advance\Tcount@\tw@\else\ifnum\tcount@=13
   \advance\Tcount@\tw@\else\advance\Tcount@\tcount@\fi\fi
  \ifnum\scount@=\z@\else\ifnum\scount@=13 \advance\Scount@\thr@@\else
   \advance\Scount@\scount@\fi\fi
 \else
  \ifcase\tcount@\advance\Tcount@\thr@@\or\or\advance\Tcount@\thr@@\or
  \advance\Tcount@\tw@\or\advance\Tcount@6 \or\advance\Tcount@7
  \or\or\or\or\or\advance\Tcount@8 \or\advance\Tcount@9 \or
  \advance\Tcount@12 \or\advance\Tcount@\thr@@\fi
  \ifcase\scount@\or\or\advance\Scount@\thr@@\or\advance\Scount@\tw@\or
  \or\or\advance\Scount@4 \or\advance\Scount@5 \or\advance\Scount@\ten@
  \or\advance\Scount@11 \or\or\or\advance\Scount@12 \or\advance
  \Scount@\tw@\fi
 \fi
 \ifcase\arrcount@\or\or\global\advance\angcount@\@ne\else\fi
 \ifN@\shifted@\firsty@\else\shifted@-\firsty@\fi
 \ifE@\else\advance\shifted@\charht@\fi
 \goal@\secondy@\advance\goal@-\firsty@
 \ifN@\else\multiply\goal@\m@ne\fi
 \setbox\shaft@\hbox{\arrowfont@\char\angcount@}%
 \ifnum\arrcount@=\thr@@
  \getcos@{1.5\p@}%
  \setbox\shaft@\hbox to\wd\shaft@{\arrowfont@
   \rlap{\hskip\dimen@ii
    \smash{\ifNESW@\let\next@\lower\else\let\next@\raise\fi
     \next@\dimen@\hbox{\arrowfont@\char\angcount@}}}%
   \rlap{\hskip-\dimen@ii
    \smash{\ifNESW@\let\next@\raise\else\let\next@\lower\fi
      \next@\dimen@\hbox{\arrowfont@\char\angcount@}}}\hfil}%
 \fi
 \rlap{\smash{\hskip\tocenter@\hskip\firstx@
  \ifnum\arrcount@=\m@ne
  \else
   \ifnum\arrcount@=\thr@@
   \else
    \ifnum\scount@=\m@ne
    \else
     \ifnum\scount@=\z@
     \else
      \setbox\ZER@\hbox{\ifnum\angcount@=117 \arrow@v\else\arrowfont@\fi
       \char\Scount@}%
      \ifNESW@
       \ifnum\scount@=\tw@
        \dimen@\shifted@\advance\dimen@-\charht@
        \ifN@\hskip-\wd\ZER@\fi
        \Nnext@
        \next@\dimen@\copy\ZER@
        \ifN@\else\hskip-\wd\ZER@\fi
       \else
        \Nnext@
        \ifN@\else\hskip-\wd\ZER@\fi
        \next@\shifted@\copy\ZER@
        \ifN@\hskip-\wd\ZER@\fi
       \fi
       \ifnum\scount@=12
        \advance\shifted@\charht@\advance\goal@-\charht@
        \ifN@\hskip\wd\ZER@\else\hskip-\wd\ZER@\fi
       \fi
       \ifnum\scount@=13
        \getcos@{\thr@@\p@}%
        \ifN@\hskip\dimen@\else\hskip-\wd\ZER@\hskip-\dimen@\fi
        \adjust@\shifted@\advance\adjust@\dimen@ii
        \Nnext@
        \next@\adjust@\copy\ZER@
        \ifN@\hskip-\dimen@\hskip-\wd\ZER@\else\hskip\dimen@\fi
       \fi
      \else
       \ifN@\hskip-\wd\ZER@\fi
       \ifnum\scount@=\tw@
        \ifN@\hskip\wd\ZER@\else\hskip-\wd\ZER@\fi
        \dimen@\shifted@\advance\dimen@-\charht@
        \Nnext@
        \next@\dimen@\copy\ZER@
        \ifN@\hskip-\wd\ZER@\fi
       \else
        \Nnext@
        \next@\shifted@\copy\ZER@
        \ifN@\else\hskip-\wd\ZER@\fi
       \fi
       \ifnum\scount@=12
        \advance\shifted@\charht@\advance\goal@-\charht@
        \ifN@\hskip-\wd\ZER@\else\hskip\wd\ZER@\fi
       \fi
       \ifnum\scount@=13
        \getcos@{\thr@@\p@}%
        \ifN@\hskip-\wd\ZER@\hskip-\dimen@\else\hskip\dimen@\fi
        \adjust@\shifted@\advance\adjust@\dimen@ii
        \Nnext@
        \next@\adjust@\copy\ZER@
        \ifN@\hskip\dimen@\else\hskip-\dimen@\hskip-\wd\ZER@\fi
       \fi
      \fi
  \fi\fi\fi\fi
  \ifnum\arrcount@=\m@ne
  \else
   \loop
    \ifdim\goal@>\charht@
    \ifE@\else\hskip-\charwd@\fi
    \Nnext@
    \next@\shifted@\copy\shaft@
    \ifE@\else\hskip-\charwd@\fi
    \advance\shifted@\charht@\advance\goal@-\charht@
    \repeat
   \ifdim\goal@>\z@
    \dimen@\charht@\advance\dimen@-\goal@
    \divide\dimen@\tan@i\multiply\dimen@\tan@ii
    \ifE@\hskip-\dimen@\else\hskip-\charwd@\hskip\dimen@\fi
    \adjust@\shifted@\advance\adjust@-\charht@\advance\adjust@\goal@
    \Nnext@
    \next@\adjust@\copy\shaft@
    \ifE@\else\hskip-\charwd@\fi
   \else
    \adjust@\shifted@\advance\adjust@-\charht@
   \fi
  \fi
  \ifnum\arrcount@=\m@ne
  \else
   \ifnum\arrcount@=\thr@@
   \else
    \ifnum\tcount@=\m@ne
    \else
     \setbox\ZER@
      \hbox{\ifnum\angcount@=117 \arrow@v\else\arrowfont@\fi\char\Tcount@}%
     \ifnum\tcount@=\thr@@
      \advance\adjust@\charht@
      \ifE@\else\ifN@\hskip-\charwd@\else\hskip-\wd\ZER@\fi\fi
     \else
      \ifnum\tcount@=12
       \advance\adjust@\charht@
       \ifE@\else\ifN@\hskip-\charwd@\else\hskip-\wd\ZER@\fi\fi
      \else
       \ifE@\hskip-\wd\ZER@\fi
     \fi\fi
     \Nnext@
     \next@\adjust@\copy\ZER@
     \ifnum\tcount@=13
      \hskip-\wd\ZER@
      \getcos@{\thr@@\p@}%
      \ifE@\hskip-\dimen@\else\hskip\dimen@\fi
      \advance\adjust@-\dimen@ii
      \Nnext@
      \next@\adjust@\box\ZER@
     \fi
  \fi\fi\fi}}%
 \iflabel@i
  \rlap{\hskip\tocenter@
  \dimen@\firstx@\advance\dimen@\secondx@\divide\dimen@\tw@
  \advance\dimen@\ldimen@i
  \dimen@ii\firsty@\advance\dimen@ii\secondy@\divide\dimen@ii\tw@
  \global\multiply\ldimen@i\tan@i\global\divide\ldimen@i\tan@ii
  \ifNESW@\advance\dimen@ii\ldimen@i\else\advance\dimen@ii-\ldimen@i\fi
  \setbox\ZER@\hbox{\ifNESW@\else\ifnum\arrcount@=\thr@@\hskip4\p@\else
   \hskip\tw@\p@\fi\fi
   $\m@th\tsize@@\label@i$\ifNESW@\ifnum\arrcount@=\thr@@\hskip4\p@\else
   \hskip\tw@\p@\fi\fi}%
  \ifnum\arrcount@=\m@ne
   \ifNESW@\advance\dimen@.5\wd\ZER@\advance\dimen@\p@\else
    \advance\dimen@-.5\wd\ZER@\advance\dimen@-\p@\fi
   \advance\dimen@ii-.5\ht\ZER@
  \else
   \advance\dimen@ii\dp\ZER@
   \ifnum\slcount@<6 \advance\dimen@ii\tw@\p@\fi
  \fi
  \hskip\dimen@
  \ifNESW@\let\next@\llap\else\let\next@\rlap\fi
  \next@{\smash{\raise\dimen@ii\box\ZER@}}}%
 \fi
 \iflabel@ii
  \ifnum\arrcount@=\m@ne
  \else
   \rlap{\hskip\tocenter@
   \dimen@\firstx@\advance\dimen@\secondx@\divide\dimen@\tw@
   \ifNESW@\advance\dimen@\ldimen@ii\else\advance\dimen@-\ldimen@ii\fi
   \dimen@ii\firsty@\advance\dimen@ii\secondy@\divide\dimen@ii\tw@
   \global\multiply\ldimen@ii\tan@i\global\divide\ldimen@ii\tan@ii
   \advance\dimen@ii\ldimen@ii
   \setbox\ZER@\hbox{\ifNESW@\ifnum\arrcount@=\thr@@\hskip4\p@\else
    \hskip\tw@\p@\fi\fi
    $\m@th\tsize@\label@ii$\ifNESW@\else\ifnum\arrcount@=\thr@@\hskip4\p@
    \else\hskip\tw@\p@\fi\fi}%
   \advance\dimen@ii-\ht\ZER@
   \ifnum\slcount@<9 \advance\dimen@ii-\thr@@\p@\fi
   \ifNESW@\let\next@\rlap\else\let\next@\llap\fi
   \hskip\dimen@\next@{\smash{\raise\dimen@ii\box\ZER@}}}%
  \fi
 \fi
}
\def\outCD@#1{\def#1{\Err@{\noexpand#1must not be used within \string\CD}}}
\newskip\preCDskip@
\newskip\postCDskip@
\preCDskip@\z@
\postCDskip@\z@
\def\preCDspace#1{\RIfMIfI@
 \onlydmatherr@\preCDspace\else\advance\preCDskip@#1\relax\fi\else
 \onlydmatherr@\preCDspace\fi}
\def\postCDspace#1{\RIfMIfI@
 \onlydmatherr@\postCDspace\else\advance\postCDskip@#1\relax\fi\else
 \onlydmatherr@\postCDspace\fi}
\def\predisplayspace#1{\RIfMIfI@
 \onlydmatherr@\predisplayspace\else
 \advance\abovedisplayskip#1\relax
 \advance\abovedisplayshortskip#1\relax\fi
 \else\onlydmatherr@\preCDspace\fi}
\def\postdisplayspace#1{\RIfMIfI@
 \onlydmatherr@\postdisplayspace\else
 \advance\belowdisplayskip#1\relax
 \advance\belowdisplayshortskip#1\relax\fi
 \else\onlydmatherr@\postdisplayspace\fi}
\def\PreCDSpace#1{\global\preCDskip@#1\relax}
\def\PostCDSpace#1{\global\postCDskip@#1\relax}
\def\CD#1\endCD{%
 \outCD@\cgaps\outCD@\rgaps\outCD@\Cgaps\outCD@\Rgaps
 \preCD@#1\endCD
 \advance\abovedisplayskip\preCDskip@
 \advance\abovedisplayshortskip\preCDskip@
 \advance\belowdisplayskip\postCDskip@
 \advance\belowdisplayshortskip\postCDskip@
 \vcenter{\offinterlineskip
  \vskip\preCDskip@\Let@\global\colcount@\@ne\global\rowcount@\z@
  \everycr{%
   \noalign{%
    \ifnum\rowcount@=\Rowcount@
    \else
     \getrgap@\rowcount@\vskip\getdim@
     \global\advance\rowcount@\@ne\global\colcount@\@ne
    \fi}}%
  \tabskip\z@
  \halign{&\global\xoff@\z@\global\yoff@\z@
   \getcgap@\colcount@\hskip\getdim@
   \hfil\vrule\height\ten@\p@\width\z@\depth\z@
   $\m@th\displaystyle{##}$\hfil
   \global\advance\colcount@\@ne\cr
   #1\crcr}\vskip\postCDskip@}%
 \preCDskip@\z@\postCDskip@\z@
 \def\getcgap@##1{\ifcase##1\or\getdim@\z@\else\getdim@\standardcgap\fi}%
 \def\getrgap@##1{\ifcase##1\getdim@\z@\else\getdim@\standardrgap\fi}%
 \let\Width@\relax\let\Height@\relax\let\Depth@\relax\let\Rowheight@\relax
 \let\Rowdepth@\relax\let\Colwidth@\relax
}
\let\enddocument\bye
\def\alloc@#1#2#3#4#5{\global\advance\count1#1by\@ne
  \ch@ck#1#4#2%
  \allocationnumber=\count1#1%
  \global#3#5=\allocationnumber
  \wlog{\string#5=\string#2\the\allocationnumber}}
\catcode`\@=\active

\catcode`\"=12 \font\black=cmbx10 \font\sblack=cmbx7 \font\ssblack=cmbx5
\font\blackital=cmmib10  \skewchar\blackital='177 \font\sblackital=cmmib7
\skewchar\sblackital='177 \font\ssblackital=cmmib5
\skewchar\ssblackital='177 \font\sanss=cmss10 \font\ssanss=cmss8 scaled 900
\font\sssanss=cmss8 scaled 600  
  \font\blackboard=msbm10
\font\sblackboard=msbm7 \font\ssblackboard=msbm5 \font\caligr=eusm10
\font\scaligr=eusm7 \font\sscaligr=eusm5 
\font\fraktur=eufm10 \font\sfraktur=eufm7 \font\ssfraktur=eufm5

%\font\tencyr=wncyr10
%\font\tencyb=wncyb10

\def\tx#1{{\fam0\relax#1}}

\newfam\blfam
\textfont\blfam=\black \scriptfont\blfam=\sblack
\scriptscriptfont\blfam=\ssblack

\newfam\bifam
\textfont\bifam=\blackital \scriptfont\bifam=\sblackital
\scriptscriptfont\bifam=\ssblackital
\def\bi#1{{\fam\bifam\relax#1}}

\newfam\bbfam
\textfont\bbfam=\blackboard \scriptfont\bbfam=\sblackboard
\scriptscriptfont\bbfam=\ssblackboard
\def\bb#1{{\fam\bbfam\relax#1}}

\newfam\ssfam
\textfont\ssfam=\sanss \scriptfont\ssfam=\ssanss
\scriptscriptfont\ssfam=\sssanss
\def\ss#1{{\fam\ssfam\relax#1}}

\newfam\clfam
\textfont\clfam=\caligr \scriptfont\clfam=\scaligr
\scriptscriptfont\clfam=\sscaligr

\newfam\frfam
\textfont\frfam=\fraktur \scriptfont\frfam=\sfraktur
\scriptscriptfont\frfam=\ssfraktur

\newfam\gpfam

\def\vpb#1{\setbox0=\hbox{${#1}$}
    \copy0 \kern-\wd0 \raise.08pt \box0}
\def\hpb#1{\setbox0=\hbox{${#1}$}
    \copy0 \kern-\wd0 \kern.2pt \box0}

\font\bsymb=cmsy10 scaled\magstep2
\def\all#1{\setbox0=\hbox{\lower1.5pt\hbox{\bsymb \char"38}}\setbox1=\hbox{$_{#1}$} \box0\lower2pt\box1}
\def\exi#1{\setbox0=\hbox{\lower1.5pt\hbox{\bsymb \char"39}}\setbox1=\hbox{$_{#1}$} \box0\lower2pt\box1}

\def\ud#1{\setbox0=\hbox{#1} \dimen0=\wd0 \box0 \kern-\dimen0
    \vrule height-1.5pt width\dimen0 depth2pt}

\def\ub#1{\setbox0=\hbox{${#1}$} \mathsurround=0pt \dimen0=\wd0 \box0
    \kern-.9\dimen0 \vrule height-1.5pt width.8\dimen0 depth2pt \kern.1\dimen0}

\mathchardef\za="710B  %\alpha
\mathchardef\zb="710C  %\beta
\mathchardef\zg="710D  %\gamma
\mathchardef\zd="710E  %\delta
\mathchardef\zve="710F %\epsilon
\mathchardef\zz="7110  %\zeta
\mathchardef\zh="7111  %\eta
\mathchardef\zvy="7112 %\theta
\mathchardef\zi="7113  %\iota
\mathchardef\zk="7114  %\kappa
\mathchardef\zl="7115  %\lambda
\mathchardef\zm="7116  %\mu
\mathchardef\zn="7117  %\nu
\mathchardef\zx="7118  %\xi
\mathchardef\zp="7119  %\pi
\mathchardef\zr="711A  %\rho
\mathchardef\zs="711B  %\sigma
\mathchardef\zt="711C  %\tau
\mathchardef\zu="711D  %\upsilon
\mathchardef\zvf="711E %\phi
\mathchardef\zq="711F  %\chi
\mathchardef\zc="7120  %\psi
\mathchardef\zw="7121  %\omega
\mathchardef\ze="7122  %\varepsilon
\mathchardef\zy="7123  %\vartheta
\mathchardef\zvp="7124  %\varpi
\mathchardef\zvr="7125 %\varrho
\mathchardef\zvs="7126 %\varsigma
\mathchardef\zf="7127  %\varphi
\mathchardef\zG="7000  %\Gamma
\mathchardef\zD="7001  %\Delta
\mathchardef\zY="7002  %\Theta
\mathchardef\zL="7003  %\Lambda
\mathchardef\zX="7004  %\Xi
\mathchardef\zP="7005  %\Pi
\mathchardef\zS="7006  %\Sigma
\mathchardef\zU="7007  %\Upsilon
\mathchardef\zF="7008  %\Phi
\mathchardef\zC="7009  %\Psi
\mathchardef\zW="700A  %\Omega

\catcode`\"=\active \loadmsam
\newsymbol\leqslant 1336

\TagsOnRight

\def\*{{\textstyle *}}
\def\s*{{\scriptstyle *}}

\def\proof{\demo{Proof}}
\def\endproof{\hfill \vrule height4pt width6pt depth2pt \enddemo}

\def\R{{\bb R}}
\def\N{{\bb N}}

\def\Z{{\bb Z}}

\def\Rp{{\bb R_{\scriptscriptstyle +}}}

\def\sT{{\ss T}}

\def\sV{{\ss V}}

\def\st{{\ss t}}

\def\+{{\scriptscriptstyle +}}

\def\xd{\tx{d}}
\def\xi{\tx{i}}
\def\xD{\tx{D}}

\def\gr{\text{graph}}

\def\im{\text{im}}

\input paper.st\relax
\hsize=33pc \hoffset=30pt \vsize=54pc \voffset=6pt \TagsOnRight
 \overfullrule 0pt
\document
\def\zza{{\scriptscriptstyle\star}}
%\line{TU.TEX \hfill \today}
\line{"Gravitation, Electromagnetism and Geometric Structures", Pitagora
Editrice 1996, pp. 91--136 \hfill}

    \title
        Homogenous Lagrangian systems
    \endtitle

    \author
        Wlodzimierz M. Tulczyjew \\
        Dipartimento di Matematica e Fisica \\
        Universit\`a di Camerino \\
        62032 Camerino, Italy \\
        Istituto Nazionale di Fisica Nucleare,
        Sezione di Napoli, Italy
                        \\
                        \\
                Pawe\l\ Urba\'nski \\
                Division of Mathematical Methods in Physics \\
                University of Warsaw \\
                Ho\D{z}a 74, 00-682 Warszawa
    \endauthor

    \maketitle

                \hl1{Introduction}

        The application of the Legendre transformation to a hyperregular Lagrangian
system results in a Hamiltonian vector field generated by a Hamiltonian defined on
the phase space of the mechanical system.  The Legendre transformation in its usual
interpretation can not be applied to homogeneous Lagrangians found in relativistic
mechanics.  The dynamics of relativistic systems must be formulated in terms of
implicit differential equations in the phase space and not in terms of Hamiltonian
vector fields.  The constrained Hamiltonian systems introduced by Dirac [1] are not
general enough to cover some important cases.  We formulate a geometric framework
which permits Lagrangian and Hamiltonian descriptions of the dynamics of a wide
class of mechanical systems.  This framework extends the applicability of earlier
definitions of the Legendre transformation [2], [3].  Lagrangians and Hamiltonians
are presented as families of functions.  The Legendre transformation and the inverse
Legendre transformation are described as transitions between these families.  Two
simple examples are given.

               \hl1{Iterated tangent and cotangent functors}

    For each differential manifold $Q$ the fibrations
    \vskip1mm
        $$\Rgaps{1.2}\Cgaps{1.4}\CD
    \sT Q @()\L{\zt_Q}@(0,-1) && \sT^\* Q @()\L{\zp_Q}@(0,-1) \\
        Q && Q
        \endCD
                                                                                \tag \label{Git1}$$
    \vskip1mm \noindent are a dual pair of vector fibrations.  We have operations
        $$+ \colon \sT Q \times_Q \sT Q \rightarrow \sT Q,
                                                                                \tag \label{Fit1}$$
        $$\cdot\, \colon \R \times \sT Q \rightarrow \sT Q,
                                                                                \tag \label{Fit2}$$
        $$+ \colon \sT^\* Q \times_Q \sT^\* Q \rightarrow \sT^\* Q,
                                                                                \tag \label{Fit3}$$
        $$\cdot\, \colon \R \times \sT^\* Q \rightarrow \sT^\* Q,
                                                                                \tag \label{Fit4}$$
    and the canonical pairing
        $$\langle \,, \rangle \colon \sT^\* Q \times_Q \sT Q \rightarrow \R.
                                                                                \tag \label{Fit5}$$

    Substituting the tangent bundle $\sT Q$ for the differential manifold $Q$ we
obtain the fibrations
    \vskip1mm
        $$\Rgaps{1.2}\Cgaps{1.4}\CD
    \sT\sT Q @()\L{\zt_{\sT Q}}@(0,-1) && \sT^\*\sT Q @()\L{\zp_{\sT Q}}@(0,-1) \\
        \sT Q && \sT Q
        \endCD
                                                                                \tag \label{Git2}$$
    \vskip1mm \noindent with operations
        $$+ \colon \sT\sT Q \times_{\sT Q} \sT\sT Q \rightarrow \sT\sT Q,
                                                                                \tag \label{Fit6}$$
        $$\cdot\, \colon \R \times \sT\sT Q \rightarrow \sT\sT Q,
                                                                                \tag \label{Fit7}$$
        $$+ \colon \sT^\* \sT Q \times_{\sT Q} \sT^\*\sT Q \rightarrow \sT^\*\sT Q,
                                                                                \tag \label{Fit8}$$
        $$\cdot\, \colon \R \times \sT^\*\sT Q \rightarrow \sT^\*\sT Q,
                                                                                \tag \label{Fit9}$$
    and the canonical pairing
        $$\langle \,, \rangle \colon \sT^\*\sT Q \times_{\sT Q} \sT\sT Q \rightarrow \R.
                                                                                \tag \label{Fit10}$$
    \vskip1mm \noindent Sets $\sT\sT Q \times_{\sT Q} \sT\sT Q$, $\sT^\* \sT Q
\times_{\sT Q} \sT^\*\sT Q$, and $\sT^\*\sT Q \times_{\sT Q} \sT\sT Q$ have
the following definitions:
        $$\sT\sT Q \times_{\sT Q} \sT\sT Q = \{(u,v) \in \sT\sT Q \times \sT\sT Q
;\; \zt_{\sT Q}(u) = \zt_{\sT Q}(v) \},
                                                                                \tag \label{Fit11}$$
        $$\sT^\*\sT Q \times_{\sT Q} \sT^\*\sT Q = \{(f,g) \in \sT^\*\sT Q \times
\sT^\*\sT Q ;\; \zp_{\sT Q}(f) = \zp_{\sT Q}(g) \},
                                                                                \tag \label{Fit12}$$
    and
        $$\sT^\*\sT Q \times_{\sT Q} \sT\sT Q = \{(f,u) \in \sT^\*\sT Q \times
\sT\sT Q ;\; \zp_{\sT Q}(f) = \zt_{\sT Q}(u) \}.
                                                                                \tag \label{Fit13}$$

    By applying the tangent functor to the fibrations \Ref{Git1} we obtain a dual
pair of vector fibrations
    \vskip1mm
        $$\Rgaps{1.2}\Cgaps{1.4}\CD
    \sT\sT Q @()\L{\sT\zt_Q}@(0,-1) && \sT\sT^\* Q @()\L{\sT\zp_Q}@(0,-1) \\
        \sT Q && \sT Q
        \endCD
                                                                                \tag \label{Git3}$$
    \vskip1mm \noindent with operations
        $$\sT+ \colon \sT\sT Q \times_{\sT Q} \sT\sT Q \rightarrow \sT\sT Q,
                                                                                \tag \label{Fit14}$$
        $$\zza \colon \R \times \sT\sT Q \rightarrow \sT\sT Q,
                                                                                \tag \label{Fit15}$$
        $$\sT+ \colon \sT\sT^\* Q \times_{\sT Q} \sT\sT^\* Q \rightarrow \sT\sT^\* Q,
                                                                                \tag \label{Fit16}$$
        $$\zza \colon \R \times \sT\sT^\* Q \rightarrow \sT\sT^\* Q,
                                                                                \tag \label{Fit17}$$
    and the pairing
        $$\langle \,, \rangle^\zza \colon \sT\sT^\* Q \times_{\sT Q} \sT\sT Q \rightarrow \R.
                                                                                \tag \label{Fit18}$$
    The operations $\zza$ and the pairing $\langle \,, \rangle^\zza$ are extracted
from the tangent operations
        $$\sT\cdot\, \colon \sT\R \times \sT\sT Q \rightarrow \sT\sT Q,
                                                                                \tag \label{Fit19}$$
        $$\sT\cdot\, \colon \sT\R \times \sT\sT^\* Q \rightarrow \sT\sT^\* Q,
                                                                                \tag \label{Fit20}$$
    and the tangent pairing
        $$\sT\langle \,, \rangle \colon \sT\sT^\* Q \times_{\sT Q} \sT\sT Q \rightarrow \sT\R.
                                                                                \tag \label{Fit21}$$
    Sets $\sT\sT Q \times_{\sT Q} \sT\sT Q$, $\sT\sT^\* Q \times_{\sT Q} \sT\sT^\*
Q$, and $\sT\sT^\* Q \times_{\sT Q} \sT\sT Q$ in formulae \Ref{Fit1},
\Ref{Fit3}, and \Ref{Fit5} are defined by
        $$\sT\sT Q \times_{\sT Q} \sT\sT Q = \{(u,v) \in \sT\sT Q \times \sT\sT Q
;\; \sT\zt_Q(u) = \sT\zt_Q(v) \},
                                                                                \tag \label{Fit22}$$
        $$\sT\sT^\* Q \times_{\sT Q} \sT\sT^\* Q = \{(x,y) \in \sT\sT^\* Q \times
\sT\sT^\* Q ;\; \sT\zp_Q(x) = \sT\zp_Q(y) \},
                                                                                \tag \label{Fit23}$$
    and
        $$\sT\sT^\* Q \times_{\sT Q} \sT\sT Q = \{(x,u) \in \sT\sT^\* Q \times
\sT\sT Q ;\; \sT\zp_Q(x) = \sT\zt_Q(u) \}.
                                                                                \tag \label{Fit24}$$

    Pairings \Ref{Fit10} and \Ref{Fit5} permit the introduction of the vector
fibration isomorphism
    \vskip1mm
        $$\Rgaps{1.2}\Cgaps{1.2}\CD
    \sT\sT^\* Q @()\L{\sT\zp_Q}@(0,-1) @()\L{\za_Q}@(1,0)
                & \sT^\*\sT Q @()\L{\zp_{\sT Q}}@(0,-1) \\
    \sT Q @()\a=@(1,0)      & \sT Q
        \endCD
                                                                                \tag \label{Git4}$$
    \vskip1mm \noindent dual to the vector fibration isomorphism
    \vskip1mm
        $$\Rgaps{1.2}\Cgaps{1.2}\CD
    \sT\sT Q @()\L{\sT\zt_Q}@(0,-1)
                & \sT\sT Q @()\L{\zt_{\sT Q}}@(0,-1) @()\L{\zk_Q}@(-1,0)\\
    \sT Q @()\a=@(1,0)      & \sT Q
        \endCD
                                                                                \tag \label{Git5}$$
    \vskip1mm \noindent constructed with the canonical involution
        $$\zk_Q \colon \sT\sT Q \rightarrow \sT\sT Q.
                                                                                \tag \label{Fit25}$$

    For each manifold $Q$ we have the mapping
        $$\align
    \zq_{\sT Q} &\colon \sT Q \times_Q \sT Q \rightarrow \sT\sT Q \\
    &\colon (v,v') \mapsto \st\zg(0),
                                                                                \tag \label{Fit26}\endalign$$
    where $\zg$ is the curve
        $$\align
    \zg &\colon \R \rightarrow \sT Q \\
    &\colon s \mapsto v + sv'
                                                                                \tag \label{Fit27}\endalign$$
    and the symbol $\st\zg(0)$ is used to denote the vector tangent to $\zg$.  There
is also the mapping
        $$\align
    \zq_{\sT^\* Q} &\colon \sT^\* Q \times_Q \sT^\* Q \rightarrow \sT\sT^\* Q \\
    &\colon (p,p') \mapsto \st\zg(0),
                                                                                \tag \label{Fit28}\endalign$$
    where $\zg$ is the curve
        $$\align
    \zg &\colon \R \rightarrow \sT^\* Q \\
    &\colon s \mapsto p + sp'.
                                                                                \tag \label{Fit29}\endalign$$

    Mappings $\zq_{\sT Q}$ and $\zq_{\sT^\* Q}$ are two examples of a general mapping
$\zq_E$ defined for each vector fibration
    \vskip1mm
        $$\Rgaps{1.2}\Cgaps{1.2}\CD
    E @()\L{\ze}@(0,-1) \\ Q
        \endCD
                                                                                \tag \label{Git6}$$
    \vskip1mm \noindent by
        $$\align
    \zq_E &\colon E \times_Q E \rightarrow E \\
    &\colon (e,e') \mapsto \st\zg(0),
                                                                                \tag \label{Fit30}\endalign$$
    where $\zg$ is the curve
        $$\align
    \zg &\colon \R \rightarrow \sT^\* Q \\
    &\colon s \mapsto e + se'.
                                                                                \tag \label{Fit31}\endalign$$

        \hl1{Derivations}
    Let $P$ be a differential manifold.  We denote by $\zF(P)$ and $\zF(\sT P)$ the
exterior algebras of differential forms on $P$ and $\sT P$ respectively.  A
{\it derivation} of degree $k$ relative to $\zt_P \colon \sT P \rightarrow P$
is a linear operator $a \colon \zF(P) \rightarrow \zF(\sT P)$ such that
        \list
    \item degree($a\zm$) = degree($\zm$) + k for each form $\zm$ on $P$,
    \item $a(\zm \wedge \zn) = a\zm \wedge \zt_P{}^\*\zn +
(-1)^{k\text{degree}(\zm)}\zt_P{}^\*\zm \wedge a\zn$ for all forms $\zm$ and
$\zn$ on $P$.
        \endlist
    The derivation $a$ is said to be of {\it type} $\xi_\*$ if $af = 0$ for each
function $f$ on $P$.  The derivation $a$ is said to be of {\it type} $\xd_\*$
if $a\xd - (-1)^k\xd a = 0$.

    We associate a derivation $\xi_X$ of type $\xi_\*$ and degree $k$ and a
derivation $\xd_X$ of type $\xd_\*$ and degree $k+1$ with each mapping
        $$X \colon \wedge^{k+1}\sT(\sT P) \rightarrow \sT P
                                                                                \tag \label{Fdr1}$$
    such that
        $$\zt_P(X(v^1 \wedge v^2 \wedge \ldots \wedge v^{k+1})) = \zt_P(\zt_{\sT P}(v^1)).
                                                                                \tag \label{Fdr2}$$
    The derivation $\xi_X$ is characterized by the relation
        $$\langle \xi_X\xd f, v^1 \wedge v^2 \wedge \ldots \wedge v^{k+1} \rangle =
\langle X(v^1 \wedge v^2 \wedge \ldots \wedge v^{k+1}), \xd f \rangle
                                                                                \tag \label{Fdr3}$$
    for each function $f$ on $P$ and each sequence of vectors $v^1, v^2, \ldots
v^{k+1}$ in $\sT\sT P$ such that $\zt_{\sT P}(v^1) = \zt_{\sT P}(v^2) =
\ldots = \zt_{\sT P}(v^{k+1})$.  The dervation $\xd_X$ is defined by
        $$\xd_X = \xi_X\xd - (-1)^k\xd\xi_X.
                                                                                \tag \label{Fdr4}$$
    The mapping \Ref{Fdr1} is called an {\it infinitesimal deformation} of the
projection $\zt_P \colon \sT P \rightarrow P$.

    Let $T \colon \sT P \rightarrow \sT P$ denote the identity mapping of $\sT P$
interpreted as an infinitesimal deformation of $\zt_P \colon \sT P
\rightarrow P$. The derivations $\xi_T$ and $\xd_T$ of degrees $-1$ and 0
respectively are used in the definitions of Lagrangian systems and
Hamiltonian systems.  If $f$ is a function on $P$, then $\xi_T\xd f = \xd_T
f$ is the function
        $$\align
    \xi_T\xd f &\colon \sT P \rightarrow \R \\
    &\colon v \mapsto \langle \xd f, v \rangle.
                                                                                \tag \label{Fdr5}\endalign$$

              \hl1{Special symplectic structures}

    Let $(P,\zw)$ be a symplectic manifold.  A {\it special symplectic structure}
for $(P,\zw)$ is a diagram
    \vskip1mm
        $$\Rgaps{1.2}\Cgaps{1.4}\CD
    (P,\zy) @()\L{\zp}@(0,-1) \\ Q
        \endCD
                                                                                \tag \label{Gss1}$$
    \vskip1mm \noindent where $\zp \colon P \rightarrow Q$ is a vector fibration and
$\zy$ is a vertical one-form on $P$ such that $\xd\zy = \zw$.  An additional
requirement is the existence of a vector fibration morphism
    \vskip1mm
        $$\Rgaps{1.2}\Cgaps{1.2}\CD
    P @()\L{\zp}@(0,-1) @()\L{\zc}@(1,0)
                & \sT^\*Q @()\L{\zp_Q} @(0,-1) \\
    Q @()\a= @(1,0)      & Q
        \endCD
                                                                                \tag \label{Gss2}$$
    \vskip1mm \noindent such that $\zy = \zc^\* \zy_Q$.  We denote by $\zy_Q$ the
Liouville one-form on $\sT^\* Q$.

    Let $Q$ be differential manifold.  The diagram
    \vskip1mm
        $$\Rgaps{1.2}\Cgaps{1.2}\CD
    (\sT^\*Q,\zy_Q) @()\L{\zp_Q}@(0,-1) \\ Q
        \endCD
                                                                                \tag \label{Gss3}$$
    \vskip1mm \noindent is a special symplectic structure for $(\sT^\*Q,\xd\zy_Q)$.

    A {\it special symplectic structure morphism} is a diagram
    \vskip1mm
        $$\Rgaps{1.2}\Cgaps{1.2}\CD
    (P_1,\zy_1) @()\L{\zp_1}@(0,-1) @()\L{\zc}@(1,0)
                & (P_2,\zy_2) @()\L{\zp_2} @(0,-1) \\
    Q_1 @()\L{\zf}@(1,0)      & Q_2
        \endCD
                                                                                \tag \label{Gss4}$$
    \vskip1mm \noindent where
    \vskip1mm
        $$\Rgaps{1.2}\Cgaps{1.2}\CD
    P_1 @()\L{\zp_1}@(0,-1) @()\L{\zc}@(1,0)
                & P_2 @()\L{\zp_2} @(0,-1) \\
    Q_1 @()\L{\zf}@(1,0)      & Q_2
        \endCD
                                                                                \tag \label{Gss5}$$
    \vskip1mm \noindent is a vector fibration morphism and $\zy_1 = \zc^\*\zy_2$.  A
special symplectic structure morphism is necessarily an isomorphism and the
mapping $\zc$ is a symplectomorphism from $(P_1,\xd\zy_1)$ to
$(P_2,\xd\zy_2)$.

    The definition of a special symplectic structure postulates the existence of a
special symplectic structure morphism
    \vskip1mm
        $$\Rgaps{1.2}\Cgaps{1.2}\CD
    (P,\zy) @()\L{\zp}@(0,-1) @()\L{\zc}@(1,0)
                & (\sT^\*Q,\zy_Q) @()\L{\zp_Q} @(0,-1) \\
    Q @()\a= @(1,0)      & Q
        \endCD
                                                                                \tag \label{Gss6}$$
    \vskip1mm \noindent for each special symplectic structure \Ref{Gss1}.

        \claim \c{p}{Proposition}{}                                                  \label{Pss1}
    The special symplectic structure morphism {\rm\Ref{Gss6}} is unique for each special
symplectic structure {\rm\Ref{Gss1}}.
        \endclaim
        \proof For each $w \in \sT P$ we have
        $$\align
    \langle \zy, w \rangle &= \langle \zc^\*\zy_Q, w \rangle \\
            &= \langle \zy_Q, \sT\zc(w) \rangle \\
            &= \langle \zt_{\sT^\*Q}(\sT\zc(w)), \sT\zp_Q(\sT\zc(w)) \rangle \\
            &= \langle \zc(\zt_P(w)), \sT(\zp_Q \circ \zc)(w) \rangle \\
            &= \langle \zc(\zt_P(w)), \sT\zp(w) \rangle
                                                                                \tag \label{Fss1}\endalign$$
    it follows that the mapping $\zc \colon P \rightarrow \sT^\*Q$ is completely
characterized by
        $$\langle \zc(p), v \rangle = \langle \zy, w \rangle,
                                                                                \tag \label{Fss2}$$
    where $v \in \sT_{\zp(p)}Q$ and $w$ is any vector in $\sT_p P$ such that
$\sT\zp(w) = v$.  Hence, the mapping $\zc$ is unique.
        \endproof

    The unique special symplectic structure morphism \Ref{Gss6} is said to be the
{\it canonical special symplectic structure morphism} for the special
symplectic structure \Ref{Gss1}.

    Let
    \vskip1mm
        $$\Rgaps{1.2}\Cgaps{1.2}\CD
    (P_1,\zy_1) @()\L{\zp_1}@(0,-1) && (P_1,\zy_1) @()\L{\zp_1}@(0,-1) @()\L{\zc_1}@(1,0)
                & (\sT^\*Q_1,\zy_{Q_1}) @()\L{\zp_{Q_1}} @(0,-1) \\
    Q_1 && Q_1 @()\a= @(1,0)      & Q_1
        \endCD
                                                                                \tag \label{Gss7}$$
    \vskip1mm \noindent and
    \vskip1mm
        $$\Rgaps{1.2}\Cgaps{1.2}\CD
    (P_2,\zy_2) @()\L{\zp_2}@(0,-1) && (P_2,\zy_2) @()\L{\zp_2}@(0,-1) @()\L{\zc_2}@(1,0)
                & (\sT^\*Q_2,\zy_{Q_2}) @()\L{\zp_{Q_2}} @(0,-1) \\
    Q_2 && Q_2 @()\a= @(1,0)      & Q_2
        \endCD
                                                                                \tag \label{Gss8}$$
    \vskip1mm \noindent be special symplectic structures for symplectic
manifolds $(P_1,\zw_1)$ and $(P_2,\zw_2)$ respectively together with their
canonical special symplectic structure morphisms.  If
    \vskip1mm
        $$\Rgaps{1.2}\Cgaps{1.2}\CD
    (P_1,\zy_1) @()\L{\zp_1}@(0,-1) @()\L{\zc}@(1,0)
                & (P_2,\zy_2) @()\L{\zp_2} @(0,-1) \\
    Q_1 @()\L{\zf}@(1,0)      & Q_2
        \endCD
                                                                                \tag \label{Gss9}$$
    \vskip1mm \noindent is a special symplectic structure morphism, then
        $$\zc = \zc_2^{-1} \circ \sT^\*\zf^{-1} \circ \zc_1.
                                                                                \tag \label{Fss3}$$

               \hl1{Examples of special symplectic structures}

    In the following series of examples each special symplectic structure is
presented together with its canonical special symplectic structure morphism.

        \claim \c{e}{Example}{}{\rm                                                  \label{Eex1}
    If
    \vskip1mm
        $$\Rgaps{1.2}\Cgaps{1.2}\CD
    (P,\zy) @()\L{\zp}@(0,-1) && (P,\zy) @()\L{\zp}@(0,-1) @()\L{\zc}@(1,0)
                & (\sT^\*Q,\zy_Q) @()\L{\zp_Q} @(0,-1) \\
    Q && Q @()\a= @(1,0)      & Q
        \endCD
                                                                                \tag \label{Gex1}$$
    \vskip1mm \noindent is a special symplectic structure for a symplectic manifold
$(P,\zw)$, then the diagrams
    \vskip1mm
        $$\Rgaps{1.2}\Cgaps{1.2}\CD
    (P,-\zy) @()\L{\zp}@(0,-1) && (P,-\zy) @()\L{\zp}@(0,-1) @()\L{-\zc}@(1,0)
                & (\sT^\*Q,\zy_Q) @()\L{\zp_Q} @(0,-1) \\
    Q && Q @()\a= @(1,0)      & Q
        \endCD
                                                                                \tag \label{Gex2}$$
    \vskip1mm \noindent  represent a special symplectic structure for the symplectic
manifold $(P,-\zw)$.
        }\endclaim

        \claim \c{e}{Example}{}{\rm                                                  \label{Eex2}
    Let
    \vskip1mm
        $$\Rgaps{1.2}\Cgaps{1.2}\CD
    (P_1,\zy_1) @()\L{\zp_1}@(0,-1) && (P_1,\zy_1) @()\L{\zp_1}@(0,-1) @()\L{\zc_1}@(1,0)
                & (\sT^\*Q_1,\zy_{Q_1}) @()\L{\zp_{Q_1}} @(0,-1) \\
    Q_1 && Q_1 @()\a= @(1,0)      & Q_1
        \endCD
                                                                                \tag \label{Gex3}$$
    \vskip1mm \noindent and
    \vskip1mm
        $$\Rgaps{1.2}\Cgaps{1.2}\CD
    (P_2,\zy_2) @()\L{\zp_2}@(0,-1) && (P_2,\zy_2) @()\L{\zp_2}@(0,-1) @()\L{\zc_2}@(1,0)
                & (\sT^\*Q_2,\zy_{Q_2}) @()\L{\zp_{Q_2}} @(0,-1) \\
    Q_2 && Q_2 @()\a= @(1,0)      & Q_2
        \endCD
                                                                                \tag \label{Gex4}$$
    \vskip1mm \noindent represent special symplectic structures for symplectic
manifolds $(P_1,\zw_1)$ and $(P_2,\zw_2)$ respectively.  We denote by $\zw_2
\oplus \zw_1$ the two-form on $P_2 \times P_1$ defined by
        $$\zw_2 \oplus \zw_1 = pr_2{}^\* \zw_2 + pr_1{}^\* \zw_1,
                                                                                \tag \label{Fex1}$$
    where $pr_1 \colon P_2 \times P_1 \rightarrow P_1$ and $pr_2 \colon P_2 \times
P_1 \rightarrow P_2$ are the canonical projections.   The symbol $\zy_2
\oplus \zy_1$ will denote the one-form on $P_2 \times P_1$ defined by
        $$\zy_2 \oplus \zy_1 = pr_2{}^\* \zy_2 + pr_1{}^\* \zy_1.
                                                                                \tag \label{Fex2}$$
    We introduce a mapping
        $$\zc_2 \oplus \zc_1 \colon P_2 \times P_1 \rightarrow \sT^\*(Q_2 \times Q_1)
                                                                                \tag \label{Fex3}$$
    characterized by
        $$\langle \zc_2 \oplus \zc_1(p_2,p_1), w \rangle = \langle \zc_2(p_2), \sT pr_2(w)
\rangle + \langle \zc_1(p_1), \sT pr_1(w) \rangle
                                                                                \tag \label{Fex4}$$
    for each $w \in \sT(Q_2 \times Q_1)$ such that $\zt_{Q_2 \times Q_1}(w) =
(\zp_2(p_2),\zp_1(p_1))$.  Symbols $pr_1$ and $pr_2$ this time denote the
canonical projections $pr_1 \colon Q_2 \times Q_1 \rightarrow Q_1$ and $pr_2
\colon Q_2 \times Q_1 \rightarrow Q_2$.  Diagrams
    \vskip1mm
        $$\Rgaps{1.2}\Cgaps{1.2}\CD
    (P_2 \times P_1,\zy_2 \oplus \zy_1) @()\L{\zp_2 \times \zp_1}@(0,-1) & (P_2
\times P_1,\zy_2 \oplus \zy_1) @()\L{\zp_2 \times \zp_1}@(0,-1) @()\L{\zc_2
\oplus \zc_1}@(1,0)
                & (\sT^\*(Q_2 \times Q_1), \zy_{Q_2 \times Q_1}) @()\L{\zp_{Q_2
\times Q_1}} @(0,-1) \\
    Q_2 \times Q_1 & Q_2 \times Q_1 @()\a= @(1,0)      & Q_2 \times Q_1
        \endCD
                                                                                \tag \label{Gex5}$$
    \vskip1mm \noindent represent a special symplectic structure for the symplectic
manifold $(P_2 \times P_1,\zw_2 \oplus \zw_1)$.  Let $\zw_2 \ominus \zw_1 =
\zw_2 \oplus (-\zw_1)$, $\zy_2 \ominus \zy_1 = \zy_2 \oplus (-\zy_1)$, and
$\zc_2 \ominus \zc_1 = \zc_2 \oplus (-\zc_1)$.  Diagrams
    \vskip1mm
        $$\Rgaps{1.2}\Cgaps{1.2}\CD
    (P_2 \times P_1,\zy_2 \ominus \zy_1) @()\L{\zp_2 \times \zp_1}@(0,-1) & (P_2
\times P_1,\zy_2 \ominus \zy_1) @()\L{\zp_2 \times \zp_1}@(0,-1) @()\L{\zc_2
\ominus \zc_1}@(1,0)
                & (\sT^\*(Q_2 \times Q_1), \zy_{Q_2 \times Q_1}) @()\L{\zp_{Q_2
\times Q_1}} @(0,-1) \\
    Q_2 \times Q_1 & Q_2 \times Q_1 @()\a= @(1,0)      & Q_2 \times Q_1
        \endCD
                                                                                \tag \label{Gex6}$$
    \vskip1mm \noindent represent a special symplectic structure for the symplectic
manifold $(P_2 \times P_1,\zw_2 \ominus \zw_1)$.
        }\endclaim

        \claim \c{e}{Example}{}{\rm                                                  \label{Eex3}
    If
    \vskip1mm
        $$\Rgaps{1.2}\Cgaps{1.2}\CD
    (P,\zy) @()\L{\zp}@(0,-1) && (P,\zy) @()\L{\zp}@(0,-1) @()\L{\zc}@(1,0)
                & (\sT^\*Q,\zy_Q) @()\L{\zp_Q} @(0,-1) \\
    Q && Q @()\a= @(1,0)      & Q
        \endCD
                                                                                \tag \label{Gex7}$$
    \vskip1mm \noindent represents a special symplectic structure for a symplectic
manifold $(P,\zw)$, then the diagrams
    \vskip1mm
        $$\Rgaps{1.2}\Cgaps{1.2}\cgaps{.9;.9;1.2}\CD
    (\sT P,\xd_T\zy) @()\L{\sT\zp}@(0,-1) && (\sT P,\xd_T\zy) @()\L{\sT\zp}@(0,-1)
@()\L{\za_Q \circ \sT\zc}@(1,0)
                & (\sT^\*\sT Q,\zy_{\sT Q}) @()\L{\zp_{\sT Q}} @(0,-1) \\
    \sT Q && \sT Q @()\a= @(1,0)      & \sT Q
        \endCD
                                                                                \tag \label{Gex8}$$
    \vskip1mm \noindent  represent a special symplectic structure for the symplectic
manifold $(\sT P,\xd_T\zw)$.
        }\endclaim

        \claim \c{e}{Example}{}{\rm                                                  \label{Eex4}
    Let $(P,\zw)$ be a symplectic manifold.  The diagrams
    \vskip1mm
        $$\Rgaps{1.2}\Cgaps{1.2}\CD
    (\sT P,\xi_T \zw) @()\L{\zt_P}@(0,-1) &&(\sT P,\xi_T\zw) @()\L{\zt_P}@(0,-1)
@()\L{\zb_{(P,\zw)}}@(1,0)
                & (\sT^\*P,\zy_P) @()\L{\zp_P} @(0,-1) \\
    P && P @()\a= @(1,0)      & P
        \endCD
                                                                                \tag \label{Gex9}$$
    \vskip1mm \noindent represent a special symplectic structure for the symplectic
manifold $(\sT P,\xd_T\zw)$.  The mapping $\zb_{(P,\zw)} \colon \sT P
\rightarrow \sT^\* P$ is characterized by
        $$\langle \zb_{(P,\zw)}(u), v \rangle = \langle \zw, u \wedge v \rangle
                                                                                \tag \label{Fex5}$$
    with $u \in \sT P$ and $v \in \sT P$ such that $\zt_P(u) = \zt_P(v)$.
        }\endclaim

              \hl1{Families of functions}

    Let
    \vskip1mm
        $$\Rgaps{1.2}\Cgaps{.8}\CD
    R @()\L{\zr} @(0,-1) \\ X
        \endCD
                                                                                \tag \label{Gff1}$$
    \vskip1mm \noindent be a differential fibration.  We denote by $\sV R$ the {\it
bundle of vertical vectors} defined by
        $$\sV R = \left\{v \in \sT R;\; \sT\zr(v) = 0 \right\}
                                                                                \tag \label{Fff1}$$
    We denote by $\sV^\circ R$ the {\it polar} of the vertical bundle defined by
        $$\sV^\circ R = \left\{s \in \sT^\* R;\; \all{v \in \sV R} \zt_R(v) =
\zp_R(s) \Rightarrow \langle s, v \rangle = 0 \right\}
                                                                                \tag \label{Fff2}$$
    A function $F \colon R \rightarrow \R$ can be considered a {\it family of functions}
defined on fibres of the fibration $\zr$.  We will represent a family of
functions by a diagram
    \vskip1mm
        $$\Rgaps{1.2}\Cgaps{.8}\CD
    R @()\L{\zr} @(0,-1) @()\L{F} @(1,0) & \R \\
    X  &
        \endCD
                                                                                \tag \label{Gff2}$$
    \vskip1mm

    The {\it critical set} for a family of functions $F \colon R \rightarrow \R$ is
the set
        $$S(F,\zr) = \left\{r \in R ;\; \all{v \in \sV_r R} \langle \xd F, v \rangle
= 0 \right\}.
                                                                                \tag \label{Fff3}$$

    At each point $r \in S(F,\zr)$ we define a bilinear mapping
        $$\align
    W(F,r) &\colon \sV_r R \times \sT_r R \rightarrow \R \\
    &\colon (v,w) \mapsto \xD^{(1,1)}(F \circ \zq)(0,0),
                                                                                \tag \label{Fff4}\endalign$$
    where $\zq$ is a mapping from $\R^2$ to $R$ such that $v = \st\zq(\cdot,0)(0)$
and $w = \st\zq(0,\cdot)(0)$.  The family $F$ is said to be {\it regular} if
the rank of $W(F,r)$ is the same at each $r \in S(F,\zr)$.  The family $F$ is
called a {\it Morse family} if the rank of $W(F,r)$ is maximal at each $r \in
S(F,\zr)$.

    It is known that if $F$ is a regular family, then the set
        $$\split
    N = \left\{p \in \sT^\* X ;\; \exi{r \in R}\,\zr(r) = q = \zp_X(p) \vphantom{\all{v
\in \sT_p \sT^\*X}} \right.  \\
                  \all{v \in \sT_q X} \all{w \in \sT_r R}& \left.\sT\zr(w) =
v \Rightarrow  \langle p, v \rangle = \langle \xd F, w
\rangle\vphantom{\all{v \in \sT_p \sT^\*X}} \right\}
                                                                        \endsplit\tag \label{Fff5}$$
    is an immersed Lagrangian submanifold of the symplectic space $(\sT^\*X,\zy_X)$.
This Lagrangian submanifold is said to be {\it generated} by the family
\Ref{Gff2}.
    \vskip1mm

    For each regular family $F$ we have the mapping $\zk \colon S(F,\zr) \rightarrow
\sT^\*X$ characterized by
        $$\langle \zk(r), v \rangle = \langle \xd F, w \rangle
                                                                                \tag \label{Fff6}$$
    for each $v \in \sT X$ such that $\zt_X(v) = \zr(r)$ and each $w \in \sT R$ such
that $\sT\zr(w) = v$.  The mapping $\zk$ is a subimmersion.  If $F$ is a
Morse family, then $\zk$ is an immersion.  The immersed Lagrangian
submanifold \Ref{Fff5} is the image $\im(\zk)$ of the mapping $\zk$.

    There is a surjective submersion $\zl : \sV^\circ R \rightarrow \sT^\* X$
characterized by
        $$\langle \zl(q), v \rangle = \langle q, w \rangle
                                                                                \tag \label{Fff7}$$
    for each $v \in \sT X$ such that $\zt_X(v) = \zr(\zp_R(q))$ and each $w \in \sT
R$ such that $\sT\zr(w) = v$.  The intersection $\im(\xd F) \cap \sV^\circ R$
is clean if $F$ is a regular family.  This intersection is transverse if $F$
is a Morse family.  The immersed Lagrangian submanifold \Ref{Fff5} is the set
$\zl(\im(\xd F) \cap \sV^\circ R)$.

    If the fibration $\zr$ is the identity morphism $1_X \colon X \rightarrow X$,
then the family of functions is a function
        $$F \colon X \rightarrow \R
                                                                                \tag \label{Fff8}$$
    and the Lagrangian submanifold $N$ is the set
        $$\align
    N &= \left\{p \in \sT^\* X ;\; \zp_X(p) = q,\, \all{v \in \sT_q X} \langle
p, v \rangle = \langle \xd F, v \rangle \right\} \\
            &= \im(\xd F).
                                                                                \tag \label{Fff9}\endalign$$

        \hl1{Generating objects}
    Let
    \vskip1mm
        $$\Rgaps{1.2}\Cgaps{1.2}\CD
    (P,\zy) @()\L{\zp}@(0,-1) && (P,\zy) @()\L{\zp}@(0,-1) @()\L{\zc}@(1,0)
                & (\sT^\*Q,\zy_Q) @()\L{\zp_Q} @(0,-1) \\
    Q && Q @()\a= @(1,0)      & Q
        \endCD
                                                                                \tag \label{Ggo1}$$
    \vskip1mm \noindent be a special symplectic structure for a symplectic manifold
$(P,\zw)$ with its canonical special symplectic structure morphism, let
        $$\zi_X \colon X \rightarrow Q
                                                                                \tag \label{Fgo1}$$
     be the canonical injection of a submanifold $X \subset Q$,
let
    \vskip1mm
        $$\Rgaps{1.2}\Cgaps{.8}\CD
    R @()\L{\zr} @(0,-1) \\ X
        \endCD
                                                                                \tag \label{Ggo2}$$
    \vskip1mm \noindent be a differential fibration, let
        \vskip1mm
        $$\Rgaps{1.2}\Cgaps{.8}\CD
    R @()\L{\zr} @(0,-1) @()\L{F} @(1,0) & \R \\
    X  &
        \endCD
                                                                                \tag \label{Ggo3}$$
    \vskip1mm\noindent be a regular family of functions defined on fibres of $\zr$,
and let $N \subset \sT^\*X$ be the immersed Lagrangian submanifold generated
by this family.  We denote by $\zx$ the mapping from $\sT^\*_X Q =
\zp_Q{}^{-1}(X)$ to $\sT^\*X$ characterized by
        $$\langle \zx(p), v \rangle = \langle p, v \rangle
                                                                                \tag \label{Fgo2}$$
    for each $v \in \sT X \subset \sT Q$ such that $\zt_Q(v) = \zp_Q(p)$.  Let $\zi
\colon \sT^\*_X Q \rightarrow \sT^\* Q$ be the canonical injection.  The set
        $$\align
    N' &= \zi(\zx^{-1}(N)) \\
            &= \left\{p \in \sT^\* Q ;\; \exi{r \in R}\,\zr(r) = \zp_Q(p) = q \in X
\vphantom{\all{v \in \sT_p \sT^\*X}} \right.  \\
            &\hskip30mm   \all{v \in \sT_q X \subset \sT_q Q} \all{w \in \sT_r R}
\left.\sT\zr(w) = v \Rightarrow  \langle p, v \rangle = \langle \xd F, w
\rangle\vphantom{\all{v \in \sT_p \sT^\*X}} \right\}
                                                                                \tag \label{Fgo3}\endalign$$
    is a Lagrangian submanifold of $(\sT^\* Q,\xd\zy_Q)$ and the set
        $$\align
    A &= \zc^{-1}(N') \\
            &= \left\{p \in P;\; q = \zp(p) \in X, \exi{r \in R_q} \all{w \in \sT_p
P}\right. \\
            & \hskip30mm \left. \all{z \in \sT_r R} \sT\zr(z) = \sT\zp(w)
\Rightarrow  \langle \zy, w \rangle = \langle \xd F, z \rangle \right\}
                                                                                \tag \label{Fgo4}\endalign$$
    is a Lagrangian submanifold of $(P,\xd\zy)$.  The diagram
    \vskip1mm
        $$\Rgaps{1.2}\Cgaps{.8}\CD
    (P,\zy) @()\L{\zp}@(0,-1)
                & R @()\L{\zr} @(0,-1) @()\L{F} @(1,0) & \R \\
    Q  & X @()\L{\zi_X} @(-1,0) &
        \endCD
                                                                                \tag \label{Ggo4}$$
    \vskip1mm \noindent is called a {\it generating object} for the immersed
Lagrangian submanifold $A$.

    The diagram
    \vskip1mm
        $$\rgaps{1.2}\cgaps{.8}\CD
    (P,\zy) @()\L{\zp}@(0,-1)
                &    \\
    Q  @()\L{F} @(1,0) & \R
        \endCD
                                                                                \tag \label{Ggo5}$$
    \vskip1mm \noindent is the simplest case of a generating object.  The Lagrangian
submanifold
        $$A = \left\{p \in P;\; \all{w \in \sT_p P} \langle \zy, w \rangle  =
\langle \xd F, \sT\zp(w) \rangle \right\}
                                                                                \tag \label{Fgo5}$$
    generated by this object is the image of the section $\zc^{-1} \circ \xd F$ of
the fibration $\zp$.

    If $X$ is a submanifold of $Q$ and the fibration $\zr$ is the identity morphism
$1_X \colon \rightarrow X$, then we have a generating object
    \vskip1mm
        $$\rgaps{1.2}\cgaps{.8}\CD
    (P,\zy) @()\L{\zp}@(0,-1)
                &  &  \\
    Q       & X @()\L{\zi_X} @(-1,0) @()\L{F} @(1,0) & \R
        \endCD
                                                                                \tag \label{Ggo6}$$
    \vskip1mm \noindent and the Lagrangian submanifold generated by this object is
the set
        $$A = \left\{p \in P;\; q = \zp(p) \in X, \all{w \in \sT_p P} v = \sT\zp(w)
\in \sT_q X \subset \sT_q Q \Rightarrow \langle \zy, w \rangle = \langle \xd
F, v\rangle \right\}.
                                                                                \tag \label{Fgo6}$$

    Another special case is obtained with $X = Q$ and a non trivial fibration $\zr$.
The generating object has the form
    \vskip1mm
        $$\rgaps{1.2}\cgaps{.8}\CD
    (P,\zy) @()\L{\zp}@(0,-1)
                & R @()\L{\zr} @(0,-1) @()\L{F} @(1,0) & \R \\
    Q @()\a= @(1,0)      & Q &
        \endCD
                                                                                \tag \label{Ggo7}$$
    \vskip1mm \noindent and the Lagrangian submanifold generated by this object is
the set
        $$A = \left\{p \in P;\; \exi{r \in R} \zr(r) = \zp(p) \all{w \in \sT_p P}
\all{z \in \sT_r R} \sT\zr(z) = \sT\zp(w) \Rightarrow \langle \zy, w \rangle
= \langle \xd F, z \rangle \right\}.
                                                                                \tag \label{Fgo7}$$

        \hl1{Reduction of generating objects}

    Let
    \vskip1mm
        $$\Rgaps{1.2}\Cgaps{.8}\CD
    R @()\L{\zr} @(0,-1) @()\L{F} @(1,0) & \R \\
    X  &
        \endCD
                                                                                \tag \label{Ggo8}$$
    \vskip1mm \noindent be a family of functions generating a Lagrangian submanifold
$N$ of $(\sT^\*X,\xd\zy_x)$.
        \claim \c{p}{Proposition}{}                                                  \label{Pgo1}
    If the family {\rm\Ref{Ggo8}} satisfies conditions
        \list
    \item the image $\widetilde{X} = \zr(S(F,\zr))$ of the critical set $S(F,\zr)$ by
the projection $\zr$ is a submanifold of $X$,
    \item the mapping
        $$\align
    \overline{\zr} &\colon S(F,\zr) \rightarrow \widetilde{X} \\
    &\colon r \mapsto \zr(r)
                                                                                \tag \label{Fgo8}\endalign$$
    induced by $\zr$ is a differential fibration with connected fibres,
        \endlist
    then there is a function $\widetilde{F} \colon \widetilde{X} \rightarrow \R$ such that
$F|S(F,\zr) = \widetilde{F} \circ \overline{\zr}$ and $N \subset
\widetilde{N}$, where $\widetilde{N}$ is the Lagrangian submanifold of
$(\sT^\*X,\xd\zy_X)$ generated by the generating object
    \vskip1mm
        $$\rgaps{1.2}\cgaps{.8}\CD
    (\sT^\*X,\zy_X) @()\L{\zp_X}@(0,-1)
                &  &  \\
    X       & \widetilde{X} @()\L{\zi_{\widetilde{X}}} @(-1,0) @()\L{\widetilde{F}} @(1,0) & \R
        \endCD
                                                                                \tag \label{Ggo9}$$
    \vskip1mm
        \endclaim
        \proof The function $\widetilde{F}$ can be defined since $F$ is constant on
fibres of $\overline{\zr}$.  We have
        $$N = \left\{p \in \sT^\* X;\; \exi{r \in R} \;\zr(r) = \zp_X(p), \all{z \in
\sT_r R} \langle p, \sT\zr(z) \rangle = \langle \xd F, z \rangle \right\}
                                                                                \tag \label{Fgo9}$$
    and
        $$\widetilde{N} = \left\{p \in \sT^\*X;\; q = \zp_X(p) \in \widetilde{X}, \all{v \in \sT_q \widetilde{X}
\subset \sT_q X} \langle p, v \rangle = \langle \xd \widetilde{F}, v \rangle
\right\}.
                                                                                \tag \label{Fgo10}$$
    Let $p \in N$ and let $q = \zp_X(p)$.  There is a point $r \in R$ such that
$\zr(r) = q$ and $\langle p,\sT\zr(z)\rangle = \langle \xd F, z \rangle$ for
each $z \in \sT_r R$. If $\sT\zr(z) = 0$, then $\langle \xd F, z \rangle =
0$.  Hence, $r \in S(F,\zr)$ and $q = \zp_X(p) \in \widetilde{X}$. If $v \in
\sT_q \widetilde{X} \subset \sT_q X$, $z \in \sT_r R$ and $\sT\zr(z) = v$,
then $z \in \sT_r S$ and
        $$\align
    \langle p,v \rangle &= \langle \xd F, z \rangle \\
            &= \langle \xd(F|S(F,\zr)), z \rangle \\
            &= \langle \xd(\widetilde{F} \circ \overline{\zr}), z \rangle \\
            &= \langle \overline{\zr}{}^\* \xd\widetilde{F}, z \rangle \\
            &= \langle \xd\widetilde{F}, v \rangle.
                                                                                \tag \label{Fgo11}\endalign$$
    It follows that $p \in \widetilde{N}$.
        \endproof

        \claim \c{p}{Proposition}{}                                                  \label{Pgo2}
    If a Lagrangian submanifold $N$ of $(\sT^\*X,\tx d\zy_X)$ is generated by a family
{\rm\Ref{Ggo8}} such that the critical set $S(F,\zr)$ is the image of a
section $\zs \colon X \rightarrow R$ of $\zr$, then $N$ is generated by the
function
    \vskip1mm
        $$\rgaps{1.2}\cgaps{1.4}\CD
        X @()\L{\widetilde{F}} @(1,0) & \R
        \endCD
                                                                                \tag \label{Ggo10}$$
    \vskip1mm \noindent with $\widetilde{F} = F \circ \zs$.
        \endclaim
        \proof  Let $\widetilde{N}$ be the Lagrangian submanifold generated by the
function $\widetilde{F}$.  For each $q \in X$ we have
        $$\align
    N_q &= N \cap \sT_q X \\
            &= \left\{p \in \sT^\* X;\; \exi{r \in R_q} \all{z \in
\sT_r R} \langle p, \sT\zr(z) \rangle = \langle \xd F, z \rangle \right\}
                                                                                \tag \label{Fgo12}\endalign$$
    and
        $$\align
    \widetilde{N}_q &= \widetilde{N} \cap \sT_q X \\
            &= \left\{p \in \sT^\* X;\; \all{v \in
\sT_q X} \langle p, v \rangle = \langle \xd F,  \rangle \right\}.
                                                                                \tag \label{Fgo13}\endalign$$
    The inclusion $N_q \subset \widetilde{N}_q$ is proved as in Proposition \Ref{Pgo1}.
The set $N_q$ is not empty and $\widetilde{N} = \left\{\xd\widetilde{F}(q)
\right\}$. Hence, $N_q = \widetilde{N}$ for each $q \in X$.  It follows that
$N = \widetilde{N}$.
        \endproof

    The following two propositions are easily derivable from Proposition \Ref{Pgo1}
and Proposition \Ref{Pgo2}.

        \claim \c{p}{Proposition}{}                                                  \label{Pgo3}
    If $N$ is generated by a family {\rm\Ref{Ggo8}} such that
        \list
    \item the fibration $\zr$ is the composition $\zr'' \circ \zr'$ of fibrations
$\zr' \colon R \rightarrow R'$ and $\zr'' \colon R' \rightarrow X$,
    \item the images $\widetilde{R} = \zr'(S(F,\zr'))$ and $\widetilde{X} =
\zr(S(F,\zr'))$ of the critical set $S(F,\zr')$ are submanifolds of $R'$ and
$X$ respectively,
    \item the induced mapping
        $$\align
    \widetilde{\zr} &\colon \widetilde{R} \rightarrow \widetilde{X} \\
    &\colon r' \mapsto \zr''(r')
                                                                                \tag \label{Fgo14}\endalign$$
    is a differential fibration
    \item the mapping
        $$\align
    \overline{\zr} &\colon S(F,\zr') \rightarrow \widetilde{R} \\
    &\colon r \mapsto \zr(r)
                                                                                \tag \label{Fgo15}\endalign$$
    is a differential fibration with connected fibres,
        \endlist
    then there is a function $\widetilde{F} \colon \widetilde{R} \rightarrow \R$
such that $F|S(F,\zr') = \widetilde{F} \circ \overline{\zr}$ and $N \subset
\widetilde{N}$, where $\widetilde{N}$ is generated by the generating object
    \vskip1mm
        $$\Rgaps{1.2}\Cgaps{.8}\CD
    (\sT^\*X,\zy_X) @()\L{\zp_X}@(0,-1)
                & \widetilde{R} @()\L{\widetilde{\zr}} @(0,-1) @()\L{\widetilde{F}}
@(1,0) & \R \\
    X & \widetilde{X} @()\L{\zi_{\widetilde{X}}} @(-1,0) &
        \endCD
                                                                                \tag \label{Ggo11}$$
    \vskip1mm \noindent with $\widetilde{F} = F \circ \zs$.
        \endclaim

        \claim \c{p}{Proposition}{}                                                  \label{Pgo4}
    If $N$ is generated by a family {\rm\Ref{Ggo8}} such that the fibration $\zr$ is
the composition $\widetilde{\zr} \circ \zr'$ of fibrations $\zr' \colon R
\rightarrow \widetilde{R}$ and $\widetilde{\zr} \colon \widetilde{R}
\rightarrow X$ and the critical set $S(F,\zr')$ is the image of a section
$\zs \colon \widetilde{R} \rightarrow R$ of $\zr'$, then $N$ is generated by
the generating object
    \vskip1mm
        $$\Rgaps{1.2}\Cgaps{.8}\CD
    (\sT^\*X,\zy_X) @()\L{\zp_X}@(0,-1)
                & \widetilde{R} @()\L{\widetilde{\zr}} @(0,-1) @()\L{\widetilde{F}} @(1,0) & \R \\
    X @()\a= @(1,0) & X &
        \endCD
                                                                                \tag \label{Ggo12}$$
    \vskip1mm \noindent with $\widetilde{F} = F \circ \zs$.
        \endclaim

    The last four propositions have the following obvious counterparts in the more
general setting obtained by replacing the family of functions \Ref{Ggo8} by a
generating object
    \vskip1mm
        $$\Rgaps{1.2}\Cgaps{.8}\CD
    (P,\zy) @()\L{\zp}@(0,-1)
                & R @()\L{\zr} @(0,-1) @()\L{F} @(1,0) & \R \\
    Q  & X @()\L{\zi_X} @(-1,0) &
        \endCD
                                                                                \tag \label{Ggo13}$$
    \vskip1mm

        \claim \c{p}{Proposition}{}                                                  \label{Pgo5}
    If a Lagrangian submanifold $A$ of $(P,\tx d\zy)$ is generated by a generating
object {\rm\Ref{Ggo13}} and
        \list
    \item the image $\widetilde{X} = \zr(S(F,\zr))$ of the critical set $S(F,\zr)$ by
the projection $\zr$ is a submanifold of $X$,
    \item the mapping
        $$\align
    \widetilde{\zr} &\colon S(F,\zr) \rightarrow \widetilde{X} \\
    &\colon r \mapsto \zr(r)
                                                                                \tag \label{Fgo16}\endalign$$
    induced by $\zr$ is a differential fibration with connected fibres,
        \endlist
    then $A \subset \widetilde{A}$, where $\widetilde{A}$ is the Lagrangian submanifold of
$(P,\xd\zy)$ generated by the generating object
    \vskip1mm
        $$\rgaps{1.2}\cgaps{.8}\CD
    (P,\zy) @()\L{\zp}@(0,-1)
                &  &  \\
    Q       & \widetilde{X} @()\L{\zi_{\widetilde{X}}} @(-1,0) @()\L{\widetilde{F}} @(1,0) & \R
        \endCD
                                                                                \tag \label{Ggo14}$$
    \vskip1mm
        \endclaim

        \claim \c{p}{Proposition}{}                                                  \label{Pgo6}
    If a Lagrangian submanifold $A$ of $(P,\tx d\zy)$ is generated by a generating
object {\rm\Ref{Ggo13}} and the critical set $S(F,\zr)$ is the image of a
section $\zs \colon X \rightarrow R$ of $\zr$, then $A$ is generated by the
generating object
    \vskip1mm
        $$\rgaps{1.2}\cgaps{.8}\CD
    (P,\zy) @()\L{\zp}@(0,-1)
                &  &  \\
    Q       & X @()\L{\zi_X} @(-1,0) @()\L{\widetilde{F}} @(1,0) & \R
        \endCD
                                                                                \tag \label{Ggo15}$$
    \vskip1mm \noindent with $\widetilde{F} = F \circ \zs$.
        \endclaim

        \claim \c{p}{Proposition}{}                                                  \label{Pgo7}
    If $A$ is generated by a generating object {\rm\Ref{Ggo13}} such that
        \list
    \item the fibration $\zr$ is the composition $\zr'' \circ \zr'$ of fibrations
$\zr' \colon R \rightarrow R'$ and $\zr'' \colon R' \rightarrow X$,
    \item the images $\widetilde{R} = \zr'(S(F,\zr'))$ and $\widetilde{X} =
\zr(S(F,\zr'))$ of the critical set $S(F,\zr')$ are submanifolds of $R'$ and
$X$ respectively,
    \item the induced mapping
        $$\align
    \widetilde{\zr} &\colon \widetilde{R} \rightarrow \widetilde{X} \\
    &\colon r' \mapsto \zr''(r')
                                                                                \tag \label{Fgo17}\endalign$$
    is a differential fibration,
    \item the mapping
        $$\align
    \overline{\zr} &\colon S(F,\zr') \rightarrow \widetilde{R} \\
    &\colon r \mapsto \zr(r)
                                                                                \tag \label{Fgo18}\endalign$$
    is a differential fibration with connected fibres,
        \endlist
    then there is a function $\widetilde{F} \colon \widetilde{R} \rightarrow \R$
such that $F|S(F,\zr') = \widetilde{F} \circ \overline{\zr}$ and $A \subset
\widetilde{A}$, where $\widetilde{A}$ is generated by the generating object
    \vskip1mm
        $$\Rgaps{1.2}\Cgaps{.8}\CD
    (P,\zy) @()\L{\zp}@(0,-1)
                & \widetilde{R} @()\L{\widetilde{\zr}} @(0,-1) @()\L{\widetilde{F}}
@(1,0) & \R \\
    Q & \widetilde{X} @()\L{\zi_{\widetilde{X}}} @(-1,0) &
        \endCD
                                                                                \tag \label{Ggo16}$$
    \vskip1mm \noindent with $\widetilde{F} = F \circ \zs$.
        \endclaim

        \claim \c{p}{Proposition}{}                                                  \label{Pgo8}
    If $A$ is generated by a generating object {\rm\Ref{Ggo13}} such that the
fibration $\zr$ is the composition $\widetilde{\zr} \circ \zr'$ of fibrations
$\zr' \colon R \rightarrow \widetilde{R}$ and $\widetilde\zr \colon
\widetilde{R} \rightarrow X$, and the critical set $S(F,\zr')$ is the image
of a section $\zs' \colon \widetilde{R} \rightarrow R$, then $A$ is generated
by the generating object
    \vskip1mm
        $$\Rgaps{1.2}\Cgaps{.8}\CD
    (P,\zy) @()\L{\zp}@(0,-1)
                & \widetilde{R} @()\L{\widetilde{\zr}} @(0,-1) @()\L{\widetilde{F}} @(1,0) & \R \\
    Q  & X @()\L{\zi_X} @(-1,0) &
        \endCD
                                                                                \tag \label{Ggo17}$$
    \vskip1mm \noindent with $\widetilde{F} = F \circ \zs'$.
        \endclaim

        \hl1{Composition of generating objects}

    The Lagrangian submanifold $A_{21}$ generated by a generating object
    \vskip1mm
        $$\rgaps{1.2}\cgaps{.8}\CD
    (P_2 \times P_1,\zy_2 \ominus \zy_1) @()\L{\zp_2 \times \zp_1}@(0,-1)
                & R_{21} @()\L{\zr_{21}} @(0,-1) @()\L{F_{21}} @(1,0) & \R \\
    Q_2 \times Q_1  & X_{21} @()\L{\zi_{X_{21}}} @(-1,0) &
        \endCD
                                                                                \tag \label{Ggo18}$$
    \vskip1mm \noindent is interpreted as the graph of a symplectic relation $\zP_{21}$
from $(P_1,\xd\zy_1)$ to  $(P_2,\xd\zy_2)$.  Let $A_1$ be a Lagrangian
submanifold of $(P_1,\xd\zy_1)$ generated by a generating object
    \vskip1mm
        $$\rgaps{1.2}\cgaps{.8}\CD
    (P_1,\zy_1) @()\L{\zp_1}@(0,-1)
                & R_1 @()\L{\zr_1} @(0,-1) @()\L{F_1} @(1,0) & \R \\
    Q_1  & X_1 @()\L{\zi_{X_1}} @(-1,0) &
        \endCD
                                                                                \tag \label{Ggo19}$$
    \vskip1mm \noindent If the image
        $$\zP_{21}(A_1) = \{p_2 \in P_2;\; \exi{p_1 \in A_1} (p_2,p_1) \in A_{21}\}
                                                                                \tag \label{Fgo19}$$

    of $A_1$ by the relation $\zP_{21}$ is a Lagrangian submanifold $A_2$ of
$(P_2,\xd\zy_2)$, then it is generated by the generating object
    \vskip1mm
        $$\rgaps{1.2}\cgaps{.8}\CD
    (P_2,\zy_2) @()\L{\zp_2}@(0,-1)
                & R_2 @()\L{\zr_2} @(0,-1) @()\L{F_2} @(1,0) & \R \\
    Q_2  & X_2 @()\L{\zi_{X_2}} @(-1,0) &
        \endCD
                                                                                \tag \label{Ggo20}$$
    \vskip1mm \noindent described below.

    We introduce the sets
        $$\align
    X_2 &= X_{21} \circ X_1 \\
            &= \left\{q_2 \in Q_2;\; \exi{q_1 \in X_1}
(q_2,q_1) \in X_{21} \right\},
                                                                                \tag \label{Fgo22}\endalign$$
        $$X_{2[11]} = \left\{(q_2,q_1,q'_1) \in Q_2 \times Q_1
\times Q_1;\; (q_2,q_1) \in X_{21}, q'_1 = q_1 \in X_1 \right\},
                                                                                \tag \label{Fgo20}$$
    and the projection
        $$\align
    pr^{2[11]} &\colon X_{2[11]} \rightarrow X_2 \\
    &\colon (q_2,q_1,q_1) \mapsto q_2.
                                                                                \tag \label{Fgo21}\endalign$$
    We define the set $R_2$, the fibration $\zr_2 \colon R_2 \rightarrow X_2$, and the
family of functions $F_2 \colon R_2 \rightarrow \R$ by
        $$R_2 = (\zr_{21} \times \zr_1)^{-1}(X_{2[11]}),
                                                                                \tag \label{Fgo23}$$
        $$\align
    \zr_2 &\colon R_2 \rightarrow X_2 \\
    &\colon (r_{21},r_1) \mapsto pr^{2[11]}(\zr_{21}(r_{21}),\zr_1(r_1)),
                                                                                \tag \label{Fgo24}\endalign$$
    and
        $$\align
    F_2 &\colon R_2 \rightarrow \R \\
    &\colon (r_{21},r_1) \mapsto F_{21}(r_{21}) + F_1(r_1).
                                                                                \tag \label{Fgo25}\endalign$$

               \hl1{Lagrangian systems}

    Let $Q$ be the configuration manifold of a mechanical system.  The special
symplectic structure
    \vskip1mm
        $$\Rgaps{1.2}\Cgaps{1.2}\cgaps{.9;.9;1.2}\CD
    (\sT\sT^\* Q,\xd_T\zy_Q) @()\L{\sT\zp_Q}@(0,-1) && (\sT\sT^\* Q,\xd_T\zy_Q)
@()\L{\sT\zp_Q}@(0,-1) @()\L{\za_Q}@(1,0)
                & (\sT^\*\sT Q,\zy_{\sT Q}) @()\L{\zp_{\sT Q}} @(0,-1) \\
    \sT Q && \sT Q @()\a= @(1,0)      & \sT Q
        \endCD
                                                                                \tag \label{Gls1}$$
    \vskip1mm \noindent is called the {\it Lagrangian special symplectic structure}.
A generating object
    \vskip1mm
        $$\Rgaps{1.2}\Cgaps{.8}\CD
    (\sT\sT^\* Q,\xd_T\zy_Q) @()\L{\sT\zp_Q}@(0,-1)
                & Y @()\L{\zh} @(0,-1) @()\L{L} @(1,0) & \R \\
    \sT Q  & C @()\L{\zi_C} @(-1,0) &
        \endCD
                                                                                \tag \label{Gls2}$$
    \vskip1mm \noindent is called a {\it Lagrangian system}.  The family of functions
    \vskip1mm
        $$\Rgaps{1.2}\Cgaps{.8}\CD
            Y @()\L{\zh} @(0,-1) @()\L{L} @(1,0) & \R \\  C &
        \endCD
                                                                                \tag \label{Gls3}$$
    \vskip1mm \noindent is called a {\it Lagrangian family} and the immersed
Lagrangian submanifold
        $$\split
    D = \left\{w \in \sT\sT^\*Q ;\; v = \sT\zp_Q(w) \in C, \exi{y \in Y_v}
\vphantom{\all{u \in \sT_w\sT\sT^\*Q}} \right.  \\
                  \all{x \in \sT_w\sT\sT^\*Q} \all{z \in \sT_y Y}\sT\zh(z) =
\sT\sT\zp_Q(x) \Rightarrow  \langle& \left. \xd_T\zy_Q, x \rangle = \langle
\xd L, z \rangle\vphantom{\all{u \in \sT_q C}} \right\}
                                                                        \endsplit\tag \label{Fls1}$$
    generated by the Lagrangian system \Ref{Gls2} is called the {\it dynamics} of the
mechanical system.

          \hl1{Hamiltonian systems}

    Let $(P,\zw)$ be a symplectic manifold representing the phase space of a
mechanical system.  The special symplectic structure represented by the
diagrams
    \vskip1mm
        $$\Rgaps{1.2}\Cgaps{1.2}\CD
    (\sT P,\xi_T \zw) @()\L{\zt_P}@(0,-1) &&(\sT P,\xi_T\zw) @()\L{\zt_P}@(0,-1)
@()\L{\zb_{(P,\zw)}}@(1,0)
                & (\sT^\*P,\zy_P) @()\L{\zp_P} @(0,-1) \\
    P && P @()\a= @(1,0)      & P
        \endCD
                                                                                \tag \label{Ghs1}$$
    \vskip1mm \noindent is called the {\it Hamiltonian special symplectic structure}.
A generating object
    \vskip1mm
        $$\Rgaps{1.2}\Cgaps{.8}\CD
    (\sT P,\xi_T\zw) @()\L{\zt_P}@(0,-1)
                & Z @()\L{\zz} @(0,-1) @()\L{-H} @(1,0) & \R \\
    P  & K @()\L{\zi_K} @(-1,0) &
        \endCD
                                                                                \tag \label{Ghs2}$$
    \vskip1mm \noindent is called a {\it Hamiltonian system}.  The family of functions
    \vskip1mm
        $$\Rgaps{1.2}\Cgaps{.8}\CD
            Z @()\L{\zz} @(0,-1) @()\L{H} @(1,0) & \R \\  K &
        \endCD
                                                                                \tag \label{Ghs3}$$
    \vskip1mm \noindent is called a {\it Hamiltonian family} and the immersed
Lagrangian submanifold
        $$\split
    D = \left\{w \in \sT P ;\; p = \zt_P(w) \in K, \exi{z \in Z_p}
\vphantom{\all{u \in \sT_w\sT\sT^\*Q}}\all{x \in \sT_w\sT P} \right.  \\
                   \all{u \in \sT_z Z}\sT\zz(u) = \sT\zt_P(x)
\Rightarrow  \langle \xi_T\zw, x \rangle & \left. = - \langle \xd H, u
\rangle\vphantom{\all{u \in \sT_q C}} \right\}
                                                                        \endsplit\tag \label{Fhs1}$$
    generated by the Hamiltonian system \Ref{Ghs2} is called the {\it dynamics} of
the mechanical system.

    A Hamiltonian system
    \vskip1mm
        $$\Rgaps{1.2}\Cgaps{.8}\CD
    (\sT P,\xi_T\zw) @()\L{\zt_P}@(0,-1) & & \\
    P  & K @()\L{\zi_K} @(-1,0) @()\L{-H} @(1,0) & \R
        \endCD
                                                                                \tag \label{Ghs4}$$
    \vskip1mm \noindent is called a {\it Dirac system}.  A more general Hamiltonian
system \Ref{Ghs2} is called a {\it generalized Dirac system}.

    The phase space of a mechanical system is usually the cotangent bundle $\sT^\*Q$
of the configuration manifold with its canonical symplectic structure
$\xd\zy_Q$. The Hamiltonian special symplectic structure is in this case
represented by
    \vskip1mm
        $$\Rgaps{1.2}\Cgaps{1.2}\cgaps{.9;.6;1.4}\CD
    (\sT\sT^\*Q,\xi_T\xd\zy_Q) @()\L{\zt_{\sT^\*Q}}@(0,-1)
&&(\sT\sT^\*Q,\xi_T\xd\zy_Q) @()\L{\zt_{\sT^\*Q}}@(0,-1)
@()\L{\zb_{(\sT^\*Q,\xd\zy_Q)}}@(1,0)
                & (\sT^\*\sT^\*Q,\zy_{\sT^\*Q}) @()\L{\zp_{\sT^\*Q}} @(0,-1) \\
    \sT^\*Q && \sT^\*Q @()\a= @(1,0)      & \sT^\*Q
        \endCD
                                                                                \tag \label{Ghs5}$$
    \vskip1mm \noindent and a Hamiltonian system is a generating object
    \vskip1mm
        $$\Rgaps{1.2}\Cgaps{.8}\CD
    (\sT\sT^\*Q,\xi_T\xd\zy_Q) @()\L{\zt_{\sT^\*Q}}@(0,-1)
                & Z @()\L{\zz} @(0,-1) @()\L{-H} @(1,0) & \R \\
    \sT^\*Q  & K @()\L{\zi_K} @(-1,0) &
        \endCD
                                                                                \tag \label{Ghs6}$$
    \vskip1mm \noindent The dynamics of the mechanical system is the set
        $$\split
    D = \left\{w \in \sT\sT^\*Q ;\; p = \zt_{\sT^\*Q}(w) \in K, \exi{z \in Z_p}
\vphantom{\all{u \in \sT_w\sT\sT^\*Q}} \all{x \in \sT_w\sT\sT^\*Q} \right.\\
                   \all{u \in \sT_z Z}\sT\zz(u) =
\sT\zt_{\sT^\*Q}(x) \Rightarrow  \langle \xi_T\xd\zy_Q, x \rangle &\left.=
-\langle \xd H, u \rangle\vphantom{\all{u \in \sT_q C}} \right\}.
                                                                        \endsplit\tag \label{Fhs2}$$

                \hl1{The Legendre transformation}

    Let $Q$ be the configuration manifold of a mechanical system.  The phase space
of the system is the symplectic manifold $(\sT^\*Q,\xd\zy_Q)$.  There are two
canonical special symplectic structures for the symplectic manifold
$(\sT\sT^\* Q,\xd_T\xd\zy_Q)$.  We have the Lagrangian special symplectic
structure
    \vskip1mm
        $$\Rgaps{1.2}\Cgaps{1.2}\cgaps{.9;.9;1.4}\CD
    (\sT\sT^\* Q,\xd_T\zy_Q) @()\L{\sT\zp_Q}@(0,-1) && (\sT\sT^\* Q,\xd_T\zy_Q)
@()\L{\sT\zp_Q}@(0,-1) @()\L{\za_Q}@(1,0)
                & (\sT^\*\sT Q,\zy_{\sT Q}) @()\L{\zp_{\sT Q}} @(0,-1) \\
    \sT Q && \sT Q @()\a= @(1,0)      & \sT Q
        \endCD
                                                                                \tag \label{Glt1}$$
    \vskip1mm \noindent and the Hamiltonian special symplectic structure
    \vskip1mm
        $$\Rgaps{1.2}\Cgaps{1.2}\cgaps{.9;.6;1.4}\CD
    (\sT\sT^\*Q,\xi_T\xd\zy_Q) @()\L{\zt_{\sT^\*Q}}@(0,-1)
&&(\sT\sT^\*Q,\xi_T\xd\zy_Q) @()\L{\zt_{\sT^\*Q}}@(0,-1)
@()\L{\zb_{(\sT^\*Q,\xd\zy_Q)}}@(1,0)
                & (\sT^\*\sT^\*Q,\zy_{\sT^\*Q}) @()\L{\zp_{\sT^\*Q}} @(0,-1) \\
    \sT^\*Q && \sT^\*Q @()\a= @(1,0)      & \sT^\*Q
        \endCD
                                                                                \tag \label{Glt2}$$
    \vskip1mm

    The identity morphism $1_{\sT\sT^\*Q}$ is a symplectic relation generated by the
generating object
    \vskip1mm
        $$\Rgaps{1.2}\Cgaps{.9}\CD
    (\sT\sT^\* Q \times \sT\sT^\* Q,\xi_T\xd\zy_Q \ominus \xd_T\zy_Q)
@()\L{\zt_{\sT^\*Q} \times \sT\zp_Q}@(0,-1) & &  \\
    \sT^\*Q \times \sT Q  & \sT^\*Q \times_Q \sT Q @()\L{\zi_{\sT^\*Q \times_Q \sT Q}}
@(-1,0) @()\L{-\langle \,,\rangle}@(1,0) & \R
        \endCD
                                                                                \tag \label{Glt3}$$
    \vskip1mm \noindent and by the generating object
    \vskip1mm
        $$\Rgaps{1.2}\Cgaps{.9}\CD
    (\sT\sT^\* Q \times \sT\sT^\* Q,\xd_T\zy_Q \ominus \xi_T\xd\zy_Q)
@()\L{\sT\zp_Q \times \zt_{\sT^\*Q}}@(0,-1) & &  \\
    \sT Q \times \sT^\*Q & \sT Q \times_Q \sT^\*Q @()\L{\zi_{\sT Q \times_Q \sT^\*Q}}
@(-1,0) @()\L{\langle \,,\rangle^\sim}@(1,0) & \R
        \endCD
                                                                                \tag \label{Glt4}$$
    \vskip1mm \noindent where $\langle \,,\rangle^\sim$ is the mapping
        $$\align
    \langle \,,\rangle^\sim &\colon \sT Q \times_Q \sT^\*Q \rightarrow \R \\
    &\colon (v,p) \mapsto \langle p, v \rangle.
                                                                                \tag \label{Flt1}\endalign$$

    Starting with a Lagrangian system
    \vskip1mm
        $$\Rgaps{1.2}\Cgaps{.8}\CD
    (\sT\sT^\* Q,\xd_T\zy_Q) @()\L{\sT\zp_Q}@(0,-1)
                & Y @()\L{\zh} @(0,-1) @()\L{L} @(1,0) & \R \\
    \sT Q  & C @()\L{\zi_C} @(-1,0) &
        \endCD
                                                                                \tag \label{Glt5}$$
    \vskip1mm \noindent we construct a Hamiltonian system
    \vskip1mm
        $$\Rgaps{1.2}\Cgaps{.8}\CD
    (\sT\sT^\*Q,\xi_T\xd\zy_Q) @()\L{\zt_{\sT^\*Q}}@(0,-1)
                & Z @()\L{\zz} @(0,-1) @()\L{-H} @(1,0) & \R \\
    \sT^\*Q  & K @()\L{\zi_K} @(-1,0) &
        \endCD
                                                                                \tag \label{Glt6}$$
    \vskip1mm \noindent with
        $$\align
    K &= \zp_Q{}^{-1}(\zt_Q(C)) \\
            &= \left\{p \in \sT^\*Q;\; \exi{v \in C} \zp_Q(p) = \zt_Q(v) \right\},
                                                                                \tag \label{Flt2}\endalign$$
        $$Z = \{(p,v,y) \in \sT^\*Q \times_Q \sT Q \times Y;\; v \in C, \zh(y) = v \},
                                                                                \tag \label{Flt3}$$
        $$\align
    \zz &\colon Z \rightarrow K \\
    &\colon (p,v,y) \mapsto p,
                                                                                \tag \label{Flt4}\endalign$$
    and
        $$\align
    H &\colon Z \rightarrow \R \\
    &\colon (p,v,y) \mapsto \langle p, v \rangle - L(y).
                                                                                \tag \label{Flt5}\endalign$$
    This Hamiltonian system is obtained by composing the Lagrangian system with the
generating object \Ref{Glt3}.  The two systems generate the same dynamics
since the generating object \Ref{Glt3} generates the identity relation.  The
passage from the Lagrangian system \Ref{Glt5} to the Hamiltonian system
\Ref{Glt6} is called the {\it Legendre transformation}.

    Conversely, given a Hamiltonian system
    \vskip1mm
        $$\Rgaps{1.2}\Cgaps{.8}\CD
    (\sT\sT^\*Q,\xi_T\xd\zy_Q) @()\L{\zt_{\sT^\*Q}}@(0,-1)
                & Z @()\L{\zz} @(0,-1) @()\L{-H} @(1,0) & \R \\
    \sT^\*Q  & K @()\L{\zi_K} @(-1,0) &
        \endCD
                                                                                \tag \label{Glt7}$$
    \vskip1mm \noindent we construct a
Lagrangian system
    \vskip1mm
        $$\Rgaps{1.2}\Cgaps{.8}\CD
    (\sT\sT^\* Q,\xd_T\zy_Q) @()\L{\sT\zp_Q}@(0,-1)
                & Y @()\L{\zh} @(0,-1) @()\L{L} @(1,0) & \R \\
    \sT Q  & C @()\L{\zi_C} @(-1,0) &
        \endCD
                                                                                \tag \label{Glt8}$$
    \vskip1mm \noindent with
        $$\align
    C &= \zt_Q{}^{-1}(\zp_Q(K)) \\
            &= \left\{v \in \sT^\*Q;\; \exi{p \in K} \zp_Q(p) = \zt_Q(v) \right\},
                                                                                \tag \label{Flt6}\endalign$$
        $$Y = \{(p,v,z) \in \sT^\*Q \times_Q \sT Q \times Z;\; p \in K, \zz(z) = p \},
                                                                                \tag \label{Flt7}$$
        $$\align
    \zh &\colon Y \rightarrow C \\
    &\colon (p,v,z) \mapsto v,
                                                                                \tag \label{Flt8}\endalign$$
    and
        $$\align
    L &\colon Y \rightarrow \R \\
    &\colon (p,v,z) \mapsto \langle p, v \rangle - H(z)
                                                                                \tag \label{Flt9}\endalign$$
    by composing the Hamiltonian system with the generating object \Ref{Glt4}.  This
passage is called the {\it inverse Legendre transformation}.

    In addition to the Legendre transformations we have {\it Legendre relations}.
Given a Lagrangian system \Ref{Glt5} we have the {\it first Legendre
relation}
        $$\zL_1(L) \colon Y \rightarrow \sT^\*Q,
                                                                                \tag \label{Flt10}$$
    whose graph is the set
        $$\split
    \gr(\zL_1(L)) = \left\{(p,y) \in \sT^\*Q \times Y;\; y \in S(L,\zh), \zp_Q(p) =
q = \zt_Q(\zh(y)) \vphantom{\all{u \in \sT_q C}} \right.  \\
                  \all{u \in C \cap \sT_q Q} \all{z \in \sT_y Y} \sT\zh(z) =
\zq_Q(\zh(y),u) \Rightarrow  \langle p, u \rangle = \langle \xd L, z &\left.
\rangle\vphantom{\all{u \in \sT_q C}} \right\}
                                                                        \endsplit\tag \label{Flt11}$$
    and the {\it second Legendre relation}
        $$\zL_2(L) \colon \sT Q \rightarrow \sT^\*Q,
                                                                                \tag \label{Flt12}$$
    whose graph is the set
        $$\gr(\zL_2(L)) = \left\{(p,v) \in \sT^\*Q \times \sT Q;\; \exi{y \in Y}
\zh(y) = v, (p,y) \in \gr(\zL_1(L)) \right\}.
                                                                                \tag \label{Flt13}$$
    If $D \subset \sT\sT^\*Q$ is the dynamics generated by the Lagrangian system,
then
        $$\gr(\zL_2(L)) = (\zt_{\sT^\*Q},\sT\zp_Q)(D).
                                                                                \tag \label{Flt14}$$

    A Lagrangian system is said to be {\it hyperregular} if the second Legendre
relation $\zL_2(L)$ is a diffeomorphism.

    Given a Hamiltonian system \Ref{Glt6} we introduce the {\it first inverse
Legendre relation}
        $$\zW_1(H) \colon Z \rightarrow \sT Q,
                                                                                \tag \label{Flt15}$$
    whose graph is the set
        $$\split
    \gr(\zW_1(H)) = \left\{(v,z) \in \sT Q \times Z;\; z \in S(H,\zz), \zt_Q(v) =
q = \zp_Q(\zz(z)) \vphantom{\all{u \in \sT_q C}} \right.  \\
                  \all{a \in K \cap \sT^\*_q Q} \all{u \in \sT_z Z} \sT\zz(u) =
\zq_{\sT^\* Q}(\zz(z),a) \Rightarrow  \langle a, v \rangle = \langle& \left.
\xd H, u \rangle\vphantom{\all{u \in \sT_q C}} \right\}
                                                                        \endsplit\tag \label{Flt16}$$
    and the {\it second inverse Legendre relation}
        $$\zW_2(H) \colon \sT^\* Q \rightarrow \sT Q,
                                                                                \tag \label{Flt17}$$
    whose graph is the set
        $$\gr(\zW_2(H)) = \left\{(v,p) \in \sT Q \times \sT^\* Q;\; \exi{z \in Z}
\zz(z) = p, (v,z) \in \gr(\zW_1(W)) \right\}.
                                                                                \tag \label{Flt18}$$
    If the Hamiltonian system is obtained from the Lagrangian system \Ref{Glt5} by
the Legendre transformation, then the second inverse Legendre relation
$\zW_2(H)$ is the transpose $\zL_2(L)^t$ of the second Legendre relation
$\zL_2(L)$.  The {\it transpose} of a relation $\zF \colon A \rightarrow B$
is the relation $\zF^t \colon B \rightarrow A$ with
        $$\gr(\zF^t) = \{(a,b) \in A \times B;\; (b,a) \in \gr(\zF) \}.
                                                                                \tag \label{Flt19}$$
        \hl1{Homogeneous families of functions}
    Let
    \vskip1mm
        $$\Rgaps{1.2}\Cgaps{.8}\CD
    R @()\L{\zr} @(0,-1) @()\L{F} @(1,0) & \R \\
    X  &
        \endCD
                                                                                \tag \label{Ghf1}$$
    \vskip1mm \noindent be a family of functions and let
    \vskip1mm
        $$\Rgaps{1.2}\Cgaps{1.2}\CD
    \Rp \times R @()\L{1_{\Rp} \times \zr}@(0,-1) @()\L{\zn}@(1,0)
                & R @()\L{\zr} @(0,-1) \\
    \Rp \times X @()\L{\zm} @(1,0)      & X
        \endCD
                                                                                \tag \label{Ghf2}$$
    \vskip1mm \noindent be a group action of the group $(\Rp, \cdot)$ by fibration
isomorphisms.  The family \Ref{Ghf1} is said to be {\it homogeneous} with
respect to the group action \Ref{Ghf2} if
        $$F(\zn(k,r)) = kF(r)
                                                                                \tag \label{Fhf1}$$
    for each $r \in R$ and each $k \in \Rp$.

    For each $k \in \Rp$ we have the diffeomorphism $\zn(k,\cdot) \colon R
\rightarrow R$ and for each $r \in R$ there is the linear isomorphism
$\sT_r(\zn(k,\cdot)) \colon \sT_r R \rightarrow \sT_{\zn(k,\cdot)}R$.  A
vector $z \in \sT_r R$ is vertical if and only if $\sT_r(\zn(k,\cdot))(z)$ is
vertical.  If \Ref{Ghf1} is a homogeneous family, then
        $$\zn(k,\cdot)^\* F = F \circ \zn(k,\cdot) = kF
                                                                                \tag \label{Fhf2}$$
    and
        $$\zn(k,\cdot)^\* \xd F = \xd\zn(k,\cdot)^\* F = k\xd F.
                                                                                \tag \label{Fhf3}$$
    Hence,
        $$\langle \xd F, \sT_r(\zn(k,\cdot))(z) \rangle = \langle \zn(k,\cdot)^\*
\xd F, z \rangle = k\langle \xd F, z \rangle
                                                                                \tag \label{Fhf4}$$
    for each $z \in \sT_r R$.

    A set $S \subset R$ is said to be {\it homogeneous} if $r \in S$ implies $\zn(k,r)
\in S$ for each $k \in \Rp$.

        \claim \c{p}{Proposition}{}                                                  \label{Phf1}
    The critical set $S(F,\zr)$ for a homogeneous family {\rm\Ref{Ghf1}} is homogeneous.
        \endclaim
        \proof Let $r \in S(F,\zr)$ and let $z \in \sT_{\zn(k,r)}R$ be a vertical
vector.  We have
        $$\align
    \langle \xd F, z \rangle &= \langle \xd F,
\sT_r(\zn(k,\cdot))(\sT_r(\zn(k^{-1},\cdot))(z)) \rangle \\
            &= k\langle \xd F, \sT_r(\zn(k^{-1},\cdot))(z) \rangle \\
            &= 0
                                                                                \tag \label{Fhf5}\endalign$$
    since $\sT_r(\zn(k^{-1},\cdot))(z)$ is a vertical vector in $\sT_r R$ and $r$ is
a critical point.  It follows that $\zn(k,r)$ is a critical point.
        \endproof

    The group action $\zm \colon \Rp \times X \rightarrow X$ is lifted to the action
    \vskip1mm
        $$\Rgaps{1.2}\Cgaps{1.2}\CD
    \Rp \times \sT^\*X @()\L{1_{\Rp} \times \zp_X}@(0,-1) @()\L{\overline\zm}@(1,0)
                & \sT^\*X @()\L{\zp_X} @(0,-1) \\
    \Rp \times X @()\L{\zm} @(1,0)      & X
        \endCD
                                                                                \tag \label{Ghf3}$$
    \vskip1mm \noindent with $\overline\zm \colon \Rp \times \sT^\*X \rightarrow
\sT^\*X$ defined by
        $$\overline\zm(k,f) = k \sT^\*\zm(k^{-1},f)
                                                                                \tag \label{Ghf4}$$
    for each $k \in \Rp$ and each $f \in \sT^\*X$.

    The set $N$ generated by a homogeneous family of functions is homogeneous.

        \hl1{Homogeneous generating objects}
    Let
    \vskip1mm
        $$\Rgaps{1.2}\Cgaps{1.2}\CD
    (P,\zy) @()\L{\zp}@(0,-1) && (P,\zy) @()\L{\zp}@(0,-1) @()\L{\zc}@(1,0)
                & (\sT^\*Q,\zy_Q) @()\L{\zp_Q} @(0,-1) \\
    Q && Q @()\a= @(1,0)      & Q
        \endCD
                                                                                \tag \label{Ghg1}$$
    \vskip1mm \noindent be a special symplectic structure with its canonical special
symplectic structure morphism.  A group action
        $$\bi\zm \colon \Rp \times Q \rightarrow Q
                                                                                \tag \label{Fhg1}$$
     is lifted to the action
    \vskip1mm
        $$\Rgaps{1.2}\Cgaps{1.2}\CD
    \Rp \times P @()\L{1_{\Rp} \times \zp}@(0,-1) @()\L{\widehat{\zm}}@(1,0)
                & P @()\L{\zp} @(0,-1) \\
    \Rp \times Q @()\L{\bi\zm} @(1,0)      & Q
        \endCD
                                                                                \tag \label{Ghg2}$$
    \vskip1mm \noindent with $\widehat{\zm} \colon \Rp \times P \rightarrow P$
defined by
        $$\widehat\zm(k,\cdot) = \zc^{-1} \circ \overline{\bi\zm} \circ \zc
                                                                                \tag \label{Fhg2}$$
    and
        $$\overline{\bi\zm}(k,f) = k \sT^\*\bi\zm(k^{-1},f)
                                                                                \tag \label{Fhg3}$$
    for each $k \in \Rp$ and each $f \in \sT^\*Q$.  Let
        $$\zi_X \colon X \rightarrow Q
                                                                                \tag \label{Fhg4}$$
    be the canonical injection of a homogeneous submanifold $X \subset Q$ and let
        $$\zm \colon \Rp \times X \rightarrow X
                                                                                \tag \label{Fhg5}$$
    be the restriction to $X$ of the group action \Ref{Fhg1} in the sense that
    \vskip1mm
        $$\Rgaps{1.2}\Cgaps{1.2}\CD
    \Rp \times X @()\L{1_{\Rp} \times \zi_X}@(0,-1) @()\L{\zm}@(1,0)
                & X @()\L{\zi_X} @(0,-1) \\
    \Rp \times Q @()\L{\bi\zm} @(1,0)      & Q
        \endCD
                                                                                \tag \label{Ghg3}$$
    \vskip1mm \noindent is a commutative diagram.  Let
    \vskip1mm
        $$\Rgaps{1.2}\Cgaps{.8}\CD
    R @()\L{\zr} @(0,-1) @()\L{F} @(1,0) & \R \\
    X  &
        \endCD
                                                                                \tag \label{Ghg4}$$
    \vskip1mm \noindent be a family of functions homogeneous with respect to a group
action
    \vskip1mm
        $$\Rgaps{1.2}\Cgaps{1.2}\CD
    \Rp \times R @()\L{1_{\Rp} \times \zr}@(0,-1) @()\L{\zn}@(1,0)
                & R @()\L{\zr} @(0,-1) \\
    \Rp \times X @()\L{\zm} @(1,0)      & X
        \endCD
                                                                                \tag \label{Ghg5}$$
    \vskip1mm \noindent of the group $(\Rp, \cdot)$.  The generating object
    \vskip1mm
        $$\Rgaps{1.2}\Cgaps{.8}\CD
    (P,\zy) @()\L{\zp}@(0,-1)
                & R @()\L{\zr} @(0,-1) @()\L{F} @(1,0) & \R \\
    Q  & X @()\L{\zi_X} @(-1,0) &
        \endCD
                                                                                \tag \label{Ghg6}$$
    \vskip1mm \noindent is said to be {\it homogeneous} with respect to the group
actions \Ref{Fhg1} and \Ref{Ghg5}.

    The immersed Lagrangian submanifold $A$ generated by a homogeneous generating
object is homogeneous.

        \hl1{Reduction of homogeneous generating objects}

    Reduction of a generating object
    \vskip1mm
        $$\Rgaps{1.2}\Cgaps{.8}\CD
    (P,\zy) @()\L{\zp}@(0,-1)
                & R @()\L{\zr} @(0,-1) @()\L{F} @(1,0) & \R \\
    Q  & X @()\L{\zi_X} @(-1,0) &
        \endCD
                                                                                \tag \label{Ghg7}$$
    \vskip1mm \noindent is possible if
        \list
    \item the fibration $\zr$ is the composition $\zr'' \circ \zr'$ of fibrations
$\zr' \colon R \rightarrow R'$ and $\zr'' \colon R' \rightarrow X$,
    \item the images $\widetilde{R} = \zr'(S(F,\zr'))$ and $\widetilde{X} =
\zr(S(F,\zr'))$ of the critical set $S(F,\zr')$ are submanifolds of $R'$ and
$X$ respectively,
    \item the induced mapping
        $$\align
    \widetilde{\zr} &\colon \widetilde{R} \rightarrow \widetilde{X} \\
    &\colon r' \mapsto \zr''(r')
                                                                                \tag \label{Ghg8}\endalign$$
    is a differential fibration,
    \item the mapping
        $$\align
    \overline{\zr} &\colon S(F,\zr') \rightarrow \widetilde{R} \\
    &\colon r \mapsto \zr(r)
                                                                                \tag \label{Ghg9}\endalign$$
    is a differential fibration with connected fibres.
        \endlist
    If the generating object is homogeneous with respect to group actions
        $$\bi\zm \colon \Rp \times Q \rightarrow Q
                                                                                \tag \label{Fhg6}$$
    and
    \vskip1mm
        $$\Rgaps{1.2}\Cgaps{1.2}\CD
    \Rp \times R @()\L{1_{\Rp} \times \zr}@(0,-1) @()\L{\zn}@(1,0)
                & R @()\L{\zr} @(0,-1) \\
    \Rp \times X @()\L{\zm} @(1,0)      & X     \endCD
                                                                                \tag \label{Ghg10}$$
    \vskip1mm \noindent and there is an action
        $$\zn' \colon \Rp \times R' \rightarrow R'
                                                                                \tag \label{Fhg7}$$
    such that diagrams
    \vskip1mm
        $$\Rgaps{1.2}\Cgaps{1.2}\CD
    \Rp \times R @()\L{1_{\Rp} \times \zr'}@(0,-1) @()\L{\zn}@(1,0)
                & R @()\L{\zr'} @(0,-1) \\
    \Rp \times R' @()\L{\zn'} @(1,0)      & R'
        \endCD
                                                                                \tag \label{Ghg11}$$
    \vskip1mm \noindent and
    \vskip1mm
        $$\Rgaps{1.2}\Cgaps{1.2}\CD
    \Rp \times R' @()\L{1_{\Rp} \times \zr''}@(0,-1) @()\L{\zn'}@(1,0)
                & R' @()\L{\zr''} @(0,-1) \\
    \Rp \times X @()\L{\zm} @(1,0)      & X
        \endCD
                                                                                \tag \label{Ghg12}$$
    \vskip1mm \noindent are commutative, then the result of the reduction process
is a homogeneous object.  It can be proved as in Proposition \Ref{Phf1} that
the critical set $S(F,\zr')$ is homogeneous.  It follows that $\widetilde{R}$
and $\widetilde{X}$ are homogeneous.  Consequently, we have a group action
    \vskip1mm
        $$\Rgaps{1.2}\Cgaps{1.2}\CD
    \Rp \times \widetilde{R} @()\L{1_{\Rp} \times \widetilde{\zr}}@(0,-1)
@()\L{\widetilde{\zn}}@(1,0)
                & \widetilde{R} @()\L{\widetilde{\zr}} @(0,-1) \\
    \Rp \times \widetilde{X} @()\L{\widetilde{\zm}} @(1,0)      & \widetilde{X}
        \endCD
                                                                                \tag \label{Ghg13}$$
    \vskip1mm \noindent and the function $\widetilde{F} \colon \widetilde{R}
\rightarrow \R$ satisfies the homogeneity condition
$\widetilde{F}(\widetilde{\zn}(k,r')) = k\widetilde{F}(r')$ for each $r' \in
\widetilde{R}$ and each $k \in \Rp$.  The reduced generating object
    \vskip1mm
        $$\Rgaps{1.2}\Cgaps{.8}\CD
    (P,\zy) @()\L{\zp}@(0,-1)
                & \widetilde{R} @()\L{\widetilde{\zr}} @(0,-1) @()\L{\widetilde{F}}
@(1,0) & \R \\
    Q & \widetilde{X} @()\L{\zi_{\widetilde{X}}} @(-1,0) &
        \endCD
                                                                                \tag \label{Ghg14}$$
    \vskip1mm \noindent is homogeneous.

        \hl1{Composition of homogeneous generating objects}

    Let
    \vskip1mm
        $$\Rgaps{1.2}\Cgaps{1.2}\CD
    (P_1,\zy_1) @()\L{\zp_1}@(0,-1) & & (P_2,\zy_2) @()\L{\zp_2} @(0,-1) \\
    Q_1 & & Q_2
        \endCD
                                                                                \tag \label{Ghg15}$$
    \vskip1mm \noindent be special symplectic structures and let
        $$\bi\zm_1 \colon \Rp \times Q_1 \rightarrow Q_1
                                                                                \tag \label{Fhg8}$$
    and
        $$\bi\zm_2 \colon \Rp \times Q_2 \rightarrow Q_2
                                                                                \tag \label{Fhg9}$$
    be group actions of the group $(\Rp,\cdot)$.  We consider a symplectic relation
$\zP_{21}$ from $(P_1,\zx\zy_1)$ to $(P_1,\zx\zy_1)$, whose graph $A_{21}$ is
generated by a generating object
    \vskip1mm
        $$\rgaps{1.2}\cgaps{.8}\CD
    (P_2 \times P_1,\zy_2 \ominus \zy_1) @()\L{\zp_2 \times \zp_1}@(0,-1)
                & R_{21} @()\L{\zr_{21}} @(0,-1) @()\L{F_{21}} @(1,0) & \R \\
    Q_2 \times Q_1  & X_{21} @()\L{\zi_{X_{21}}} @(-1,0) &
        \endCD
                                                                                \tag \label{Ghg16}$$
    \vskip1mm \noindent homogeneous with respect to a group action
    \vskip1mm
        $$\Rgaps{1.2}\Cgaps{1.2}\CD
    \Rp \times R_{21} @()\L{1_{\Rp} \times \zr_{21}}@(0,-1) @()\L{\zn_{21}}@(1,0)
                & R_{21} @()\L{\zr_{21}} @(0,-1) \\
    \Rp \times X_{21} @()\L{\zm_{21}} @(1,0)      & X_{21}      \endCD
                                                                                \tag \label{Ghg17}$$
    \vskip1mm \noindent where
        $$\zm_{21} \colon \Rp \times X_{21} \rightarrow X_{21}
                                                                                \tag \label{Fhg10}$$
    is the restriction of the action
        $$\align
    \bi\zm_{21} &\colon \Rp \times Q_2 \times Q_1 \rightarrow Q_2 \times Q_1 \\
    &\colon (k,q_2,q_1) \mapsto (\bi\zm_2(k,q_2),\bi\zm_1(k,q_1))
                                                                                \tag \label{Fhg11}\endalign$$
    to the submanifold $X_{21}$.  Let $A_1$ be a Lagrangian submanifold of
$(P_1,\xd\zy_1)$ generated by a generating object
    \vskip1mm
        $$\rgaps{1.2}\cgaps{.8}\CD
    (P_1,\zy_1) @()\L{\zp_1}@(0,-1)
                & R_1 @()\L{\zr_1} @(0,-1) @()\L{F_1} @(1,0) & \R \\
    Q_1  & X_1 @()\L{\zi_{X_1}} @(-1,0) &
        \endCD
                                                                                \tag \label{Ghg18}$$
    \vskip1mm \noindent homogeneous with respect to a group action
    \vskip1mm
        $$\Rgaps{1.2}\Cgaps{1.2}\CD
    \Rp \times R_1 @()\L{1_{\Rp} \times \zr_1}@(0,-1) @()\L{\zn_1}@(1,0)
                & R_1 @()\L{\zr_1} @(0,-1) \\
    \Rp \times X_1 @()\L{\zm_1} @(1,0)      & X_1       \endCD
                                                                                \tag \label{Ghg19}$$
    \vskip1mm \noindent where
        $$\zm_1 \colon \Rp \times X_1 \rightarrow X_1
                                                                                \tag \label{Fhg12}$$
    is the restriction of the action $\bi\zm_1$ to $X_1$.

\def\oT{\overset\circ\to\sT}

        \hl1{Homogeneous Lagrangian systems}

    Let $Q$ be a differential manifold and let $\zk$ be the action
        $$\align
    \bi\zk &\colon \Rp \times \sT Q \rightarrow \sT Q \\
    &\colon (k,v) \mapsto kv.
                                                                                \tag \label{Fhl1}\endalign$$
    Let
        $$\oT Q = \{v \in \sT Q;\; v \neq 0 \}
                                                                                \tag \label{Fhl2}$$
    be the tangent bundle of $Q$ with the image of the zero section removed.  The set
$\oT Q$ is an homogeneous open submanifold of $\sT Q$.

    A Lagrangian system
    \vskip1mm
        $$\Rgaps{1.2}\Cgaps{.8}\CD
    (\sT\sT^\* Q,\xd_T\zy_Q) @()\L{\sT\zp_Q}@(0,-1)
                & Y @()\L{\zh} @(0,-1) @()\L{L} @(1,0) & \R \\
    \sT Q  & C @()\L{\zi_C} @(-1,0) &
        \endCD
                                                                                \tag \label{Ghl1}$$
    \vskip1mm \noindent is said to be {\it homogeneous} with respect to an action
    \vskip1mm
        $$\Rgaps{1.2}\Cgaps{1.2}\CD
    \Rp \times Y @()\L{1_{\Rp} \times \zh}@(0,-1) @()\L{\zr}@(1,0)
                & Y @()\L{\zh} @(0,-1) \\
    \Rp \times C @()\L{\zk} @(1,0)      & C
        \endCD
                                                                                \tag \label{Ghl2}$$
    \vskip1mm \noindent if $C$ is an homogeneous submanifold of $\oT Q$,
        $$\align
    \zk &\colon \Rp \times C \rightarrow C \\
    &\colon (k,v) \mapsto kv.
                                                                                \tag \label{Fhl3}\endalign$$
    is the restriction of the action $\bi\zk$ to $C$, and
        $$L(\zr(k,y)) = kL(y)
                                                                                \tag \label{Fhl4}$$
    for each $y \in Y$ and each $k \in \Rp$.  The Lagrangian family
    \vskip1mm
        $$\Rgaps{1.2}\Cgaps{.8}\CD
            Y @()\L{\zh} @(0,-1) @()\L{L} @(1,0) & \R \\  C &
        \endCD
                                                                                \tag \label{Ghl3}$$
    \vskip1mm \noindent is said to be a {\it homogeneous Lagrangian family}.

    The action $\bi\zk$ is lifted to the action
    \vskip1mm
        $$\Rgaps{1.2}\Cgaps{1.2}\CD
    \Rp \times \sT\sT^\* Q @()\L{1_{\Rp} \times \sT\zp_Q}@(0,-1) @()\L{\widehat{\zk}}@(1,0)
                & \sT\sT^\* Q @()\L{\sT\zp_Q} @(0,-1) \\
    \Rp \times \sT Q @()\L{\bi\zk} @(1,0)      & \sT Q
        \endCD
                                                                                \tag \label{Ghl4}$$
    \vskip1mm \noindent with the action
        $$\widehat{\zk} \colon \Rp \times \sT\sT^\* Q \rightarrow \sT\sT^\* Q
                                                                                \tag \label{Fhl5}$$
    defined by
        $$\widehat{\zk}(k,\cdot) = \za_Q^{-1} \circ \overline{\bi\zk}(k,\cdot) \circ
\za_Q
                                                                                \tag \label{Fhl6}$$
     for each $k \in \Rp$.  The relation
        $$\widehat{\zk}(k,w) = kw
                                                                                \tag \label{Fhl7}$$
    holds for each $w \in \sT\sT^\*Q$ and each $k \in \Rp$.

    The dynamics $D$ generated by a homogeneous Lagrangian system is homogeneous.

        \hl1{Homogeneous Hamiltonian systems}

    Let $(P,\zw)$ be a symplectic manifold representing the phase space of a
mechanical system and let
    \vskip1mm
        $$\Rgaps{1.2}\Cgaps{.8}\CD
    (\sT P,\xi_T\zw) @()\L{\zt_P}@(0,-1)
                & Z @()\L{\zz} @(0,-1) @()\L{-H} @(1,0) & \R \\
    P  & K @()\L{\zi_K} @(-1,0) &
        \endCD
                                                                                \tag \label{Ghh1}$$
    \vskip1mm \noindent be a Hamiltonian system.  Let
        $$\zs \colon \Rp \times Z \rightarrow Z
                                                                                \tag \label{Fhh1}$$
    be an action of the group $(\Rp,\cdot)$ such that $\zz \circ \zs(k,\cdot) = \zz$
for each $k \in \Rp$.  The Hamiltonian system \Ref{Ghh1} is said to be {\it
homogeneous} with respect to the group action $\zs$ if
        $$H(\zs(k,z)) = kH(z)
                                                                                \tag \label{Fhh2}$$
    for each $z \in Z$ and each $k \in \Rp$.  The Hamiltonian family
    \vskip1mm
        $$\Rgaps{1.2}\Cgaps{.8}\CD
            Z @()\L{\zz} @(0,-1) @()\L{H} @(1,0) & \R \\  K &
        \endCD
                                                                                \tag \label{Ghh2}$$
    \vskip1mm \noindent is said to be a {\it homogeneous Hamiltonian family}.

    The same concepts of homogeneity apply to a Hamiltonian system
    \vskip1mm
        $$\Rgaps{1.2}\Cgaps{.8}\CD
    (\sT\sT^\*Q,\xi_T\xd\zy_Q) @()\L{\zt_{\sT^\*Q}}@(0,-1)
                & Z @()\L{\zz} @(0,-1) @()\L{-H} @(1,0) & \R \\
    \sT^\*Q  & K @()\L{\zi_K} @(-1,0) &
        \endCD
                                                                                \tag \label{Ghh3}$$
    \vskip1mm \noindent based on a phase space $(\sT^\*Q,\xd\zy_Q)$.

    The dynamics $D$ generated by a homogeneous Hamiltonian system is homogeneous.

        \hl1{Homogeneous Legendre transformations}

    The generating objects
    \vskip1mm
        $$\Rgaps{1.2}\Cgaps{.9}\CD
    (\sT\sT^\* Q \times \sT\sT^\* Q,\xi_T\xd\zy_Q \ominus \xd_T\zy_Q)
@()\L{\zt_{\sT^\*Q} \times \sT\zp_Q}@(0,-1) & &  \\
    \sT^\*Q \times \sT Q  & \sT^\*Q \times_Q \sT Q @()\L{\zi_{\sT^\*Q \times_Q \sT Q}}
@(-1,0) @()\L{-\langle \,,\rangle}@(1,0) & \R
        \endCD
                                                                                \tag \label{Ght1}$$
    \vskip1mm \noindent and
    \vskip1mm
        $$\Rgaps{1.2}\Cgaps{.9}\CD
    (\sT\sT^\* Q \times \sT\sT^\* Q,\xd_T\zy_Q \ominus \xi_T\xd\zy_Q)
@()\L{\sT\zp_Q \times \zt_{\sT^\*Q}}@(0,-1) & &  \\
    \sT Q \times \sT^\*Q & \sT Q \times_Q \sT^\*Q @()\L{\zi_{\sT Q \times_Q \sT^\*Q}}
@(-1,0) @()\L{\langle \,,\rangle^\sim}@(1,0) & \R
        \endCD
                                                                                \tag \label{Ght2}$$
    \vskip1mm \noindent are homogeneous with respect to the group action
        $$\align
    \bi\zk &\colon \Rp \times \sT Q \rightarrow \sT Q \\
    &\colon (k,v) \mapsto kv
                                                                                \tag \label{Fht1}\endalign$$
    combined with the trivial action of $\Rp$ in $\sT^\*Q$ since
        $$\langle p, kv \rangle = k\langle p, v \rangle
                                                                                \tag \label{Fht2}$$
    for each $k \in \Rp$.  It follows that a homogeneous Lagrangian system will be
transformed by the Legendre transformation in a homogeneous Hamiltonian
system and a homogeneous Hamiltonian system will be transformed by the
inverse Legendre transformation in a homogeneous Lagrangian system.  The
graphs of Legendre relations and inverse Legendre relations are homogeneous
with respect to the relevant group actions.
\def\oV{\overset \,\circ \to V}
\def\oT{\overset\circ\to\sT}

        \hl1{Examples of homogeneous systems}
    Two examples of homogeneous system in the affine space-time of special
relativity will be discussed.

    An {\it affine space} is a triple $(Q,V,\zs)$, where $Q$ is a set, $V$ is a real
vector space of finite dimension and $\zs$ is a mapping $\zs \colon Q \times
Q \rightarrow V$ such that
        \list
    \item $\zs(q_3,q_2) + \zs(q_2,q_1) + \zs(q_1,q_3) = 0$;
    \item the mapping $\zs(\cdot,q) \colon Q \rightarrow V$ is bijective for each $q
\in Q$.
        \endlist
    The tangent bundle $\sT Q$ of an affine space is identified with the product $Q
\times V$ and the cotangent bundle $\sT^\*Q$ is identified with $Q \times
V^\*$. The space $\oT Q$ is identified with $Q \times \oV$, where
        $$\oV = \left\{v \in V;\; v \neq 0 \right\}
                                                                                \tag \label{Fxx1}$$
Spaces $\sT\sT^\*Q$ and $\sT\sT\sT^\*Q$ are identified with $Q \times V^\*
\times V \times V^\*$ and $Q \times V^\* \times V \times V^\* \times V \times
V^\* \times V \times V^\*$ respectively.  The evaluations of the forms
$\xd_T\zy$ and $\xi_T\xd\zy$ on an element $(q,p,\dot q,\dot p,\zd q,\zd
p,\zd\dot q,\zd\dot p)$ is represented by
        $$\langle \xd_T\zy, (q,p,\dot q,\dot p,\zd q,\zd p,\zd\dot q,\zd\dot p)
\rangle = \langle \dot p, \zd q \rangle + \langle p, \zd\dot q \rangle
                                                                                \tag \label{Fxx2}$$
    and
        $$\langle \xi_T\xd\zy, (q,p,\dot q,\dot p,\zd q,\zd p,\zd\dot q,\zd\dot p)
\rangle = \langle \dot p, \zd q \rangle - \langle \zd p, \dot q \rangle.
                                                                                \tag \label{Fxx3}$$

        \claim \c{e}{Example}{\!\!\!. Dynamics of a relativistic particle}{\rm       \label{Exx1}
    Let $(Q,V,\zs,g)$ be the affine space-time of special relativity. The triple
$(Q,V,\zs)$ is an affine space and $g \colon V \rightarrow V^\*$ is the
Minkowski metric.  The dynamics of a free particle of mass $m$ is generated
by the Lagrangian system
    \vskip1mm
        $$\Rgaps{1.2}\Cgaps{.8}\CD
    (\sT\sT^\* Q,\xd_T\zy_Q) @()\L{\sT\zp_Q}@(0,-1) & & \\
    \sT Q  & C @()\L{\zi_C} @(-1,0) @()\L{L} @(1,0) & \R
        \endCD
                                                                                \tag \label{Gxx1}$$
    \vskip1mm \noindent where
        $$C = Q \times \left\{\dot q \in V;\; \langle g(\dot q), \dot q \rangle > 0 \right\}
                                                                                \tag \label{Fxx4}$$
    and
        $$\align
    L &\colon C \rightarrow \R \\
    &\colon (q,\dot q) \mapsto m\sqrt{\langle g(\dot q), \dot q \rangle}.
                                                                                \tag \label{Fxx5}\endalign$$
    The dynamics is the set
        $$D = \left\{(q,p,\dot q,\dot p) \in \sT\sT^\*Q ;\; \langle g(\dot q), \dot
q \rangle > 0, p = \frac{mg(\dot q)}{\sqrt{\langle g(\dot q), \dot q
\rangle}}, \dot p = 0 \right\}.
                                                                                \tag \label{Fxx6}$$

    The set $C$ is a homogeneous submanifold of $\oT Q$ and $L(q,k\dot q) = kL(q,\dot q)$.  It
follows that the Lagrangian system \Ref{Gxx1} is homogeneous.

    The Legendre transformation applied to the Lagrangian system \Ref{Gxx1} leads to a
Hamiltonian system
    \vskip1mm
        $$\Rgaps{1.2}\Cgaps{.8}\CD
    (\sT\sT^\*Q,\xi_T\xd\zy_Q) @()\L{\zt_{\sT^\*Q}}@(0,-1)
                & Z @()\L{\zz} @(0,-1) @()\L{-H} @(1,0) & \R \\
    \sT^\*Q  & \sT^\*Q  @()\a= @(-1,0) &
        \endCD
                                                                                \tag \label{Gxx2}$$
    \vskip1mm \noindent with
        $$Z = \sT^\*Q \times \left\{v \in V;\; \langle g(v), v
\rangle > 0 \right\},
                                                                                \tag \label{Fxx7}$$
        $$\align
    \zz &\colon Z \rightarrow \sT^\*Q \\
    &\colon (q,p,v) \mapsto (q,p),
                                                                                \tag \label{Fxx8}\endalign$$
    and
        $$\align
    H &\colon Z \rightarrow \R \\
    &\colon (q,p,v) \mapsto \langle p, v \rangle - m\sqrt{\langle g(v), v
\rangle}.
                                                                                \tag \label{Fxx9}\endalign$$

    The Hamiltonian system \Ref{Gxx2} can be simplified.  The fibration $\zz$ can be
represented as a composition $\zz'' \circ \zz'$ of fibrations
        $$\align
    \zz' &\colon Z \rightarrow Z' \\
    &\colon (q,p,v) \mapsto (q,p,\|v\|),
                                                                                \tag \label{Fxx10}\endalign$$
    and
        $$\align
    \zz'' &\colon Z' \rightarrow \sT^\*Q \\
    &\colon (q,p,\zl) \mapsto (q,p),
                                                                                \tag \label{Fxx11}\endalign$$
    where
        $$Z' = \sT^\*Q \times \Rp
                                                                                \tag \label{Fxx12}$$
    and
        $$\|v\| = \sqrt{\langle g(v), v \rangle}.
                                                                                \tag \label{Fxx13}$$
    Equating to zero the derivative of $H$ along the fibres of $\zz'$ we obtain the
relation
        $$p = \zm g(v)
                                                                                \tag \label{Fxx14}$$
    valid for some values of the Lagrange multiplier $\zm$.  It follows from this
relation that the covector $p$ is in the set
        $$\left\{p \in V^\*;\; \langle p, g^{-1}(p) \rangle > 0 \right\}
                                                                                \tag \label{Fxx15}$$
    and that
        $$\left\{(q,p,v) \in Z;\; \|v\|p = \pm \|p\|g(v) \right\}
                                                                                \tag \label{Fxx16}$$
    is the critical set $S(H,\zz')$.  We have denoted by $\|p\|$ the norm
$\sqrt{\langle p, g^{-1}(p) \rangle }$ of $p$.  The images $\widetilde K =
\zz(S(H,\zz'))$ and $\widetilde Z = \zz'(S(H,\zz'))$ of the critical set
$S(H,\zz')$ are the sets
        $$\widetilde K = Q \times \left\{p \in V^\*;\; \langle p,g^{-1}(p) \rangle >
0 \right\}
                                                                                \tag \label{Fxx17}$$
    and
        $$\widetilde Z = \widetilde K \times \Rp.
                                                                                \tag \label{Fxx18}$$
    The mapping
        $$\align
    \widetilde \zz &\colon \widetilde Z \rightarrow \widetilde K \\
    &\colon (q,p,\zl) \mapsto (q,p).
                                                                                \tag \label{Fxx19}\endalign$$
    is a differential fibration.  The critical set $S(H,\zz')$ is the union of
images of two local sections
        $$\align
    \zx_+ &\colon \widetilde Z \rightarrow Z \\
    &\colon (q,p,\zl) \mapsto \left(q,p,\frac{\zl}{\|p\|} g^{-1}(p)\right)
                                                                                \tag \label{Fxx20}\endalign$$
    and
        $$\align
    \zx_- &\colon \widetilde Z \rightarrow Z \\
    &\colon (q,p,\zl) \mapsto \left(q,p,-\frac{\zl}{\|p\|} g^{-1}(p)\right)
                                                                                \tag \label{Fxx21}\endalign$$
    of the fibration $\zz'$.  We have two families of functions
    \vskip1mm
        $$\Rgaps{1.2}\Cgaps{.8}\CD
            \widetilde Z @()\L{\widetilde\zz} @(0,-1) @()\L{\widetilde H_-} @(1,0) &
\R & & & \widetilde Z @()\L{\widetilde\zz} @(0,-1) @()\L{\widetilde H_+}
@(1,0) & \R \\ \widetilde K & & & & \widetilde K &
        \endCD
                                                                                \tag \label{Gxx3}$$
    \vskip1mm\noindent The functions
        $$\align
    \widetilde H_- &\colon \widetilde Z \rightarrow \R \\
    &\colon (q,p,\zl) \mapsto - \zl \left(\|p\| + m\right)
                                                                                \tag \label{Fxx22}\endalign$$
    and
        $$\align
    \widetilde H_+ &\colon \widetilde Z \rightarrow \R \\
    &\colon (q,p,\zl) \mapsto \zl \left(\|p\| - m\right)
                                                                                \tag \label{Fxx23}\endalign$$
    are the compositions $H \circ \zx_-$ and $H \circ \zx_+$ respectively.  It is
easily seen that the family \Ref{Gxx3} generates an empty set.  The dynamics
of the particle is generated by the reduced Hamiltonian system
    \vskip1mm
        $$\Rgaps{1.2}\Cgaps{1}\CD
    (\sT\sT^\*Q,\xi_T\xd\zy_Q) @()\L{\zt_{\sT^\*Q}}@(0,-1)
                & \widetilde Z @()\L{\widetilde\zz} @(0,-1) @()\L{-\widetilde H_+} @(1,0) & \R \\
    \sT^\*Q  & \widetilde K @()\L{\zi_{\widetilde K}} @(-1,0) &
        \endCD
                                                                                \tag \label{Gxx4}$$
    \vskip1mm

    The critical set $S(\widetilde H_+,\widetilde \zz)$ is the submanifold
        $$\left\{(q,p,\zl) \in \widetilde Z;\; \|p\| = m \right\}
                                                                                \tag \label{Fxx24}$$
    and the image $\widetilde\zz(S(\widetilde H_+,\widetilde \zz))$ is the mass shell
        $$K_m = \left\{(q,p) \in \widetilde K;\; \|p\| = m \right\}.
                                                                                \tag \label{Fxx25}$$
    The mapping
        $$\align
    \widehat\zz &\colon S(\widetilde H_+,\widetilde \zz) \rightarrow K_m \\
    &\colon (q,p,\zl) \mapsto (q,p)
                                                                                \tag \label{Fxx26}\endalign$$
    is a differential fibration with connected fibres.  It follows that the dynamics
$D$ is included in the Lagrangian submanifold $\widehat D \subset \sT\sT^\*Q$
generated by the Dirac system
    \vskip1mm
        $$\Rgaps{1.2}\Cgaps{1}\CD
    (\sT\sT^\*Q,\xi_T\xd\zy_Q) @()\L{\zt_{\sT^\*Q}}@(0,-1) & & \\
    \sT^\*Q  & K_m @()\L{\zi_{K_m}} @(-1,0) @()\L{0} @(1,0) & \R
        \endCD
                                                                                \tag \label{Gxx5}$$
    \vskip1mm  The dynamics of the particle is not the same as the Lagrangian
submanifold $\widehat D$.  Hence, no further reduction of the Hamiltonian
system \Ref{Gxx4} is possible.

    The application of the inverse Legendre transformation to the Hamiltonian system
\Ref{Gxx4} results in the Lagrangian system
    \vskip1mm
        $$\Rgaps{1.2}\Cgaps{.8}\CD
    (\sT\sT^\* Q,\xd_T\zy_Q) @()\L{\sT\zp_Q}@(0,-1)
                & \widetilde Y @()\L{\zh} @(0,-1) @()\L{\widetilde L} @(1,0) & \R \\
    \sT Q  & \widetilde C @()\L{\zi_{\widetilde C}} @(-1,0) &
        \endCD
                                                                                \tag \label{Gxx6}$$
    \vskip1mm \noindent with
        $$\widetilde C = \oT Q,
                                                                                \tag \label{Fxx27}$$
        $$\widetilde Y = \left\{(q,\dot q,r,\zl) \in Q \times \oV \times V^\* \times
\Rp;\; \langle r,g^{-1}(r) \rangle > 0 \right\}
                                                                                \tag \label{Fxx28}$$
        $$\align
    \widetilde \zh &\colon \widetilde Y \rightarrow \widetilde C \\
    &\colon (q,\dot q,r,\zl) \mapsto (q,\dot q),
                                                                                \tag \label{Fxx29}\endalign$$
    and
        $$\align
    \widetilde L &\colon \widetilde Y \rightarrow \R \\
    &\colon (q,\dot q,r,\zl) \mapsto \langle r, \dot q \rangle - \zl(\|r\| - m).
                                                                                \tag \label{Fxx30}\endalign$$
    The image $\widetilde\zh(S(\widetilde L,\widetilde\zh))$ of the critical set
        $$S(\widetilde L,\widetilde\zh) = \left\{(q,\dot q,r,\zl) \in \widetilde
Y;\; \|r\| = m, m\dot q = \zl g^{-1}(r) \right\}
                                                                                \tag \label{Fxx31}$$
    is the set
        $$C = Q \times \left\{\dot q \in V;\; \langle g(\dot q), \dot q \rangle > 0 \right\}.
                                                                                \tag \label{Fxx32}$$
    The critical set is the image of the local section
        $$\align
    \zx &\colon C \rightarrow \widetilde Y \\
    &\colon (q,\dot q) \mapsto \left(q,\dot q,\frac{m}{\|\dot q\|}g(\dot q),\|\dot q\|\right)
                                                                                \tag \label{Fxx33}\endalign$$
    The composition $\widetilde L \circ \zx$ is the function
        $$\align
    L &\colon C \rightarrow \R \\
    &\colon (q,\dot q) \mapsto m\sqrt{\langle g(\dot q), \dot q \rangle}.
                                                                                \tag \label{Fxx34}\endalign$$
    The reduced Lagrange system is the original Lagrange system \Ref{Gxx1}.
        }\endclaim

    The Hamiltonian systems \Ref{Gxx2} and \Ref{Gxx4} are homogeneous.  The graph of
the second Legendre relation $\zL_2(L)$ is the homogeneous set
        $$\gr(\zL_2(L)) = \left\{(q,p,q',\dot q) \in \sT^\*Q \times \sT Q;\; q' = q,
p = \frac{mg(\dot q)}{\sqrt{\langle g(\dot q), \dot q \rangle}} \right\}.
                                                                                \tag \label{Fxx35}$$

        \claim \c{e}{Example}{\!\!. Space-time formulation of geometric optics}{\rm  \label{Exx2}
    The dynamics of light rays in affine Minkowski space-time $(Q,V,\zs,g)$ is
generated by the Lagrangian system
    \vskip1mm
        $$\Rgaps{1.2}\Cgaps{.8}\CD
    (\sT\sT^\* Q,\xd_T\zy_Q) @()\L{\sT\zp_Q}@(0,-1)
                & Y @()\L{\zh} @(0,-1) @()\L{L} @(1,0) & \R \\
    \sT Q  & C @()\L{\zi_C} @(-1,0) &
        \endCD
                                                                                \tag \label{Gxx7}$$
    \vskip1mm \noindent with
        $$C = \oT Q,
                                                                                \tag \label{Fxx36}$$
        $$Y = C \times \Rp,
                                                                                \tag \label{Fxx37}$$
        $$\align
    \zh &\colon Y \rightarrow C \\
    &\colon (q,\dot q,\zm) \mapsto (q,\dot q),
                                                                                \tag \label{Fxx38}\endalign$$
    and
        $$\align
    L &\colon Y \rightarrow \R \\
    &\colon (q,\dot q,\zm) \mapsto \frac{1}{2\zm} \langle g(\dot q), \dot q \rangle.
                                                                                \tag \label{Fxx39}\endalign$$
    The dynamics is the set
        $$D = \left\{(q,p,\dot q,\dot p) \in \sT\sT^\*Q ;\; \langle g(\dot q), \dot
q \rangle = 0, \exi{\zm \in \Rp} p = \frac{1}{\zm} g(\dot q), \dot p = 0
\right\}.
                                                                                \tag \label{Fxx40}$$
    The Lagrangian system \Ref{Gxx7} is homogeneous.

    The result of the Legendre transformation applied to the Lagrangian system
\Ref{Gxx7} is the Hamiltonian system
    \vskip1mm
        $$\Rgaps{1.2}\Cgaps{.8}\CD
    (\sT\sT^\*Q,\xi_T\xd\zy_Q) @()\L{\zt_{\sT^\*Q}}@(0,-1)
                & Z @()\L{\zz} @(0,-1) @()\L{-H} @(1,0) & \R \\
    \sT^\*Q  & \sT^\*Q  @()\a= @(-1,0) &
        \endCD
                                                                                \tag \label{Gxx8}$$
    \vskip1mm \noindent with
        $$Z = \sT^\*Q \times \oV \times \Rp,
                                                                                \tag \label{Fxx41}$$
        $$\align
    \zz &\colon Z \rightarrow \sT^\*Q \\
    &\colon (q,p,v,\zm) \mapsto (q,p),
                                                                                \tag \label{Fxx42}\endalign$$
    and
        $$\align
    H &\colon Z \rightarrow \R \\
    &\colon (q,p,v,\zm) \mapsto \langle p, v \rangle - \frac{1}{2\zm} \langle g(v), v \rangle.
                                                                                \tag \label{Fxx43}\endalign$$

    The Hamiltonian system \Ref{Gxx8} can be simplified as in the case of a
relativistic particle with mass $m > 0$.  The fibration $\zz$ is be
represented as a composition $\zz'' \circ \zz'$ of fibrations
        $$\align
    \zz' &\colon Z \rightarrow Z' \\
    &\colon (q,p,v,\zm) \mapsto (q,p,\zm),
                                                                                \tag \label{Fxx44}\endalign$$
    and
        $$\align
    \zz'' &\colon Z' \rightarrow \sT^\*Q \\
    &\colon (q,p,\zm) \mapsto (q,p),
                                                                                \tag \label{Fxx45}\endalign$$
    where
        $$Z' = \sT^\*Q \times \Rp.
                                                                                \tag \label{Fxx46}$$
    We have
        $$S(H,\zz') = \left\{(q,p,v,\zm) \in Z;\; \zm p = g(v) \right\}
                                                                                \tag \label{Fxx47}$$
    and
        $$\widetilde Z = \zz'(S(H,\zz')) = \left\{(q,p,\zm) \in Z';\; p \neq 0 \right\}.
                                                                                \tag \label{Fxx48}$$
    The critical set $S(H,\zz')$ is the image of the local section
        $$\align
    \zx &\colon \widetilde Z \rightarrow Z \\
    &\colon (q,p,\zm) \mapsto (q,p,\zm g^{-1}(p),\zm)
                                                                                \tag \label{Fxx49}\endalign$$
    of $\zz'$.  The reduced system is the Hamiltonian system
    \vskip1mm
        $$\Rgaps{1.2}\Cgaps{1}\CD
    (\sT\sT^\*Q,\xi_T\xd\zy_Q) @()\L{\zt_{\sT^\*Q}}@(0,-1)
                & \widetilde Z @()\L{\widetilde\zz} @(0,-1) @()\L{-\widetilde H} @(1,0) & \R \\
    \sT^\*Q  & \widetilde K @()\L{\zi_{\widetilde K}} @(-1,0) &
        \endCD
                                                                                \tag \label{Gxx9}$$
    \vskip1mm \noindent with
        $$\widetilde K = \left\{(q,p) \in \sT^\*Q;\; p \neq 0 \right\},
                                                                                \tag \label{Fxx50}$$
        $$\align
    \widetilde\zz &\colon \widetilde Z \rightarrow \widetilde K \\
    &\colon (q,p,\zm) \mapsto (q,p),
                                                                                \tag \label{Fxx51}\endalign$$
    and
        $$\align
    \widetilde H &\colon \widetilde Z \rightarrow \R \\
    &\colon (q,p,\zm) \mapsto \frac{\zm}{2} \langle p, g^{-1}(p) \rangle.
                                                                                \tag \label{Fxx52}\endalign$$
    No further simplification is possible.

    The application of the inverse Legendre transformation to the Hamiltonian system
\Ref{Gxx9} results in the original Lagrangian system \Ref{Gxx7}.

    The Hamiltonian systems \Ref{Gxx8} and \Ref{Gxx9} are homogeneous.  The graph of
the first Legendre relation $\zL_1(L)$ is the homogeneous set
        $$\gr(\zL_1(L)) = \left\{(q,p,q',\dot q,\zm) \in \sT^\*Q \times Y;\; q' = q,
\langle g(\dot q), \dot q \rangle = 0, \zm p = g(\dot q) \right\}.
                                                                                \tag \label{Fxx53}$$
    The graph of the second Legendre relation $\zL_2(L)$ is the set
        $$\gr(\zL_2(L)) = \left\{(q,p,q',\dot q) \in \sT^\*Q \times \sT Q;\; q' = q,
\langle g(\dot q), \dot q \rangle = 0,\exi{\zm \in \Rp} \zm p = g(\dot q)
\right\}.
                                                                                \tag \label{Fxx54}$$
    The graph of the first inverse Legendre relation $\zW_1(\widetilde{H})$ is the
set
        $$\gr(\zW_1(\widetilde{H})) = \left\{(q,\dot q',p,\zm) \in \sT Q \times
\widetilde Z;\; q' = q, \langle p, g^{-1}(p) \rangle = 0, \dot q = \zm
g^{-1}(p) \right\}.
                                                                                \tag \label{Fxx55}$$
        }\endclaim

      \hl1{References}
    \noindent [1] P. A. M. Dirac, {\it Generalized Hamiltonian Dynamics}, Canad. J.
Math., {\bf 2} (1950), 129--148.
    \vskip3mm
    \noindent [2] W. M. Tulczyjew, {\it Hamiltonian systems, Lagrangian systems and
the Legendre transformation}, Symposia Mathematica, {\bf 14} (1974),
247--258.
    \vskip3mm
    \noindent [3] W. M. Tulczyjew, {\it The Legendre transformation}, Ann. Inst.
Henri Poincar\'e, {\bf 27} (1977), 101--114.

\enddocument